\documentclass{iopart}

\usepackage{harvard}
\usepackage{iopams} 
\usepackage{graphicx}
\usepackage[export]{adjustbox}
\usepackage{epstopdf} 
\usepackage{caption}
\usepackage[caption=false,font=normalsize,labelfont=sf,textfont=sf]{subfig}
\usepackage{array}
\newcolumntype{C}[1]{>{\centering\arraybackslash}m{#1}}

\usepackage{setspace} 
\usepackage{color}
\usepackage{cite}
\usepackage{algorithm}
\usepackage{algpseudocode}
\usepackage{algcompatible}

\expandafter\let\csname equation*\endcsname\relax 
\expandafter\let\csname endequation*\endcsname\relax 
\usepackage{amsmath}

\usepackage{bbm}
\usepackage{upgreek}

\usepackage{amsfonts,amssymb}

\usepackage{hyperref}

\begin{document}
\title[DECT with Limited-Angular-Range Data]{Dual-Energy CT Imaging with Limited-Angular-Range Data}

\author{
Buxin~Chen$^1$,
Zheng~Zhang$^1$,
Dan~Xia$^1$,
Emil~Y.~Sidky$^1$,
and~Xiaochuan~Pan$^{1,2}$,}

\address{$^1$ Department of Radiology, The University of Chicago, Chicago, IL 60637, USA} \address{$^2$ Department of Radiation and Cellular Oncology, The University of Chicago,
Chicago, IL 60637, USA} 

\ead{xpan@uchicago.edu}

\begin{abstract}
{\color{black} In dual-energy computed tomography (DECT), low- and high-kVp data are collected often over a full-angular range (FAR) of $360^\circ$. While there exists strong interest in DECT with low- and high-kVp data acquired over limited-angular ranges (LARs), there remains little investigation of image reconstruction in DECT with LAR data. {\it Objective}: We investigate image reconstruction with minimized LAR artifacts from low- and high-kVp data over LARs of $\le 180^\circ$ by using a directional-total-variation (DTV) algorithm.} {\it Methods}: Image reconstruction from LAR data is formulated as a convex optimization problem in which data-$\ell_2$ is minimized with constraints on image's DTVs along orthogonal axes. We then achieve image reconstruction by applying the DTV algorithm to solve the optimization problem. We conduct numerical studies from data generated over arcs of LARs, ranging from $14^\circ$ to $180^\circ$, and perform visual inspection and quantitative analysis of images reconstructed. {\it Results}: Monochromatic images of interest obtained with the DTV algorithm from LAR data show substantially reduced artifacts that are observed often in images obtained with existing algorithms. The improved image quality also leads to accurate estimation of physical quantities of interest, such as effective atomic number and iodine-contrast concentration. {\it Conclusion}: Our study reveals that {\color{black} from LAR data of low- and high-kVp, monochromatic images can be obtained that are visually, and physical quantities can be estimated that are quantitatively}, comparable to those obtained in FAR DECT. {\it Significance}: {\color{black} As LAR DECT} is of high practical application interest, the results acquired in the work may engender insights into the design of DECT with LAR scanning configurations of practical application significance.
\end{abstract}


\submitto{\PMB}

\maketitle

\section{Introduction} \label{sec:intro}
In dual-energy CT (DECT), data are collected with low- and high-kVp spectra over a full-scan (or at least a short-scan) range~\cite{alvarez_energy-selective_1976}; and images are reconstructed then directly or indirectly from low- and high-kVp data often by use of algorithms that are developed basing upon a linear-data model in conventional CT. The images reconstructed are used subsequently for estimation of basis images from which monochromatic images at given energies, and physical quantities of application interest, such as effective atomic numbers (simply referred to as atomic numbers hereinafter) and iodine concentrations, can be estimated~\cite{johnson_material_2007,maass_image-based_2009,goodsitt2011accuracies,chandarana2011iodine,faby2015performance,chen2018algorithm}. 

DECT capable of limited-angular-range (LAR) imaging is of practical application interest because it may allow for the reduction of radiation dose and scanning time and for the design of scanning configurations avoiding possible collisions of the moving gantry, e.g., in a C-arm DECT, with patient or other components involved in the scanning. While image reconstruction is a key to enabling LAR DECT, it remains largely unexplored as only limited effort has been reported in the literature for highly special scanning configurations in which the sum of low- and high-kVp angular ranges is generally larger than $180^\circ$ \cite{zhang2016reconstruction,zhang2020reconstruction}.

In this work, {\color{black} we investigate image reconstructions in LAR DECT by exploiting the directional-total-variation (DTV) algorithm~\cite{zhang2021dtv} recently developed for image reconstruction in conventional CT with LAR data}~\cite{batenburg2011dart,liu2016cooperative,xu2019image,zhang2021dtv}. In the investigation, for each of the low- and high-kVp data sets collected over LARs in DECT, we first formulate image reconstruction as a convex optimization problem designed in which data-$\ell_2$ is minimized under image's DTV constraints along orthogonal axes, and then use the DTV algorithm developed recently to solve the optimization problem for achieving image reconstructions. 
The DTV algorithm may allow for an efficient recovery of ``invisible boundaries''~\cite{quinto2017artifacts} along the scanning direction in reconstructed images from the dual-energy LAR data.

{\color{black}In LAR DECT, while images suffer from both LAR and beam-hardening (BH) artifacts, the LAR artifact is dominantly more significant than the BH artifact, as results in Sec.~\ref{sec:rslt} below show. The work focuses thus on LAR-artifact correction in LAR DECT without explicit BH-artifact correction. In particular, results of LAR DECT obtained are compared against those of the full-angular-range (FAR) (with full- or short-scan range) DECT without suffering from LAR artifacts.}

We carry out numerical studies with digital phantoms mimicking a suitcase and a breast, which are of relevance to industrial and clinical applications, respectively. Both noiseless and noisy data are generated from each phantom with low- and high-kVp spectra  over arcs of LARs spanning from $14^\circ$ to $180^\circ$, as well as over the FAR of $360^\circ$. Using the images reconstructed from low- and high-kVp LAR data, we estimate basis images, from which monochromatic images at energies of interest are formed. We then perform visual inspection and quantitative analysis of monochromatic images obtained, and estimate physical quantities including atomic numbers and iodine-contrast concentrations from the basis images determined. 

Following the introduction in Sec.~\ref{sec:intro}, we present materials and methods in Sec.~\ref{sec:methods}, including scanning configuration, data generation, image reconstruction, and image analysis. Image results and quantitative analysis are shown in Sec.~\ref{sec:rslt}, while discussions and conclusion are followed in Secs.~\ref{sec:discussion} and~\ref{sec:conclusion}, respectively. We include in the Appendices description of the methods for determination of basis images and for estimation of physical quantities to avoid distractions from the presentation flow in the main text.

\section{Materials and Methods} \label{sec:methods}

\subsection{Scanning configuration}\label{sec:scanning-configs}
\begin{figure}[t!]
		\centering
		\includegraphics[height=0.40\textwidth, trim={60 220 60 140}, clip]{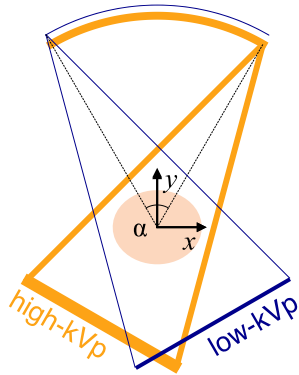}
	\caption{Scanning configuration with overlapping arcs of LAR (thin and thick curves) for collecting low- and high-kVp data. The coordinate system of the image array is set up such that the circular scanning arc is symmetric relative to the $y$-axis.}
	\label{fig:config}
\end{figure}

We consider a fan-beam-based scanning configuration, as shown in Fig.~\ref{fig:config}, while the study presented in Secs.~\ref{sec:methods-recon} and~\ref{sec:methods-analysis} below can readily be extended to cone-beam geometries.
The low- and high-kVp data are collected over two overlapping circular arcs of LAR $\alpha$, from an object that is within the field of view of the configuration. 
The image grid is set up such that the circular scanning arc is symmetric relative to the $y$-axis. This setup with overlapping arcs might be of practical interest in situations where the total scanning angular range is physically limited due to workflow or safety concerns, and it can be implemented with current DECT techniques, such as sandwiched detectors, fast-kVp-switching X-ray tubes, and sequential scans~\cite{mccollough_dual-_2015}. 

In this work, we consider LARs $\alpha=14^\circ, 20^\circ, 30^\circ, 60^\circ, 90^\circ, 120^\circ$, $150^\circ$, and $180^\circ$, with an angular interval of $1^\circ$ per view. 
Images are also obtained over a FAR of $360^\circ$ by use of the DTV and FBP algorithms from noiseless data. They are used as the {\it DTV-} and {\it FBP-reference images}, respectively, against which we investigate how image quality from LAR data decreases. 
For suitcase-phantom studies, the source-to-rotation distance (SRD) and source-to-detector distance (SDD) are 100 cm and 150 cm, and a linear detector of 32 cm consists of 512 bins, whereas for breast-phantom studies, SRD and SDD are 36 cm and 72 cm, and a linear detector of 37.5 cm consists of 512 bins. 


\subsection{Dual-energy data}\label{sec:methods-data}

\begin{figure}[t!]
	\centering
	\begin{tabular}{c c}
		\includegraphics[width=0.45\textwidth]{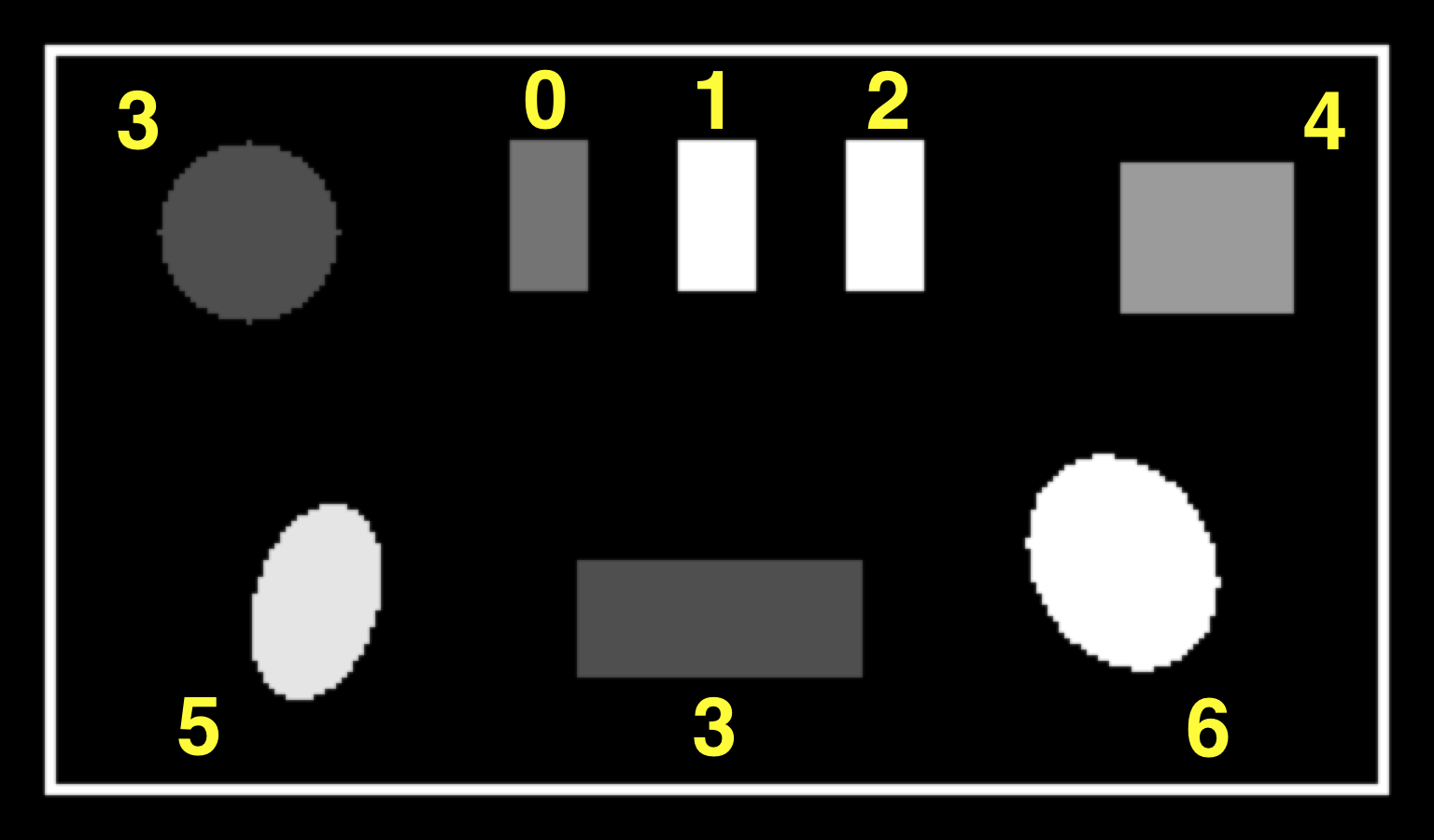} &
		\includegraphics[width=0.45\textwidth,trim={10 0 10 0}, clip]{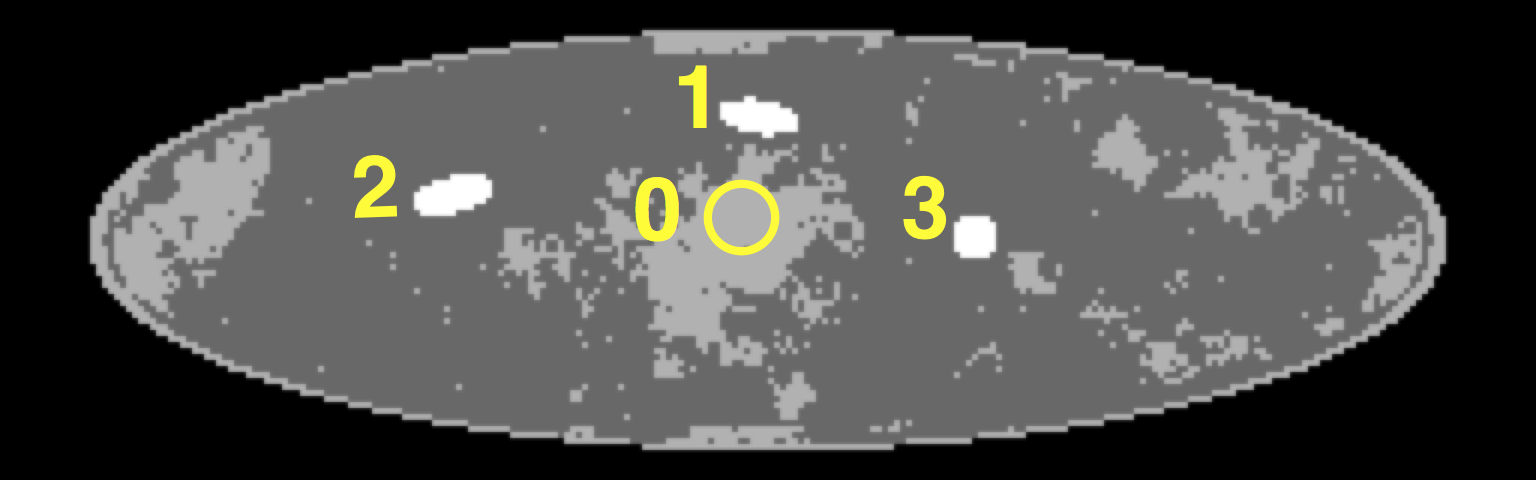} \\
		(a) suitcase phantom & (b) breast phantom \\
	\end{tabular}
	\caption{Monochromatic images for the (a) suitcase phantom at 40 keV and (b) breast phantom at 34 keV. The numbers in each of the phantoms indicate ROIs of different materials. 
	In the suitcase phantom, ROIs 0-6 contain C, Al, Ca, water, ANFO, teflon, and PVC, respectively. In the breast phantom, the darker and brighter background regions are adipose tissue and breast tissue (including skin), respectively; ROI 0 contains breast tissue, whereas ROIs 1-3 are filled with iodine contrast agent of concentration 5, 2, and 2.5 mg/ml, respectively. Display windows for the suitcase and breast phantoms are [0.1, 0.65] cm$^{-1}$ and [0.2, 0.35] cm$^{-1}$, respectively.}
	\label{fig:phan}
\end{figure}

The suitcase and breast numerical phantoms shown in Fig.~\ref{fig:phan} are used in the numerical study, as the former is of potential interest in security-scan applications such as baggage screening and the latter mimicks the cross section of a breast in contrast-enhanced imaging. In the suitcase phantom, three bar-shaped regions of interest (ROIs) contain elements C, Al, and Ca, respectively, whereas rectangular and elliptical ROIs are filled with water, ANFO (Ammonium Nitrate and Fuel Oil~\cite{ying2006dual}), teflon, and PVC. In the breast phantom, the background, mixed with adipose and breast tissue, is embedded with three ROIs mimicking iodine-contrast-enhanced vasculature and tumor of concentrations of 2, 2.5, and 5 mg/ml, which are typical values of potential clinical relevance~\cite{jong2003contrast,volterrani2020dual}. The suitcase and breast phantoms are represented, respectively, by image arrays of $150\times256$ square pixels of size 0.7 mm and of $80\times256$ square pixels of size 0.7 mm. 

For either phantom, each pixel is labeled with a material type, which is associated with a  linear attenuation coefficient at a given energy.
As such, we consider a non-linear-data-model incorporating the polychromatic X-ray spectrum~\cite{chen2017image} as
\begin{eqnarray} \label{eq:nonlinear-model}
	g^s_j = - \ln \sum_m^M q^s_{jm} \exp \left( - \sum_i^I a^s_{ji} f_{m i} \right),
\end{eqnarray}
where $g^s_j$ is the model data for ray $j$, $j=1, 2, ..., J^s$, within the low- ($s=L$) or high-kVp ($s=H$) scan; $J^s$ the total number of rays in the low- or high-kVp scan; $q^s_{jm}$ the normalized, low- or high-kVp X-ray spectrum (including detector response) for ray $j$ at energy bin $m$, $m=1,2, ..., M$; $M$ the total number of energy bins; $a^s_{ji}$ the intersection length of ray $j$, in the low- or high-kVp scan, within pixel $i$; $f_{m i}$ the linear attenuation coefficient at energy bin $m$ for the labeled material at pixel $i$, $i=1, 2, ..., I$; and $I$ the total number of image pixels. We can form matrix $\mathcal{A}^s$ of size $J^s\times I$ with element $a^s_{ji}$ and refer to $\mathcal{A}^s$ as the discrete X-ray transform (DXT). 

In the numerical studies in Sec.~\ref{sec:rslt} below, we use the non-linear model in Eq.~\eqref{eq:nonlinear-model} to generate noiseless and Poisson-noisy (corresponding to $10^7$ photons per ray in the air scan) data from the suitcase and breast phantoms with low- and high-kVp spectra, which are generated using the TASMIC model~\cite{hernandez_tungsten_2014}. 
For the suitcase phantom, the low- and high-kVp spectra are set at 80 and 140 kVp's, respectively, with an additional 5-mm Al filter for both. For the breast phantom, the low- and high-kVp spectra are set at 33 and 49 kVp's, with 8-mm Al and 0.25-mm Cu filters, respectively.

\subsection{Image reconstruction}\label{sec:methods-recon}
We use vector $\mathbf{g}^{s[\mathcal{M}]}$ of size $J^s$, where $s=L$ or $H$, to denote measured data, with element $g^{s[\mathcal{M}]}_j$ indicating low- or high-kVp measurement with ray $j$ in DECT, where $j=1, 2, ...., J^s$. In the noiseless, numerical study below, $g^{s[\mathcal{M}]}_j=g^{s}_j$ in Eq.~\eqref{eq:nonlinear-model}, as it is considered to be a reasonable data model in DECT, whereas in the noisy numerical study in the work, $g^{s[\mathcal{M}]}_j$ is obtained by addition of Poisson noise to $g^{s}_j$ as described above. 

In standard DECT, images are reconstructed often by use of an algorithm that is based upon a linear-data model instead of the non-linear data model in Eq. \eqref{eq:nonlinear-model}. We propose to use in the work a DTV algorithm that is also based upon a linear-data model for image reconstruction directly from low- and high-kVp data collected over arcs of LAR. The images reconstructed thus contain the BH effect inherent in data generated with the non-linear-data model. 

We formulate the reconstruction problem from either low- or high-kVp data over an arc of LAR as a convex optimization problem~\cite{zhang2021dtv} given by:
\begin{eqnarray} \label{eq:opt}
\begin{aligned}
	\mathbf{f}^{s\,\star} &= \underset{\mathbf{f}^s}{\mathsf{argmin}} \,\, 
		\frac{1}{2} \parallel  \mathbf{g}^{s[\mathcal{M}]} - \mathcal{A}^s\,\mathbf{f}^s \parallel_2^2 \\
		 \quad {\rm s.t.} \,\, &
		|| \mathcal{D}_x \mathbf{f}^s ||_1 \le t^s_x, \,\, || \mathcal{D}_y \mathbf{f}^s ||_1 \le t^s_y, 
		\,\, {\rm and} \,\, f^s_i \ge 0, 
\end{aligned}
\end{eqnarray}
where $s=L$ or $H$; $\parallel \cdot \parallel_2^2$ operating on a vector denotes the squared $\ell_2$-norm; vector $\mathbf{f}^s$ of size $I$ the image to be reconstructed; matrices $\mathcal{D}_x$ and $\mathcal{D}_y$ of size $I \times I$ denote two-point differences along the $x$- and $y$-axis, respectively; and $|| \mathcal{D}_x \mathbf{f}^s ||_1$ and $|| \mathcal{D}_y \mathbf{f}^s ||_1$ are $\ell_1$ norms of the image partial derivatives along the $x$- and $y$-axis, respectively, also referred to as the image directional total variations (DTVs). In the formulation of Eq.~\eqref{eq:opt}, linear model $\mathcal{A}^s\,\mathbf{f}^s$, i.e., the DXT of $\mathbf{f}^s$, is used to approximate low- or high-kVp data $\mathbf{g}^{s[\mathcal{M}]}$.

Basing upon the general primal-dual (PD) algorithm~\cite{chambolle_first-order_2010, sidky_convex_2012} solving mathematically exactly convex optimization problems, we have developed a DTV algorithm tailored to solve Eq.~\eqref{eq:opt} for reconstructing  $\mathbf{f}^s$ from low- or high-kVp data collected over an arc of LAR. 
As the detailed derivation of the DTV algorithm can be found in Appendix A of Ref.~\cite{zhang2021dtv}, we list below only the pseudo-code of the DTV algorithm.
\begin{algorithm}\leavevmode
\caption{Pseudo-code of the DTV algorithm for 
solving Eq. \eqref{eq:opt}}\label{alg:1}
\begin{algorithmic}[1]
\STATEx INPUT: $g^{s [\mathcal{M}]}$, $t^s_x$, $t^s_y$, $\mathcal{A}^s$, $b$
\STATE $L^s \leftarrow ||\mathcal{K}^s||_2$, $\tau^s \leftarrow b/L^s$, $\sigma^s \leftarrow 1/(b L^s)$, $\nu^s_1 \leftarrow ||\mathcal{A}^s||_2/||\mathcal{D}_x||_2$, $\nu^s_2 \leftarrow ||\mathcal{A}^s||_2/||\mathcal{D}_y||_2$, $\mu^s \leftarrow ||\mathcal{A}^s||_2/||\mathcal{I}||_2$
\STATE $n \leftarrow 0$
\STATE INITIALIZE: $\mathbf{f}^{(0)}$, $\mathbf{w}^{(0)}$, $\mathbf{p}^{(0)}$, $\mathbf{q}^{(0)}$, and $\mathbf{t}^{(0)}$ to zero
\STATE $\bar{\mathbf{f}}^{(0)} \leftarrow \mathbf{f}^{(0)}$
\REPEAT
\STATE  $\mathbf{w}^{(n+1)} = (\mathbf{w}^{(n)} + \sigma^s(\mathcal{A}^s\bar{\mathbf{f}}^{(n)} - \mathbf{g}^{s [\mathcal{M}]}))/(1+\sigma^s)$
\STATE $\mathbf{p}^{\prime(n)} = \mathbf{p}^{(n)} + \sigma^s \nu^s_1 \mathcal{D}_x \bar{\mathbf{f}}^{(n)}$
\STATEx \hspace{0.25cm} $\mathbf{q}^{\prime(n)} = \mathbf{q}^{(n)} + \sigma^s \nu^s_2 \mathcal{D}_y \bar{\mathbf{f}}^{(n)}$
\STATE $\mathbf{p}^{(n+1)} = \mathbf{p}^{\prime(n)} - \sigma^s \frac{\mathbf{p}^{\prime(n)}}{|\mathbf{p}^{\prime(n)}|}\ell_1 {\rm ball}_{\nu^s_1 t^s_x} (\frac{|\mathbf{p}^{\prime(n)}|}{\sigma^s})$
\STATEx \hspace{0.25cm} $\mathbf{q}^{(n+1)} = \mathbf{q}^{\prime(n)} - \sigma^s \frac{\mathbf{q}^{\prime(n)}}{|\mathbf{q}^{\prime(n)}|}\ell_1 {\rm ball}_{\nu^s_2 t^s_y} (\frac{|\mathbf{q}^{\prime(n)}|}{\sigma^s})$
\STATE $\mathbf{t}^{(n+1)} = {\rm neg}({\mathbf{t}^{(n)} + \sigma^s \mu^s \bar{\mathbf{f}}^{(n)}})$
\STATE $\mathbf{f}^{(n+1)} = \mathbf{f}^{(n)}-\tau^s ({\mathcal{A}^s}^{\top}\mathbf{w}^{(n+1)}+\nu^s_1\mathcal{D}_x^{\top}{\mathbf{p}}^{(n+1)} +\nu^s_2\mathcal{D}_y^{\top}{\mathbf{q}}^{(n+1)} + \mu^s \mathbf{t}^{(n+1)})$
\STATE $\bar{\mathbf{f}}^{(n+1)} = 2 \mathbf{f}^{(n+1)}-\mathbf{f}^{(n)}$
\STATE $n \leftarrow n+1$
\UNTIL the convergence conditions are satisfied
\STATE {OUTPUT: $\mathbf{f}^{(n)}$ as the estimate of $\mathbf{f}^{s}$} 
\end{algorithmic}
\end{algorithm}


In the pseudo-code, 
stacked matrix $\mathcal{K}^s$ is defined as $\mathcal{K}^{s \top} = (\mathcal{A}^{s \top}, \nu^s_1 \mathcal{D}_x^{\top}, \nu^s_2 \mathcal{D}_y^{\top}, \mu \mathcal{I}$), where superscript ``$\top$'' indicates a transpose operation; the nuclear norm of a matrix, indicated by $||\cdot||_2$, calculates the largest singular value of the matrix; $\mathcal{I}$ is an identity matrix of size $I \times I$; vectors $\mathbf{w}^{(n)}$ is of size $J^s$, whereas vectors $\mathbf{p}^{\prime(n)}$, $\mathbf{q}^{\prime(n)}$, $\mathbf{p}^{(n)}$, $\mathbf{q}^{(n)}$, and $\mathbf{t}^{(n)}$ are of size $I$; ${\rm neg}(\cdot)$ enforces the non-positivity constraint by thresholding; operator $\ell_1 {\rm ball}_{\beta}(\cdot)$ projects a vector onto the $\ell_1$-ball of size $\beta$; and $|\mathbf{q}^{\prime(n)}|$ denotes a vector with entry $j$ given by $(|\mathbf{q}^{\prime(n)}|)_j = |{q}^{\prime(n)}_j|$, where ${q}^{\prime(n)}_j$ is the $j$th entry of vector $\mathbf{q}^{\prime(n)}$.

In DECT, monochromatic image $\mathbf{f}_m$, i.e., the linear attenuation coefficient distribution, at energy bin $m$ is of interest, and it can be decomposed into a linear combination of two basis images $\mathbf{b}_0$ and $\mathbf{b}_1$ as
\begin{eqnarray} \label{eq:decomp-mono}
\mathbf{f}_m 
= \mu_{m0} \mathbf{b}_0 +  \mu_{m1} \mathbf{b}_1,
\end{eqnarray}
where expansion coefficients $\mu_{mk}$ ($k=0$ or $1$)
can be either calculated or looked up, and basis images $\mathbf{b}_k$ can be estimated from low- and high-kVp images ${\mathbf{f}}^L$ and ${\mathbf{f}}^H$ reconstructed, as described in \ref{app:decomp}.

Image reconstruction with the DTV algorithm, like reconstructions with any algorithms, involves constraint parameters such as $t^s_x$ and $t^s_y$ whose selection can impact reconstruction quality. In the work, we select parameters $t^s_x$ and $t^s_y$ by visual evaluation of monochromatic images obtained with minimum artifacts,  
as shown in \ref{app:para-selection}. In addition, the images are reconstructed with the FBP algorithm, along with a Hanning kernel and a cutoff frequency at 0.5, directly from the low- and high-kVp LAR data as they can provide a benchmark for the DTV reconstructions.

\subsection{Analysis of monochromatic images obtained} \label{sec:methods-analysis}

We reconstruct images $\mathbf{f}^L$ and $\mathbf{f}^H$ directly from low- and high-kVp data generated over an arc of LAR by use of the DTV algorithm, and estimate two basis images from $\mathbf{f}^L$ and $\mathbf{f}^H$ by using either the interaction- or material-based method described in \ref{app:decomp}.
With the basis images estimated, we then compose monochromatic images at energies of interest using  Eq.~\eqref{eq:decomp-mono}, and perform visual inspection and quantitative analysis of the monochromatic images. Furthermore, we analyze DTV reconstructions in tasks of estimation of atomic number and contrast-agent concentration within ROIs defined in Fig.~\ref{fig:phan}.

\paragraph{Visual inspection and quantitative analysis of monochromatic images}

We first perform visual inspection of the monochromatic images obtained at energy levels of interest to assess LAR artifacts. In addition to visual inspection, we compute two quantitative metrics, Pearson correlation coefficient (PCC) and normalized mutual information (nMI)~\cite{pearson1895notes,Viergever:2003,Bian-PMB:2010,zhang2021dtv}, for evaluating the visual correlation between a monochromatic image obtained from the LAR data and its corresponding reference image. Specifically, the maximum values of PCC and nMI are 1, and the higher the PCC and nMI values, the better the visual correlation between an image and its reference image. While the image and its reference are identical when PCC=1 and nMI=1, the image generally appears visually resembling the reference image even as ${\rm PCC}>0.8$ and ${\rm nMI} >0.6\sim0.7$. 


\paragraph{Estimation of physical quantities}
In addition to visual inspection and analysis, we also analyze the DTV reconstructions in two tasks of estimating physical quantities of interest as described below.

\noindent{\it Estimation of atomic number:} The study involving the suitcase phantom is of potential interest to industrial/security CT applications, such as baggage screening, in which estimation of the atomic number of materials is used for explosive detection~\cite{ying2006dual}. Using the interaction-based method on the DTV-reference image of the suitcase phantom, we obtain the $2\times2$ decomposition matrix in Eq.~\eqref{eq:decomp-basis}, which is used for estimating two basis images of photoelectric effect (PE) and Compton scattering (KN) components from $\mathbf{f}^L$ and $\mathbf{f}^H$ reconstructed throughout the studies with the suitcase phantom, as discussed in \ref{app:decomp}.
Using the estimated basis images in the affine relationship in Eq.~\eqref{eq:z-log} in \ref{app:tasks}, we then estimate the atomic numbers within ROIs 3-6 of the suitcase phantom, as shown in Fig.~\ref{fig:phan}a. Constants $c$ and $n$ in Eq.~\eqref{eq:z-log} are fitted and calibrated using the image values within calibration ROIs 0-2 of the suitcase phantom from the DTV-reference image. 
The three ROIs correspond to three single-element common materials, C, Al, and Ca, which are picked as their atomic numbers, $z=6$, $13$, and $20$, cover the range of atomic numbers of interest for the other materials contained in the suitcase phantom.



\noindent{\it Estimation of iodine concentration:} In mammography, digital breast tomosynthesis (DBT), and breast CT, iodine-based contrast agents can be used for enhancing tumor contrast~\cite{dromain2012dual,jochelson2013bilateral,carton2010dual,samei2011dual,de2012dual,zhang2018axillary}. Quantitative estimation of iodine-contrast concentration is of interest in breast tumor staging~\cite{volterrani2020dual} and capturing the contrast-uptake kinetics, as it may help differentiate between benign and malignant tumors~\cite{jong2003contrast}. In the study involving the breast phantom, the basis materials are selected as breast tissue and 5-mg/ml iodine contrast agent in calibration ROIs 0 and 1, respectively, as shown in Fig.~\ref{fig:phan}b. Similarly, using the material-based method on the DTV-reference image of the breast phantom, we first obtain the $2\times2$ decomposition matrix in Eq.~\eqref{eq:decomp-basis}, which is used for estimating two basis images of breast tissue and iodine contrast agent from $\mathbf{f}^L$ and $\mathbf{f}^H$ reconstructed throughout the studies with the breast phantom, as discussed in \ref{app:decomp}.
Using the estimated basis image of iodine contrast agent in Eq.~\eqref{eq:beta-linear} in \ref{app:tasks}, we estimate the concentration of iodine contrast agent within ROIs 1-3. Constants $\gamma$ and $\tau$ are fitted and calibrated using image values within ROIs 1-3 in the DTV-reference image of the breast phantom, together with their known concentrations.

\section{Results} \label{sec:rslt}

We reconstruct below $\mathbf{f}^s$ from data $\mathbf{g}^{s[\mathcal{M}]}$ of the suitcase and breast phantoms in Fig. \ref{fig:phan} from low- and high-kVp LAR data by using the DTV algorithm, where $s=L$ and $H$.
In each reconstruction, the DTV-constraint parameters are selected by use of the approach described in \ref{app:para-selection}. Subsequently, using the interaction- or material-based method, as described in \ref{app:decomp}, we estimate the basis images from which monochromatic images $\mathbf{f}_m$ at energy bin $m$ are obtained by using Eq.~\eqref{eq:decomp-mono}. In addition to visual inspection and quantitative analysis of the monochromatic images, we estimate atomic numbers and iodine-contrast concentrations within selected ROIs in the suitcase and breast phantoms, respectively, as described in \ref{app:tasks}.

\subsection{Image reconstruction from noiseless data of the suitcase phantom}\label{sec:suitcase-noiseless}
We reconstruct images from noiseless low- and high-kVp data of the suitcase phantom generated over arcs of LARs $\alpha=14^\circ, 20^\circ, 30^\circ, 60^\circ, 90^\circ, 120^\circ$, $150^\circ$, and $180^\circ$ by use of the DTV algorithm, and then estimate basis images by using the interaction-based method with the DTV-reference image, as described in \ref{app:decomp}, from the images reconstructed.
Subsequently, using the basis images estimated, we obtain monochromatic images at energy 40 keV for enhanced contrast in the images, and then compute atomic numbers within selected ROIs, as described in \ref{app:tasks}.

\begin{figure*}[t!]
	\centering
	\begin{tabular}{c c c c}
			DTV-$60^\circ$ & FBP-$60^\circ$  & DTV-$360^\circ$ & FBP-$360^\circ$ \\
			\includegraphics[width=0.22\textwidth]{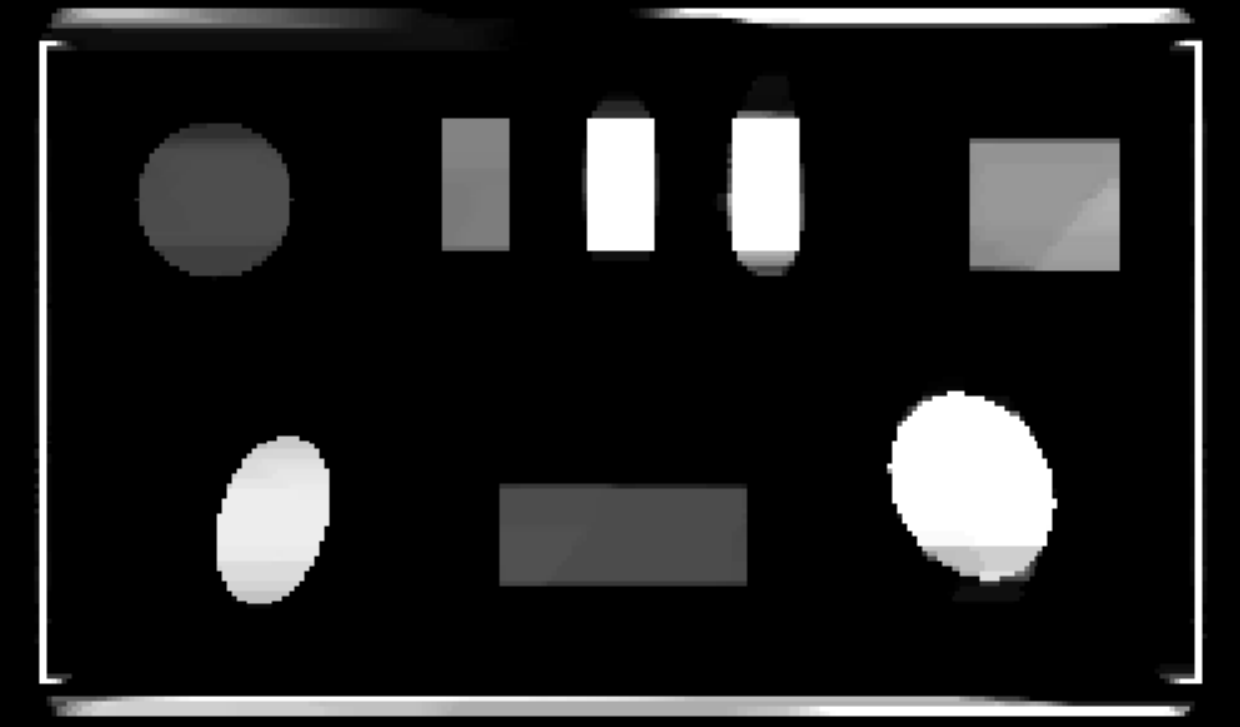}
			\hspace{-10pt} &
			\includegraphics[width=0.22\textwidth]{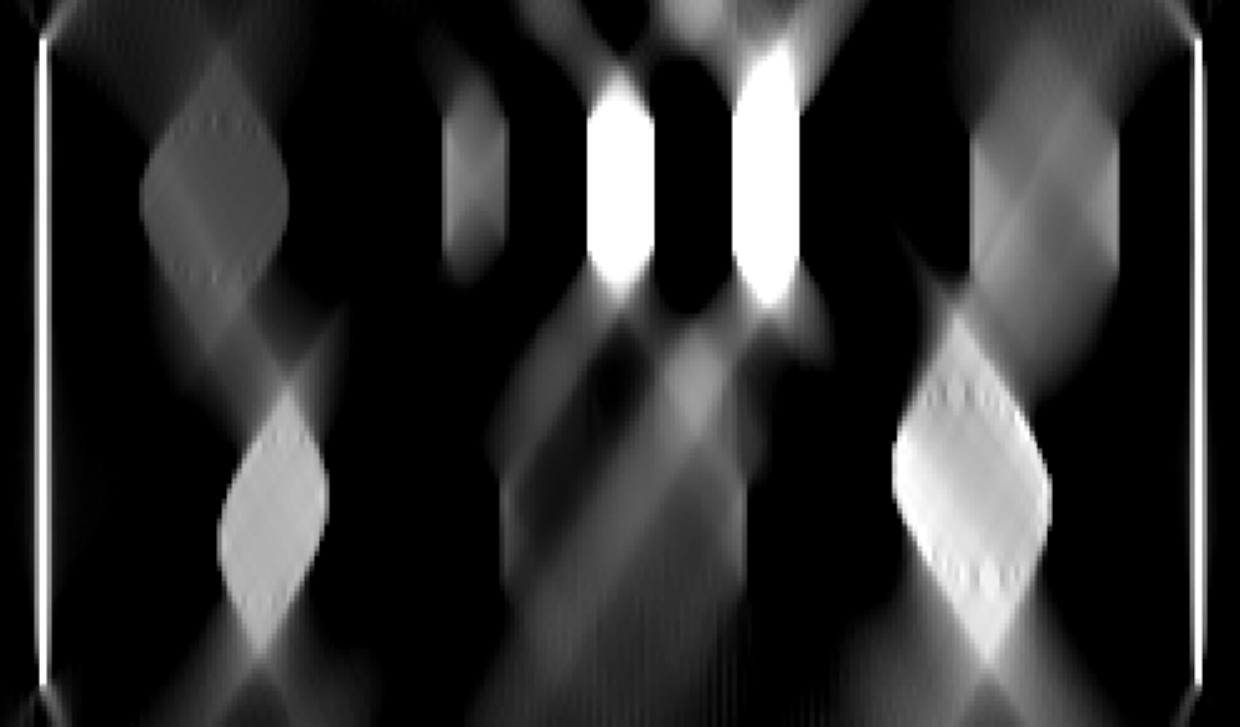}
			 \hspace{-10pt} &
			\includegraphics[width=0.22\textwidth]{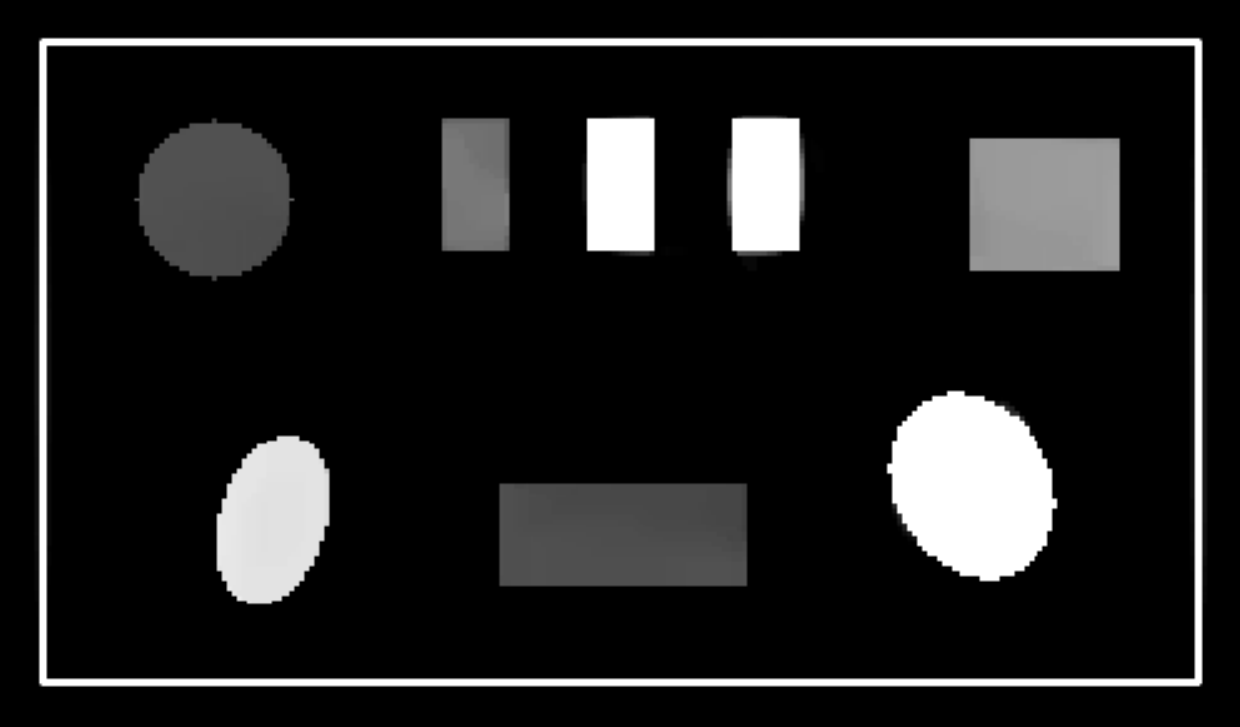}
			\hspace{-10pt} &
			\includegraphics[width=0.22\textwidth]{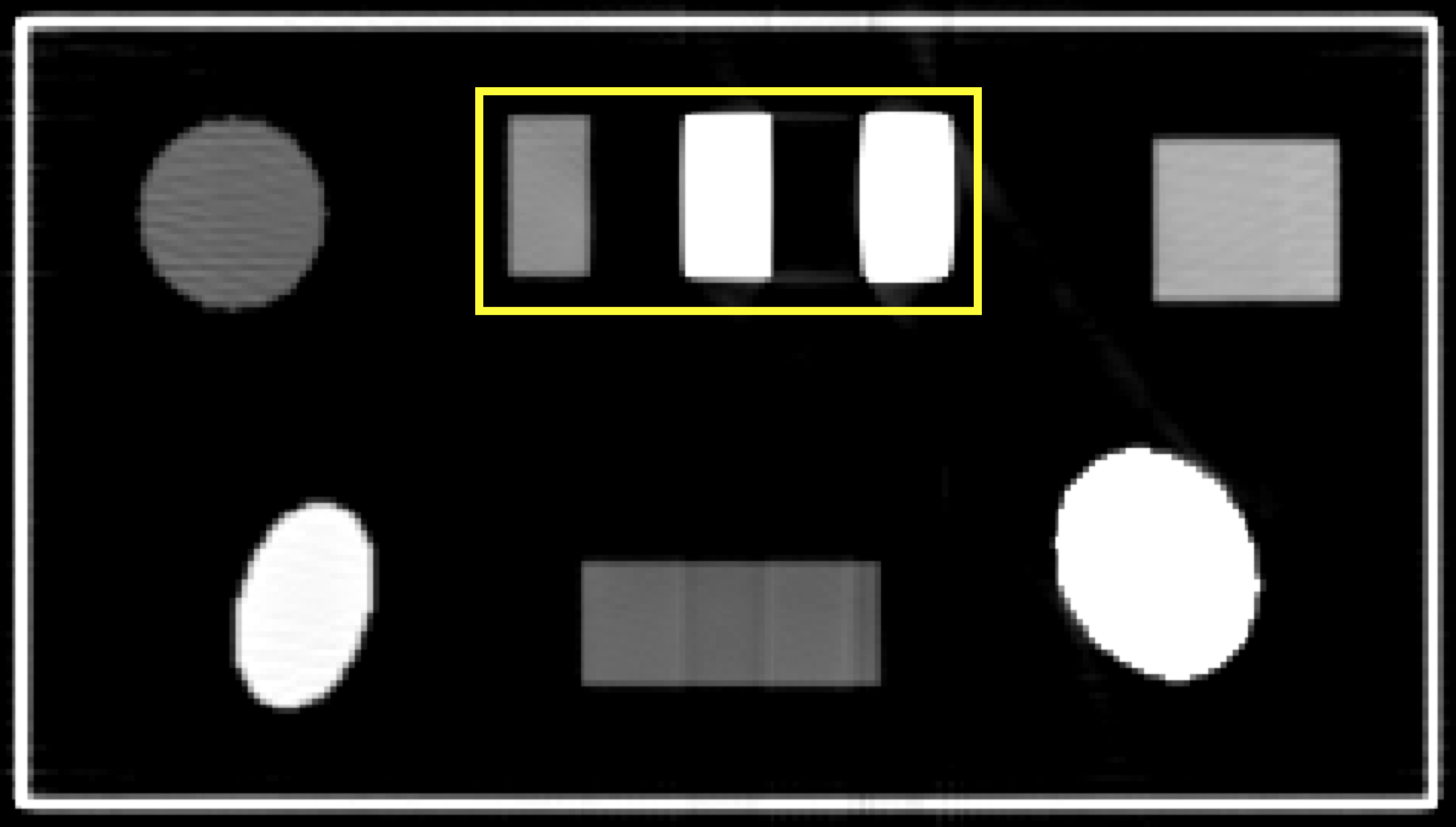}
			%
			\\
			\includegraphics[width=0.22\textwidth,trim={120 130 120 25}, clip]{figures/barPhan4/Noiseless/XY_pdf/barPhan4_XY_60D_40keV.pdf}
			\hspace{-10pt} &
			\includegraphics[width=0.22\textwidth,trim={120 130 120 25}, clip]{figures/barPhan4/Noiseless/FBP_pdf/barPhan4_FBP_60D_40keV.pdf}
			\hspace{-10pt} &
			\includegraphics[width=0.22\textwidth,trim={120 130 120 25}, clip]{figures/barPhan4/Noiseless/XY_pdf/barPhan4_XY_360D_40keV.pdf}
			\hspace{-10pt} &
			\includegraphics[width=0.22\textwidth,trim={120 130 120 25}, clip]{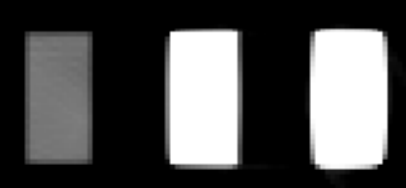}
	\end{tabular}
	\caption{Monochromatic images at 40 keV (top row) and their respective zoomed-in ROI views (bottom row) of the suitcase phantom obtained by use of the DTV (column 1) and FBP (column 2) algorithms from noiseless data generated over an arc of $60^\circ$, and the DTV-reference image (column 3) and FBP-reference image (column 4). The ROI is enclosed by the rectangular box depicted in the FBP-reference image. Display window: [0.1, 0.65] cm$^{-1}$.}
	\label{fig:suitcase-mono-60}
\end{figure*}

\begin{figure*}[t!]
		\centering
		\begin{tabular}{c c c c}
			DTV-$14^\circ$ & DTV-$20^\circ$ & DTV-$30^\circ$ & DTV-$60^\circ$ \\
			\includegraphics[width=0.22\textwidth]{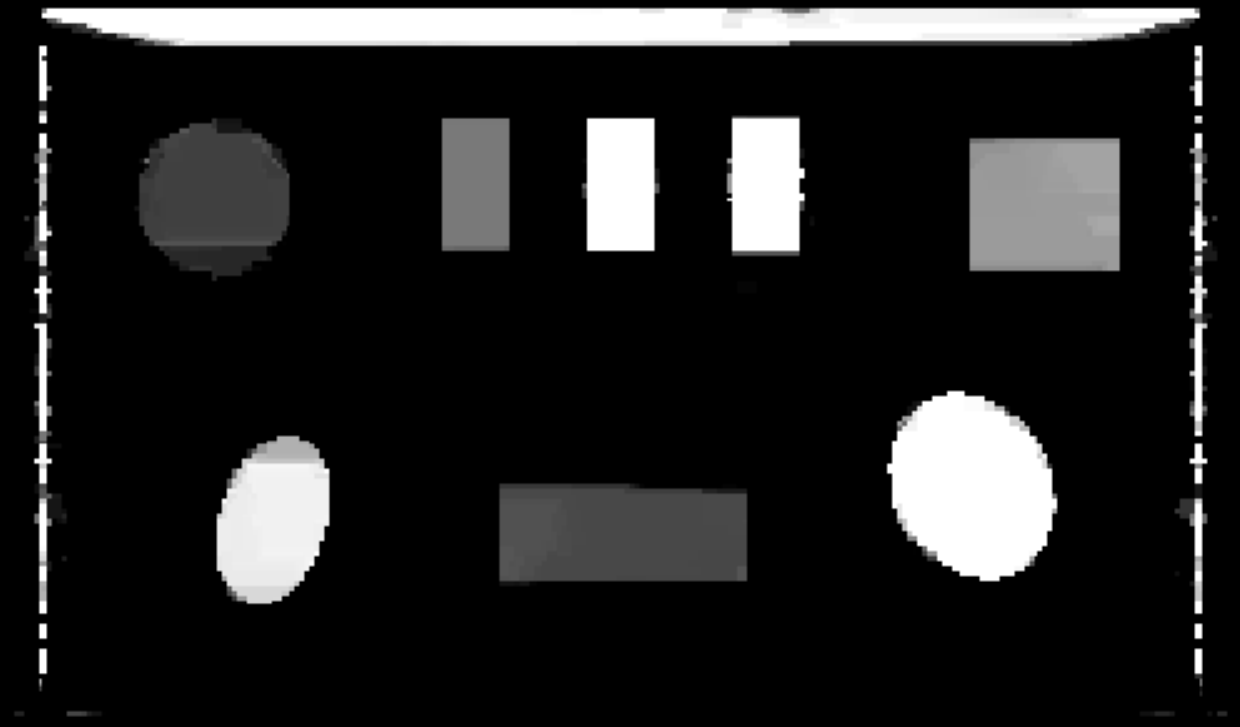}
			\hspace{-10pt} &
			\includegraphics[width=0.22\textwidth]{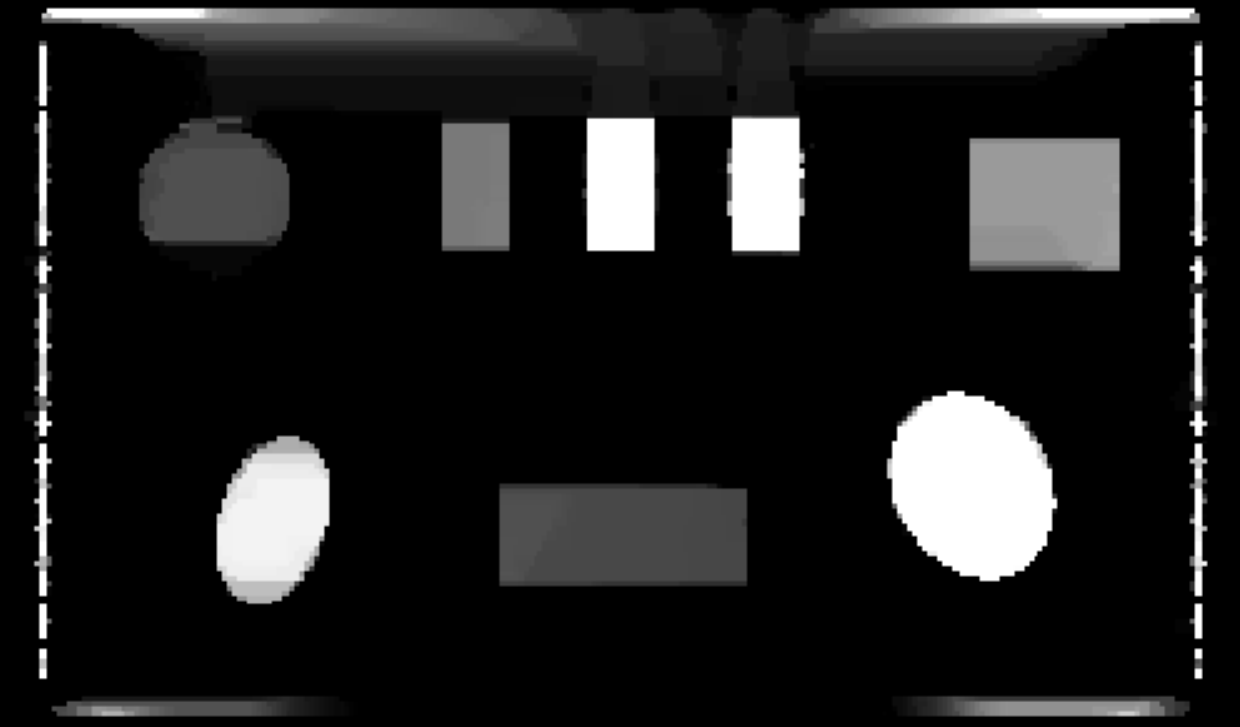}
			\hspace{-10pt} &
			\includegraphics[width=0.22\textwidth]{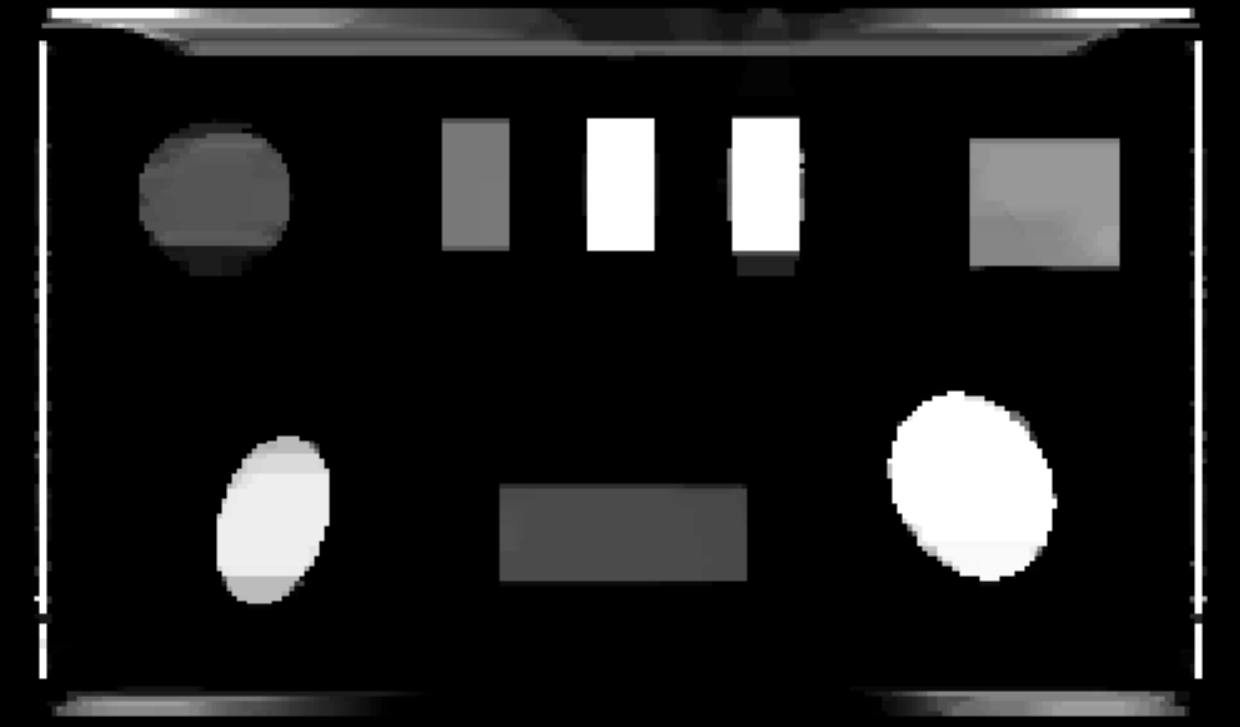}
			\hspace{-10pt} &
			\includegraphics[width=0.22\textwidth]{figures/barPhan4/Noiseless/XY_pdf/barPhan4_XY_60D_40keV.pdf} %
			\\
			\includegraphics[width=0.22\textwidth,trim={120 130 120 25}, clip]{figures/barPhan4/Noiseless/XY_pdf/barPhan4_XY_14D_40keV.pdf}
			\hspace{-10pt} &
			\includegraphics[width=0.22\textwidth,trim={120 130 120 25}, clip]{figures/barPhan4/Noiseless/XY_pdf/barPhan4_XY_20D_40keV.pdf}
			\hspace{-10pt} &
			\includegraphics[width=0.22\textwidth,trim={120 130 120 25}, clip]{figures/barPhan4/Noiseless/XY_pdf/barPhan4_XY_30D_40keV.pdf}
			\hspace{-10pt} &
			\includegraphics[width=0.22\textwidth,trim={120 130 120 25}, clip]{figures/barPhan4/Noiseless/XY_pdf/barPhan4_XY_60D_40keV.pdf} %
			\\
			DTV-$90^\circ$ & DTV-$120^\circ$ & DTV-$150^\circ$ & DTV-$360^\circ$ \\
			\includegraphics[width=0.22\textwidth]{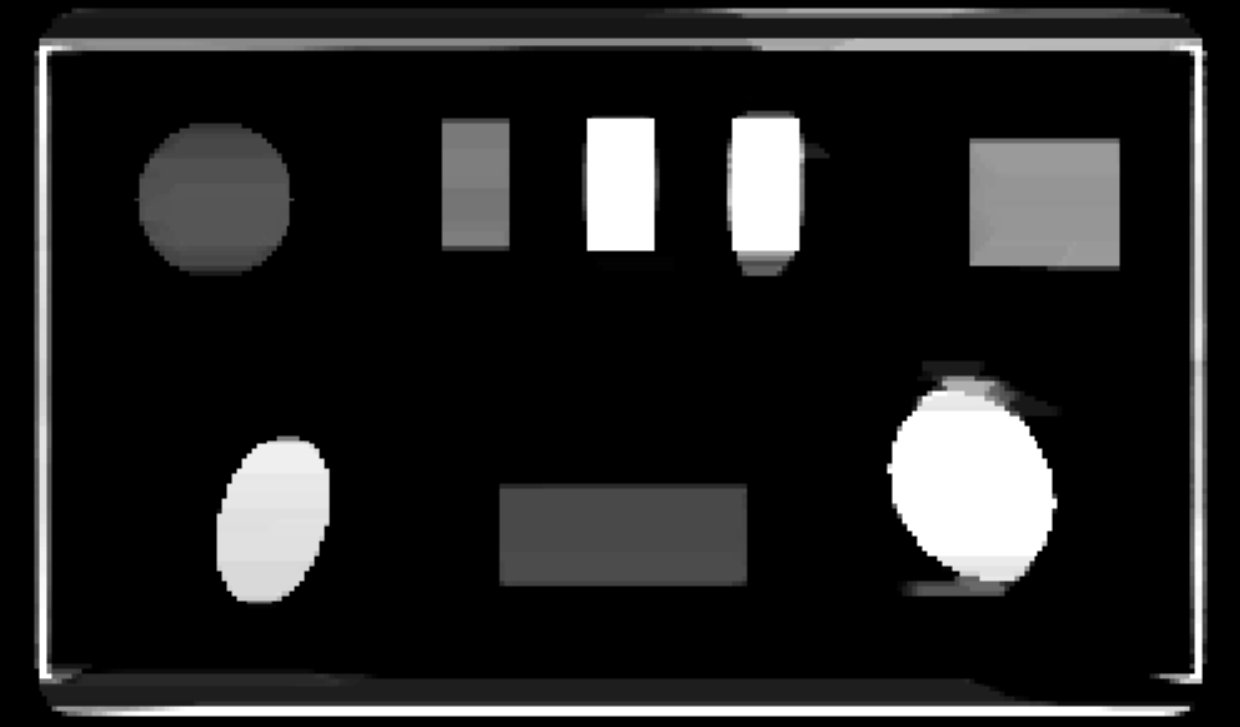}
			\hspace{-10pt} &
			\includegraphics[width=0.22\textwidth]{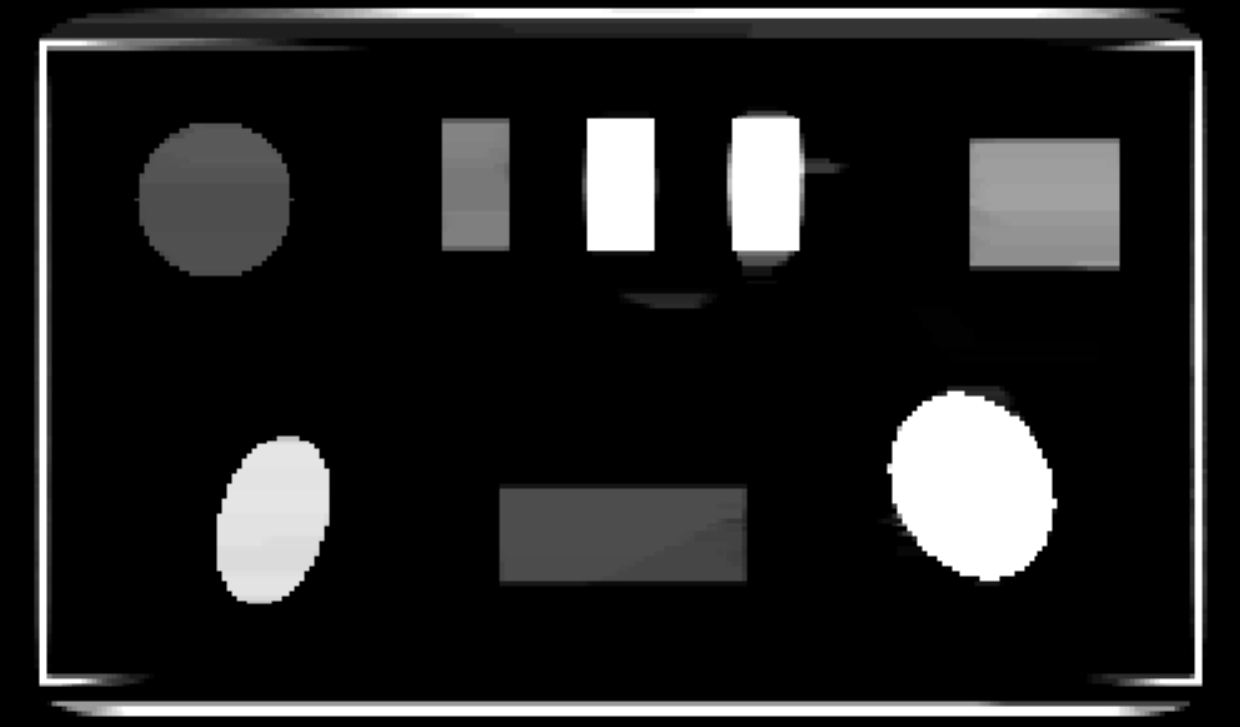}
			\hspace{-10pt} &
			\includegraphics[width=0.22\textwidth]{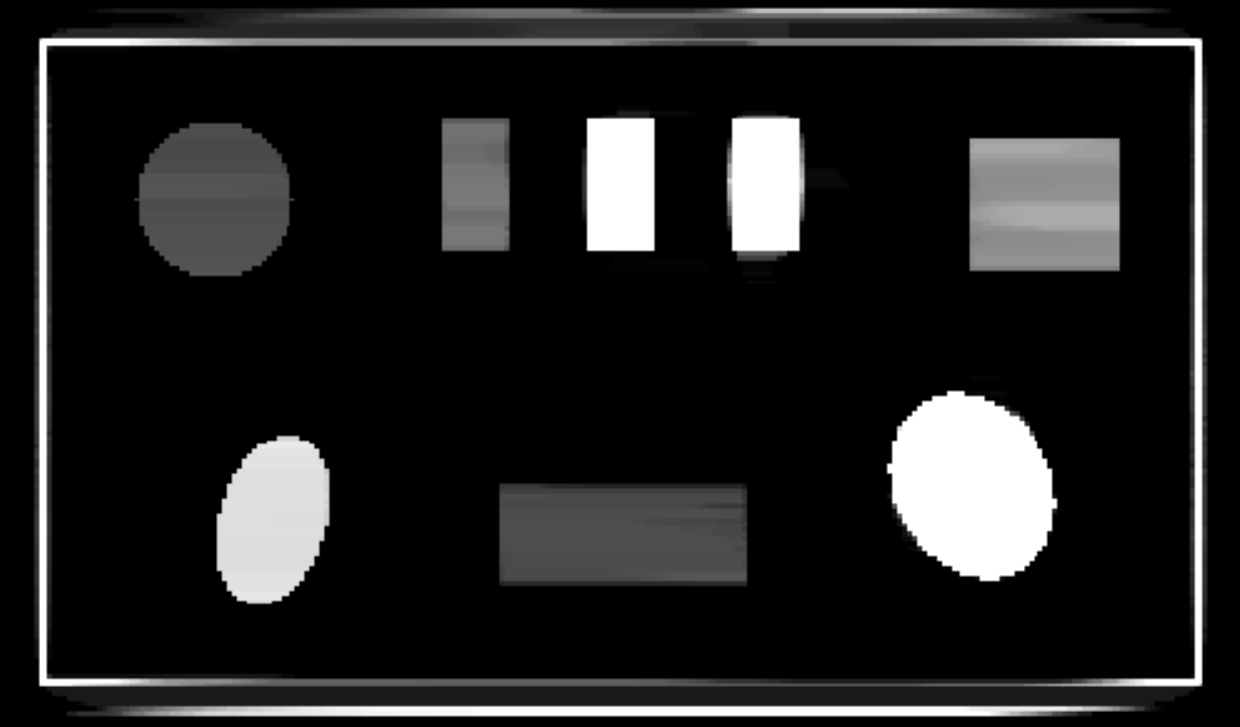}
			\hspace{-10pt} &
			\includegraphics[width=0.22\textwidth]{figures/barPhan4/Noiseless/XY_pdf/barPhan4_XY_360D_40keV.pdf} %
			\\
			\includegraphics[width=0.22\textwidth,trim={120 130 120 25}, clip]{figures/barPhan4/Noiseless/XY_pdf/barPhan4_XY_90D_40keV.pdf}
			\hspace{-10pt} &
			\includegraphics[width=0.22\textwidth,trim={120 130 120 25}, clip]{figures/barPhan4/Noiseless/XY_pdf/barPhan4_XY_120D_40keV.pdf}
			\hspace{-10pt} &
			\includegraphics[width=0.22\textwidth,trim={120 130 120 25}, clip]{figures/barPhan4/Noiseless/XY_pdf/barPhan4_XY_150D_40keV.pdf}
			\hspace{-10pt} &
			\includegraphics[width=0.22\textwidth,trim={120 130 120 25}, clip]{figures/barPhan4/Noiseless/XY_pdf/barPhan4_XY_360D_40keV.pdf}
	\end{tabular}
	\caption{Monochromatic images (rows 1 and 3) of the suitcase phantom at 40 keV obtained from noiseless data generated over arcs of LARs $14^\circ$, $20^\circ$, $30^\circ$, $60^\circ$, $90^\circ$, $120^\circ$, $150^\circ$, and $360^\circ$ by use of the DTV algorithm, along with their respective zoomed-in ROI views (rows 2 and 4). The zoomed-in ROI is enclosed by the rectangular box depicted in the FBP-reference image in Fig. \ref{fig:suitcase-mono-60}. Display window: [0.1, 0.65] cm$^{-1}$.}
	\label{fig:suitcase-mono-angles}
\end{figure*}

\paragraph{Visual inspection of monochromatic images}\label{sec:suitcase-noiseless-qual-vis}
%
We first show in Fig.~\ref{fig:suitcase-mono-60} monochromatic images and their respective zoomed-in ROI views at 40 keV obtained from the noiseless data generated over an arc of LAR $60^\circ$ by use of the DTV and FBP algorithms, along with the DTV- and FBP-reference images from the noiseless data over $360^\circ$. The zoomed-in ROI is enclosed by the rectangular box depicted in the FBP-reference image (row 1, column 4) in Fig.~\ref{fig:suitcase-mono-60}. It can be observed that the DTV image from data over $60^\circ$  displays significantly reduced LAR artifacts, which are otherwise observed and severely obscuring structures in the FBP image from the same data. The DTV image is also visually comparable to the DTV- and FBP-reference images. 
Furthermore, the contrast between water and ANFO and the edges in the three bar-shaped structures, as shown in the zoomed-in ROI views in the bottom row of Fig.~\ref{fig:suitcase-mono-60}, can be discerned in the DTV image of $60^\circ$ as clearly as that observed in the DTV- and FBP-reference images. 

We show in Fig.~\ref{fig:suitcase-mono-angles} monochromatic images and their zoomed-in ROI views obtained by use of the DTV algorithm from noiseless data over, respectively, 7 arcs of LARs $\alpha=14^\circ$, $20^\circ$, $30^\circ$, $60^\circ$, $90^\circ$, $120^\circ$, and $150^\circ$,  along with the DTV-reference image. It can be observed that while the DTV images of $\alpha \le 60^\circ$ contain some visible artifacts as a result of the compound LAR and BH effects, the edges in the bar-shaped structures in ROIs 0-2, as well as other structures, in the suitcase phantom can be discerned as clearly as that observed in the DTV- and FBP-reference images.  The remaining artifacts in the DTV images appear to be largely due to the BH effect. The monochromatic image from data over $180^\circ$ is visually similar to the DTV-reference image and is thus not shown in this study and the following studies.

\paragraph{Quantitative analysis of monochromatic images}\label{sec:suitcase-noiseless-quan-vis}

\begin{figure}[t!]
		\centering
		\includegraphics[width=0.24\textwidth]{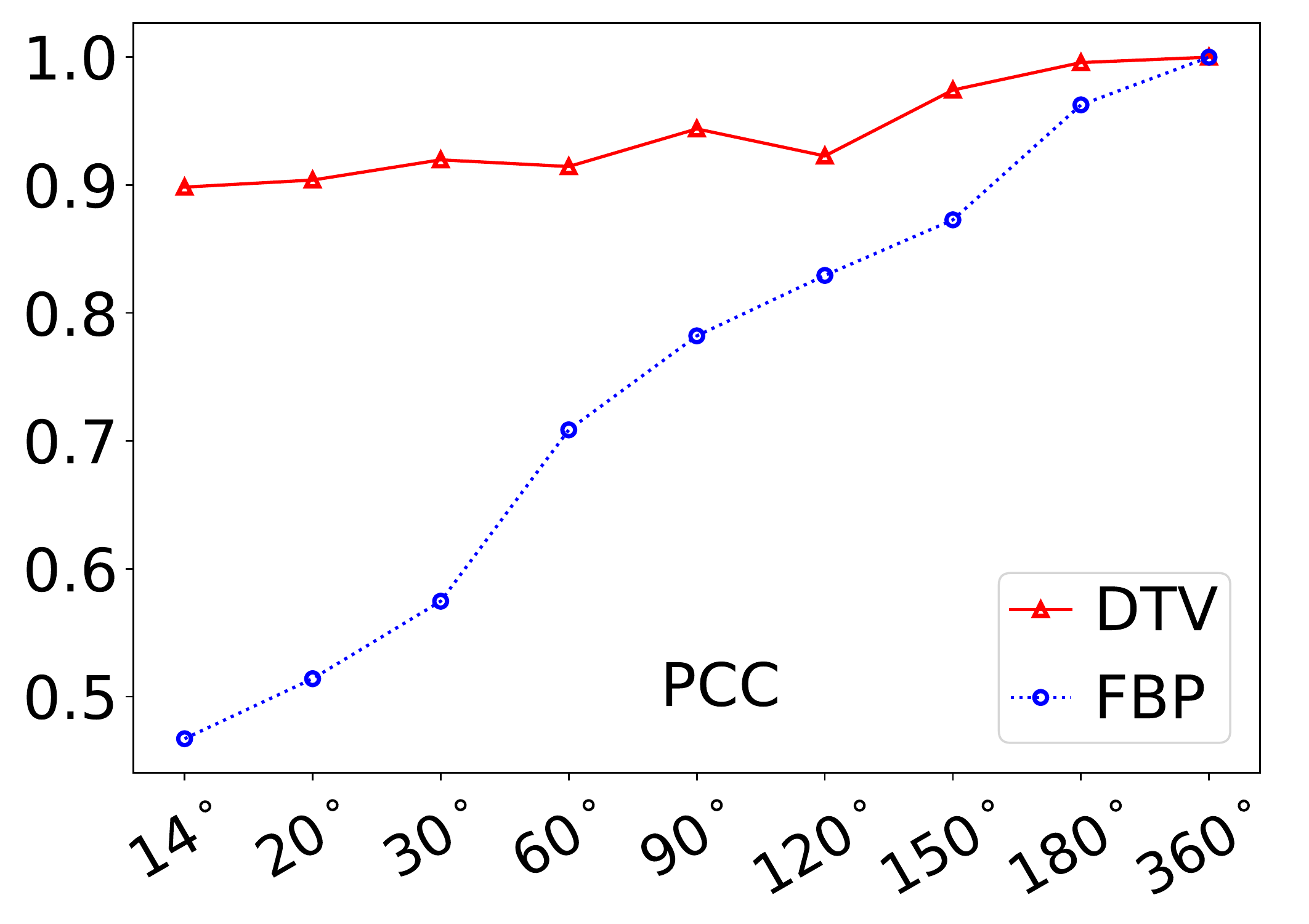}
		\includegraphics[width=0.24\textwidth]{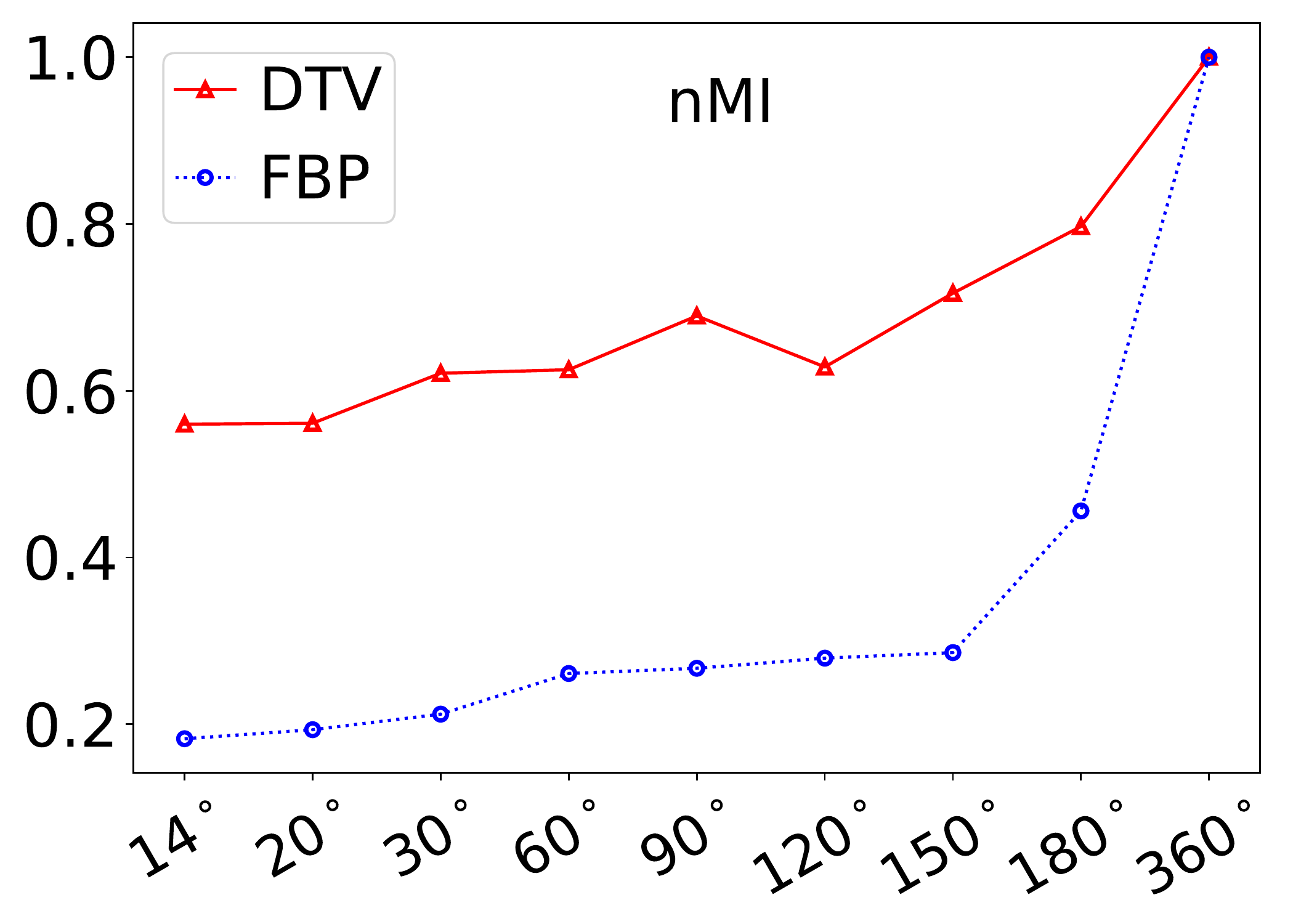}
	\caption{Metrics PCC and nMI, as functions of LAR $\alpha$, of monochromatic images of the suitcase at 40 keV obtained by use of the DTV (solid) and FBP (dotted) algorithms from noiseless data.}
	\label{fig:suitcase-metrics-tech}
\end{figure}

In addition to visual inspection, using the DTV-reference image (row 1, column 3) in Fig.~\ref{fig:suitcase-mono-60}, we compute metrics PCC and nMI of the DTV monochromatic images of the suitcase phantom, and display them in Fig.~\ref{fig:suitcase-metrics-tech} as functions of LAR $\alpha$. It can be observed that while the PCC and nMI drop understandably as $\alpha$ decreases, they remain generally above 0.9 and 0.5, respectively, suggesting that the DTV monochromatic images obtained with LAR data correlate reasonably well with the DTV-reference image.  For providing a benchmark, we also obtain monochromatic images by use of the FBP algorithm for $\alpha=14^\circ$, $20^\circ$, $30^\circ$, $60^\circ$, $90^\circ$, $120^\circ$, $150^\circ$, and $180^\circ$, but without showing them because the structures in the suitcase phantom are obscured by significant LAR artifacts in these FBP images, similar to those observed in the FBP image of $60^\circ$ shown in column 2 of Fig.~\ref{fig:suitcase-mono-60}.  Using the FBP-reference image (row 1, column 4) in Fig.~\ref{fig:suitcase-mono-60}, we compute metrics PCC and nMI of the FBP monochromatic images and plot them in Fig.~\ref{fig:suitcase-metrics-tech}. The results reveal that the FBP monochromatic images for $\alpha < 180^\circ$ correlate poorly with their reference image.


\paragraph{Estimation of atomic numbers}\label{sec:suitcase-noiseless-atomic-numbers}

\begin{figure}[t!]
		\centering
		\includegraphics[width=0.24\textwidth]{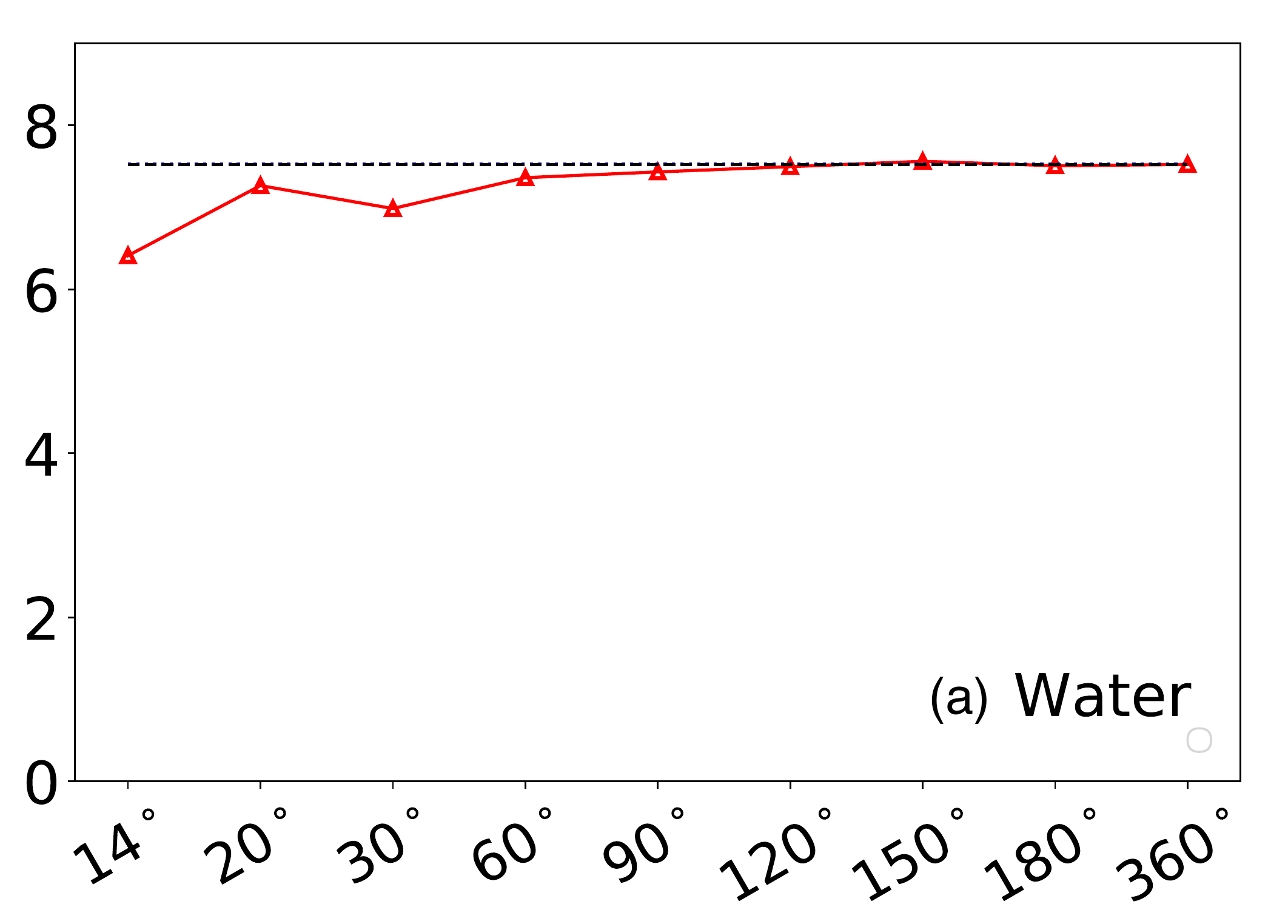}
		\includegraphics[width=0.24\textwidth]{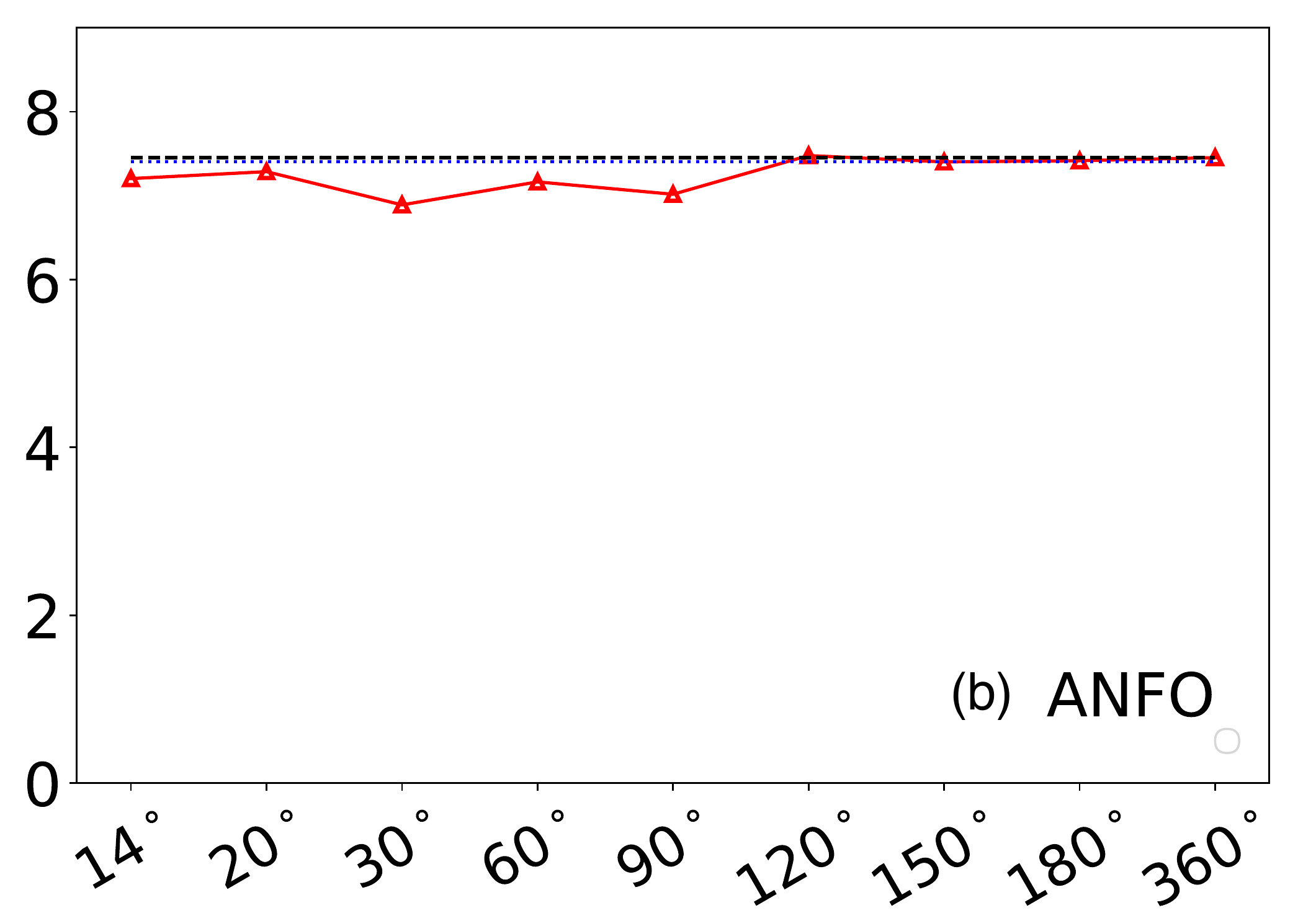}
		\includegraphics[width=0.24\textwidth]{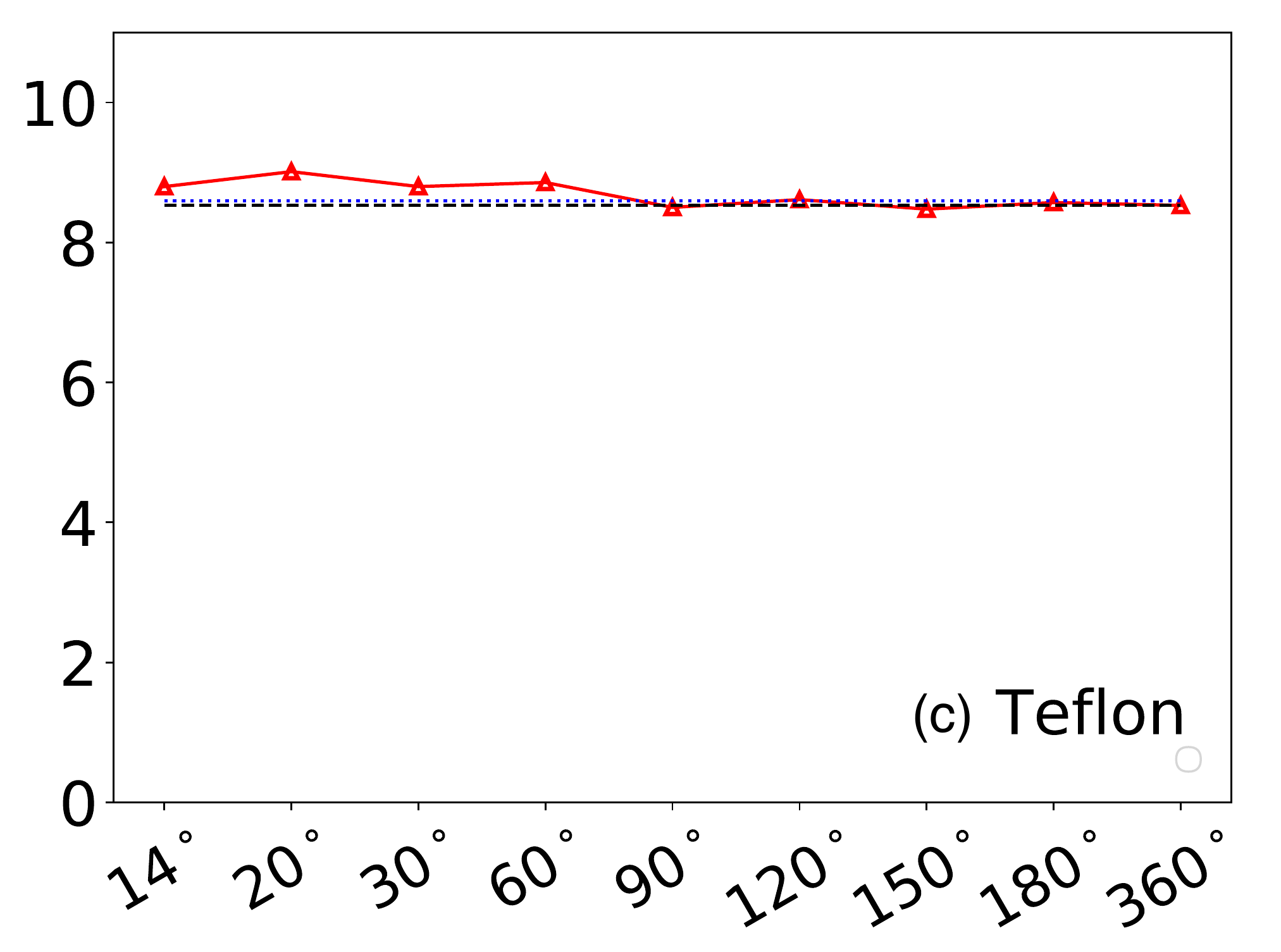}
		\includegraphics[width=0.24\textwidth]{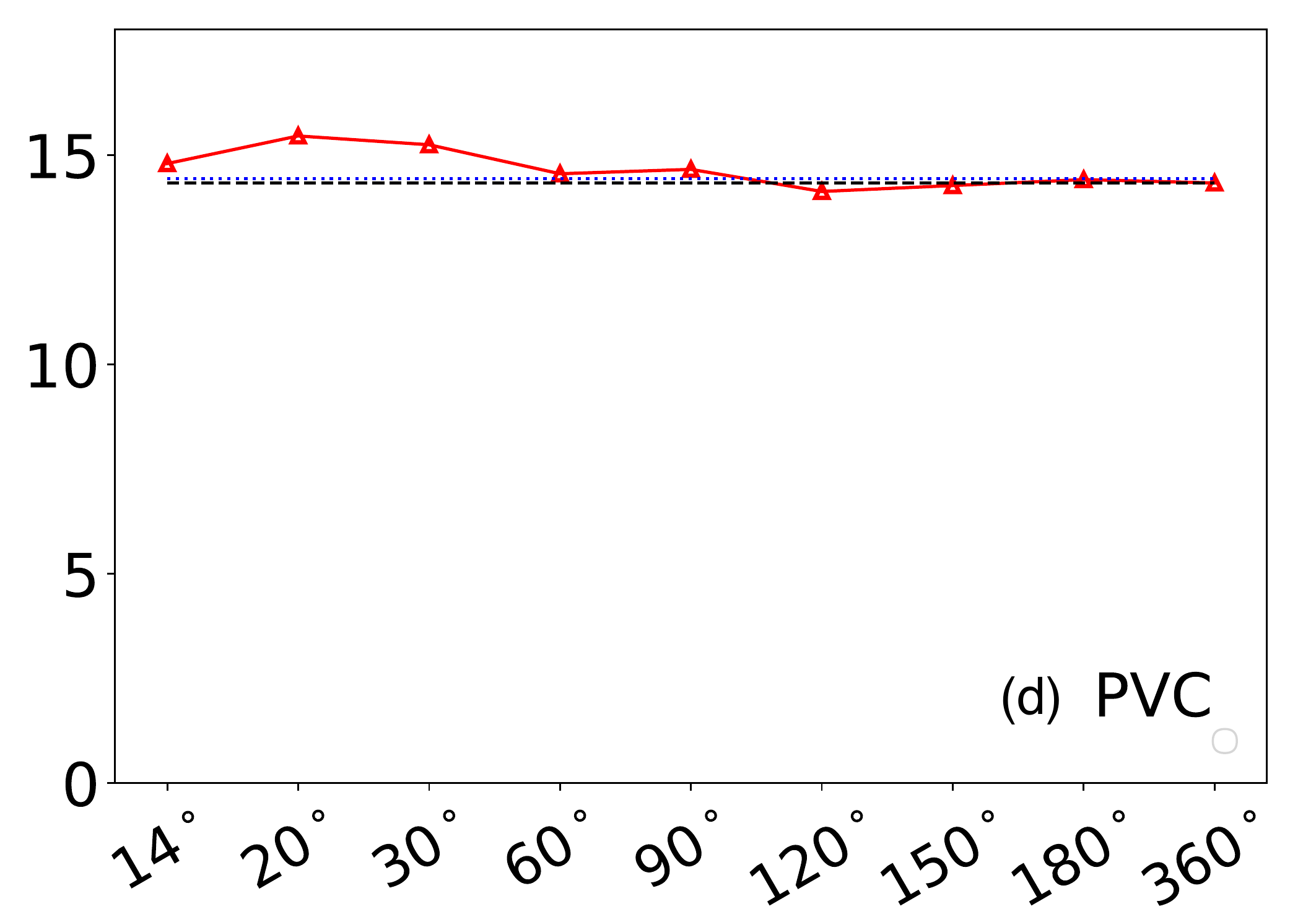}
	\caption{Atomic numbers (solid), as functions of LAR $\alpha$, for (a) water, (b) ANFO, (c) teflon, and (d) PVC in the suitcase phantom estimated from images reconstructed by use of the DTV algorithm from noiseless data. The two horizontal lines, which are very close, indicate the atomic numbers estimated from the DTV-reference (dashed)  and FBP-reference (dotted) images.}
	\label{fig:suitcase-metrics-task}
\end{figure}

Using the interaction-based method described in \ref{app:decomp} and \ref{app:tasks}, we compute atomic numbers for materials within ROIs 3-6 in the DTV images of the suitcase phantom as shown in Fig.~\ref{fig:phan}a. Specifically, using basis images estimated, and constants $c$ and $n$ in Eq.~\eqref{eq:z-log} fitted with calibration materials in ROIs 0-2 from the DTV-reference image, we obtain atomic numbers for materials water, ANFO, teflon, and PVC, respectively, in ROIs 3-6 and plot them as functions of angular range $\alpha$ in Fig.~\ref{fig:suitcase-metrics-task}, along with the atomic numbers obtained from the DTV- and FBP-reference images. The results indicate that the atomic numbers obtained with the DTV algorithm for the LARs considered appear to agree well with those obtained from their reference images, only with slight deviations observed for angular ranges less than $90^\circ$. Due to the severe LAR artifacts in the corresponding FBP images, their basis images estimated can be negative, and Eq.~\eqref{eq:z-log} thus cannot be applied because it involves the computation of a logarithmic. Therefore, no atomic numbers can be estimated from images obtained by use of the FBP algorithm for a majority of the LARs considered in the work. 


\subsection{Image reconstruction from noisy data of the suitcase phantom}\label{sec:suitcase-noisy}
We repeat the study of Sec.~\ref{sec:suitcase-noiseless} except that noisy data are now used, which are obtained by addition of Poisson noise to the corresponding noiseless data, as described in Sec. \ref{sec:methods-data}. 

\paragraph{Visual inspection of monochromatic images}\label{sec:suitcase-noisy-qual-vis}

\begin{figure*}[t!]
		\centering
		\begin{tabular}{c c c c}
			DTV-$14^\circ$ & DTV-$20^\circ$ & DTV-$30^\circ$ & DTV-$60^\circ$\\
			\includegraphics[width=0.22\textwidth]{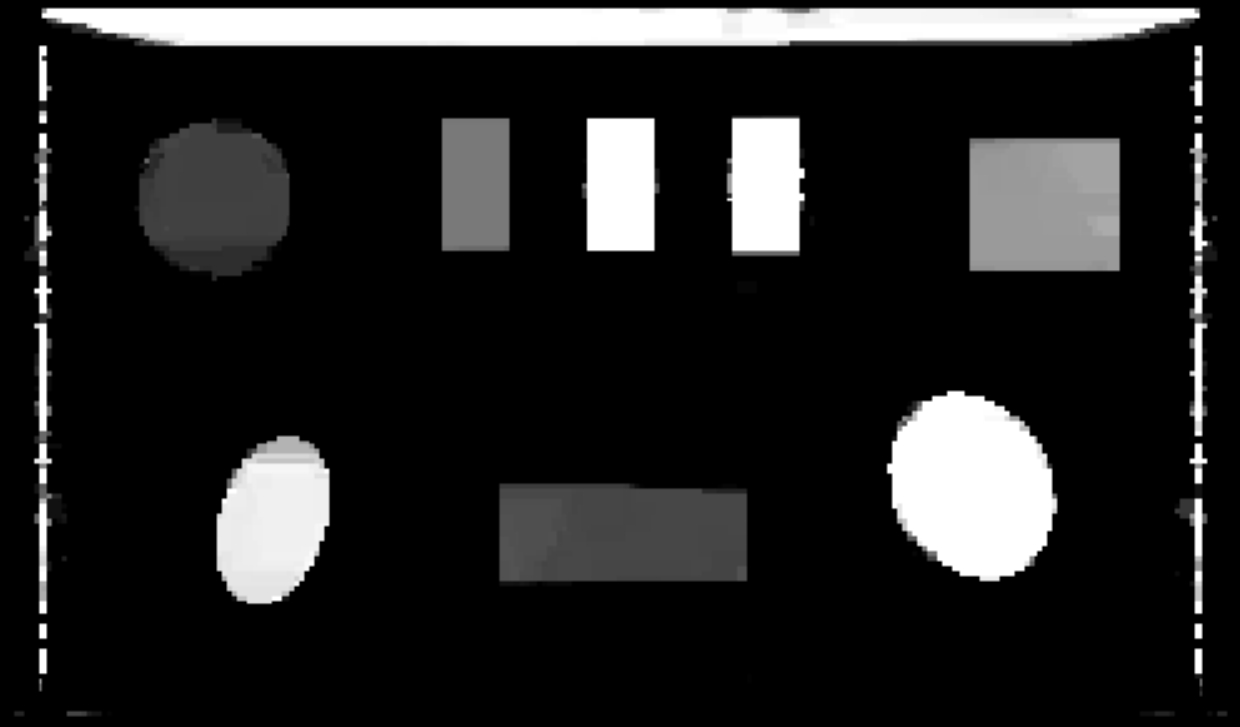}
			\hspace{-10pt} &
			\includegraphics[width=0.22\textwidth]{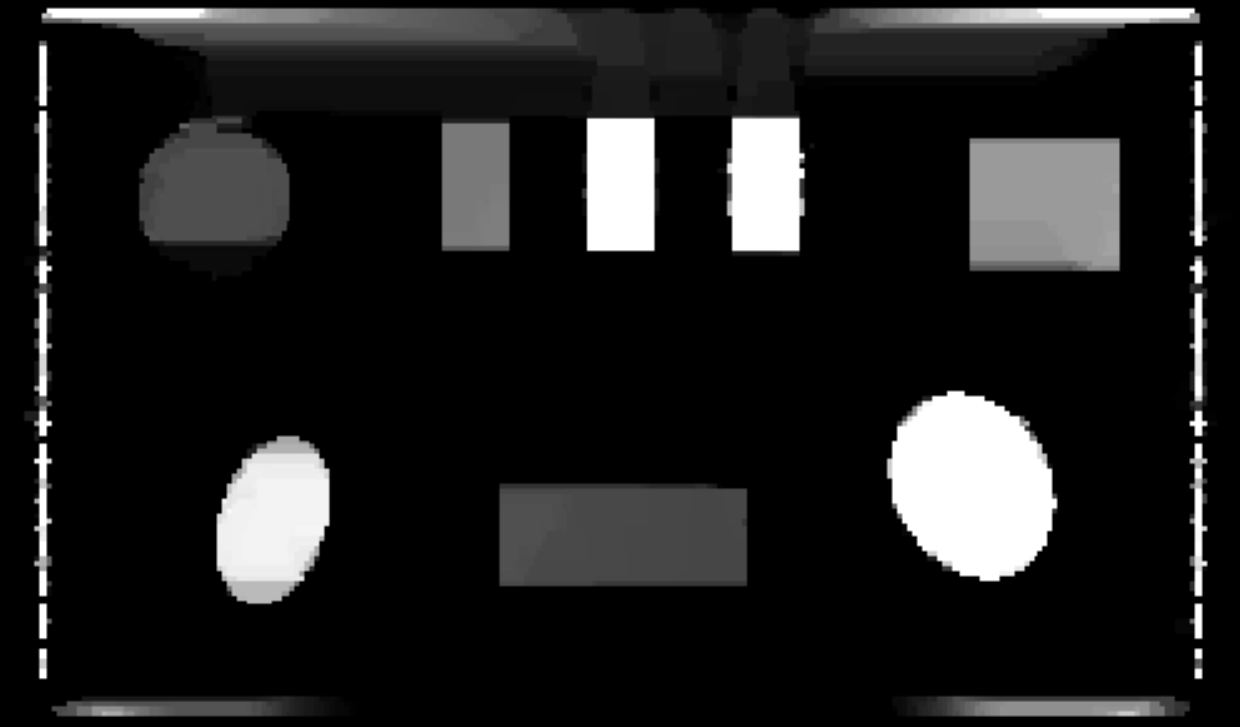}
			\hspace{-10pt} &
			\includegraphics[width=0.22\textwidth]{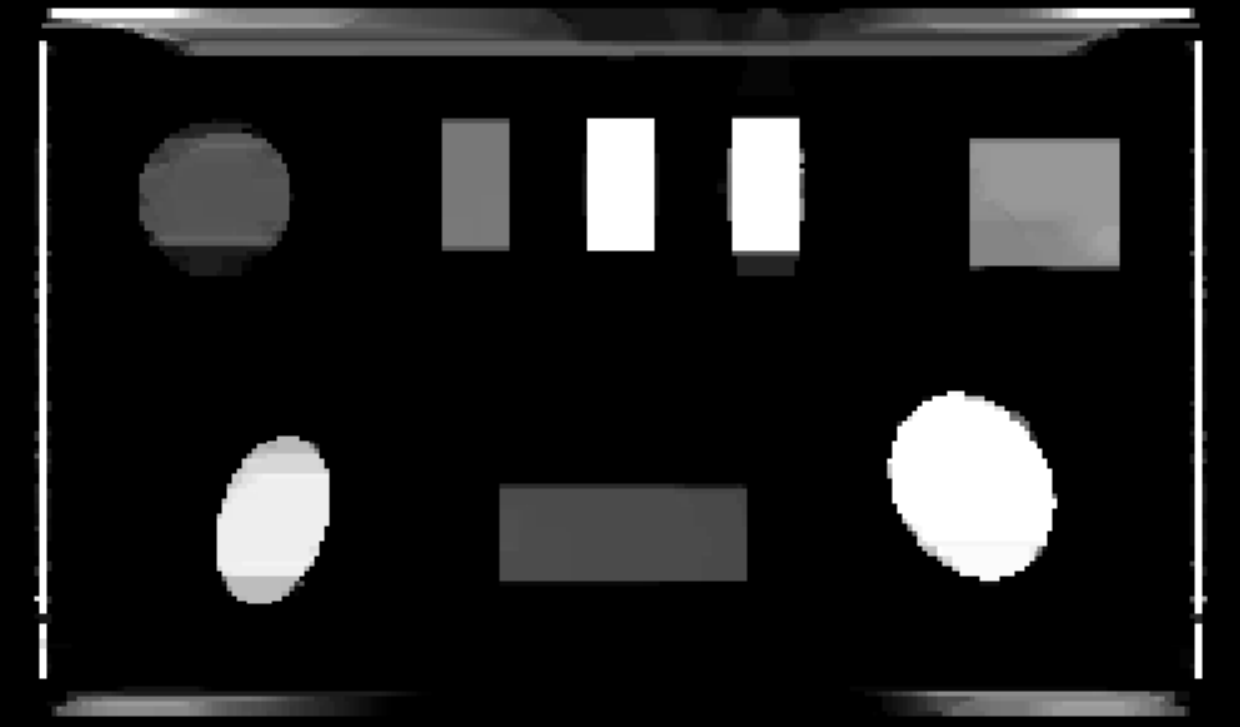}
			\hspace{-10pt} &
			\includegraphics[width=0.22\textwidth]{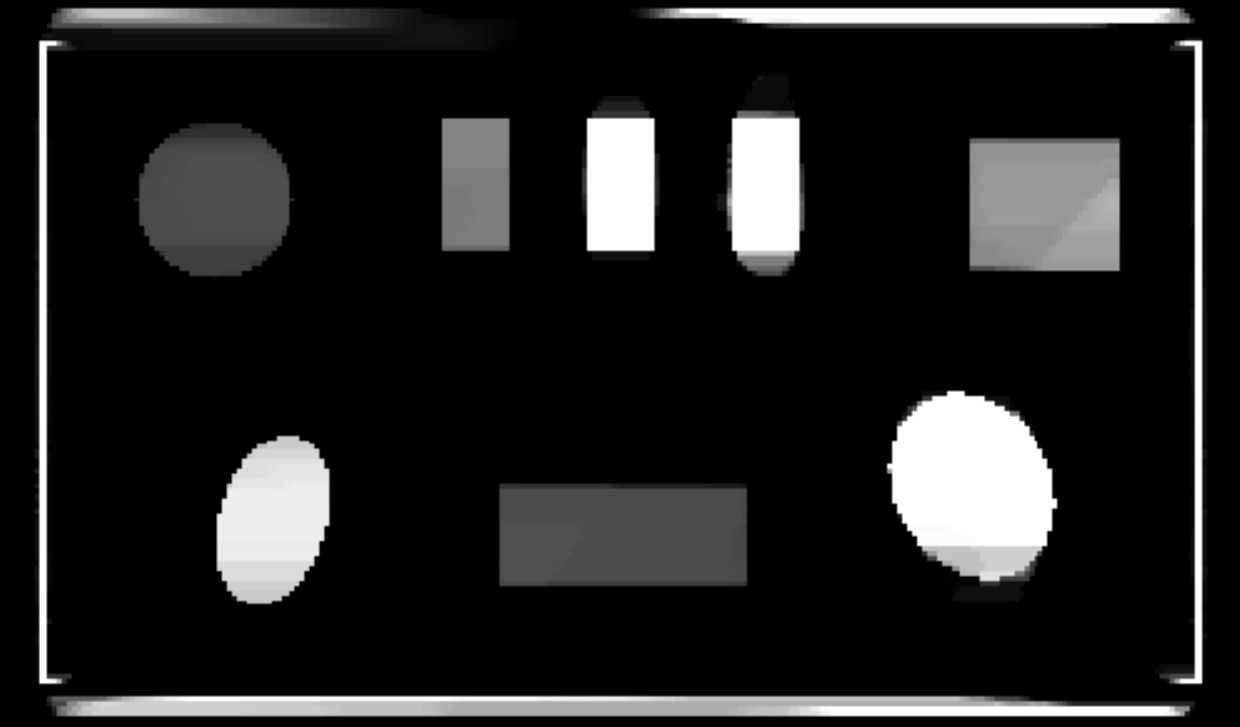} %
			\\
			\includegraphics[width=0.22\textwidth,trim={120 130 120 25}, clip]{figures/barPhan4/Noise_1e7/XY_pdf/barPhan4_noise1e7_XY_14D_40keV.pdf}
			\hspace{-10pt} &
			\includegraphics[width=0.22\textwidth,trim={120 130 120 25}, clip]{figures/barPhan4/Noise_1e7/XY_pdf/barPhan4_noise1e7_XY_20D_40keV.pdf}
			\hspace{-10pt} &
			\includegraphics[width=0.22\textwidth,trim={120 130 120 25}, clip]{figures/barPhan4/Noise_1e7/XY_pdf/barPhan4_noise1e7_XY_30D_40keV.pdf}
			\hspace{-10pt} &
			\includegraphics[width=0.22\textwidth,trim={120 130 120 25}, clip]{figures/barPhan4/Noise_1e7/XY_pdf/barPhan4_noise1e7_XY_60D_40keV.pdf} %
			\\
			DTV-$90^\circ$ & DTV-$120^\circ$ & DTV-$150^\circ$ & DTV-$360^\circ$ \\
			\includegraphics[width=0.22\textwidth]{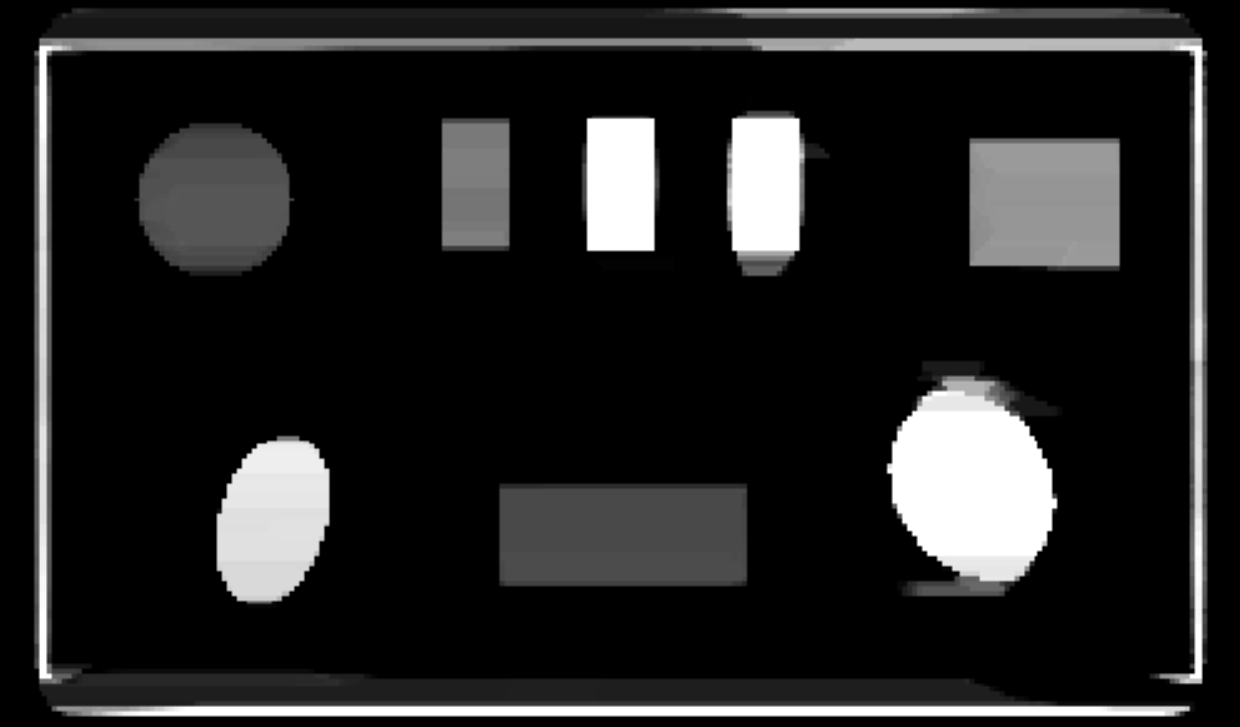}
			\hspace{-10pt} &
			\includegraphics[width=0.22\textwidth]{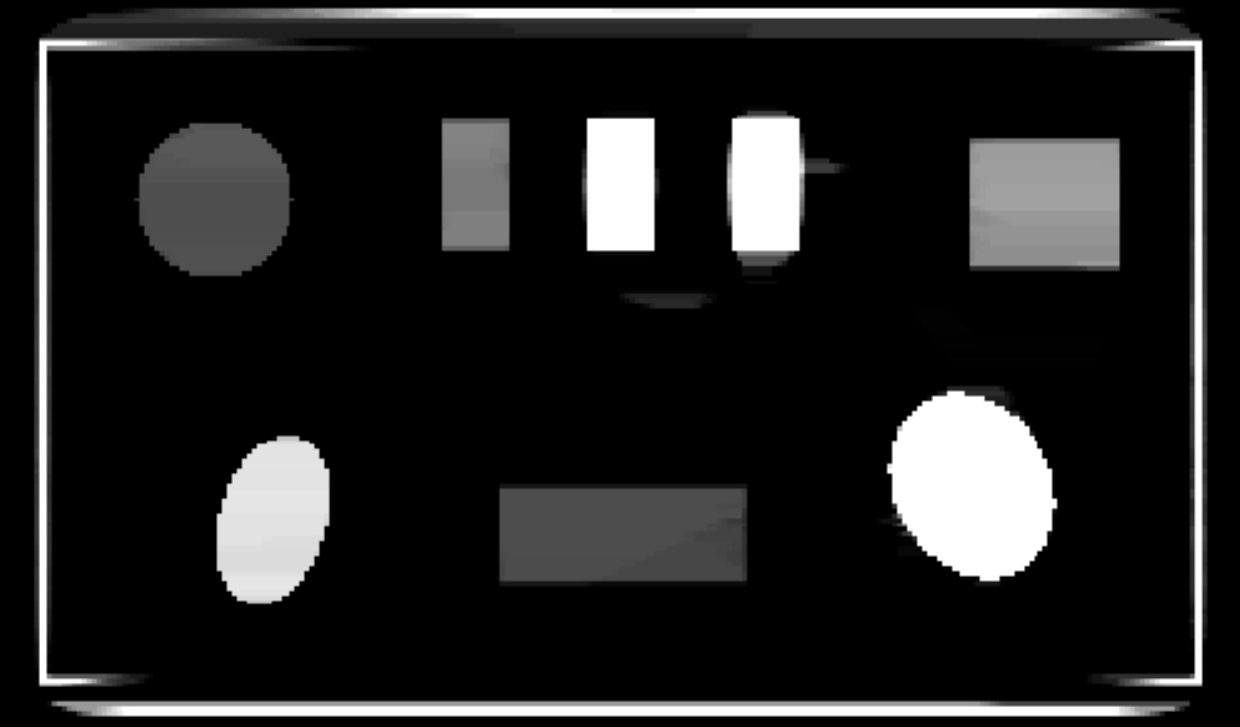}
			\hspace{-10pt} &
			\includegraphics[width=0.22\textwidth]{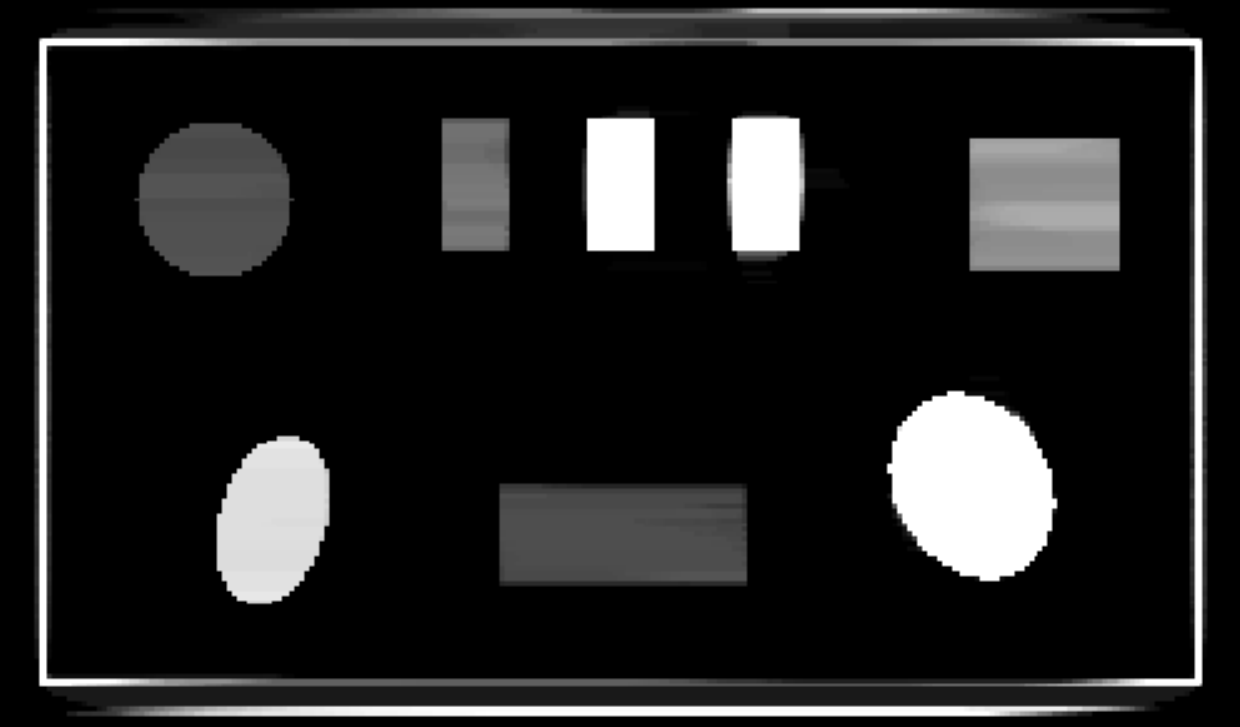}
			\hspace{-10pt} &
			\includegraphics[width=0.22\textwidth]{figures/barPhan4/Noise_1e7/XY_pdf/barPhan4_noise1e7_XY_150D_40keV.pdf} %
			\\
			\includegraphics[width=0.22\textwidth,trim={120 130 120 25}, clip]{figures/barPhan4/Noise_1e7/XY_pdf/barPhan4_noise1e7_XY_90D_40keV.pdf}
			\hspace{-10pt} &
			\includegraphics[width=0.22\textwidth,trim={120 130 120 25}, clip]{figures/barPhan4/Noise_1e7/XY_pdf/barPhan4_noise1e7_XY_120D_40keV.pdf}
			\hspace{-10pt} &
			\includegraphics[width=0.22\textwidth,trim={120 130 120 25}, clip]{figures/barPhan4/Noise_1e7/XY_pdf/barPhan4_noise1e7_XY_150D_40keV.pdf}
			\hspace{-10pt} &
			\includegraphics[width=0.22\textwidth,trim={120 130 120 25}, clip]{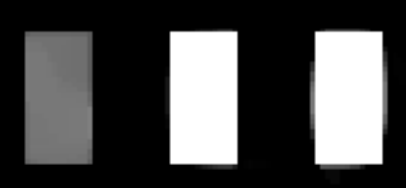}
	\end{tabular}
	\caption{Monochromatic images (rows 1 and 3) of the suitcase phantom at 40 keV obtained from noisy data generated over arcs of LARs $14^\circ$, $20^\circ$, $30^\circ$, $60^\circ$, $90^\circ$, $120^\circ$, $150^\circ$, and $360^\circ$ by use of the DTV algorithm, along with their respective zoomed-in ROI views (rows 2 and 4). The zoomed-in ROI is enclosed by the rectangular box depicted in the FBP-reference image in Fig.~\ref{fig:suitcase-mono-60}. Display window: [0.1, 0.65] cm$^{-1}$.}
	\label{fig:suitcase-mono-angles-noisy}
\end{figure*}

We show in Fig.~\ref{fig:suitcase-mono-angles-noisy} monochromatic images and their zoomed-in ROI views obtained by use of the DTV algorithm from noisy data over, respectively, 7 arcs of LARs $\alpha=14^\circ$, $20^\circ$, $30^\circ$, $60^\circ$, $90^\circ$, $120^\circ$, and $150^\circ$, along with the FAR of $360^\circ$.  It can be observed that the DTV images appear to contain only significantly reduced visual artifacts as a result of the compound LAR, BH, and noise, and that structures in the suitcase phantom can be discerned as clearly as that observed in the DTV- and FBP-reference images in Fig.~\ref{fig:suitcase-mono-60}.  

\paragraph{Quantitative analysis of monochromatic images}\label{sec:suitcase-noisy-quan-vis}

\begin{figure}[t!]
		\centering
		\includegraphics[width=0.24\textwidth]{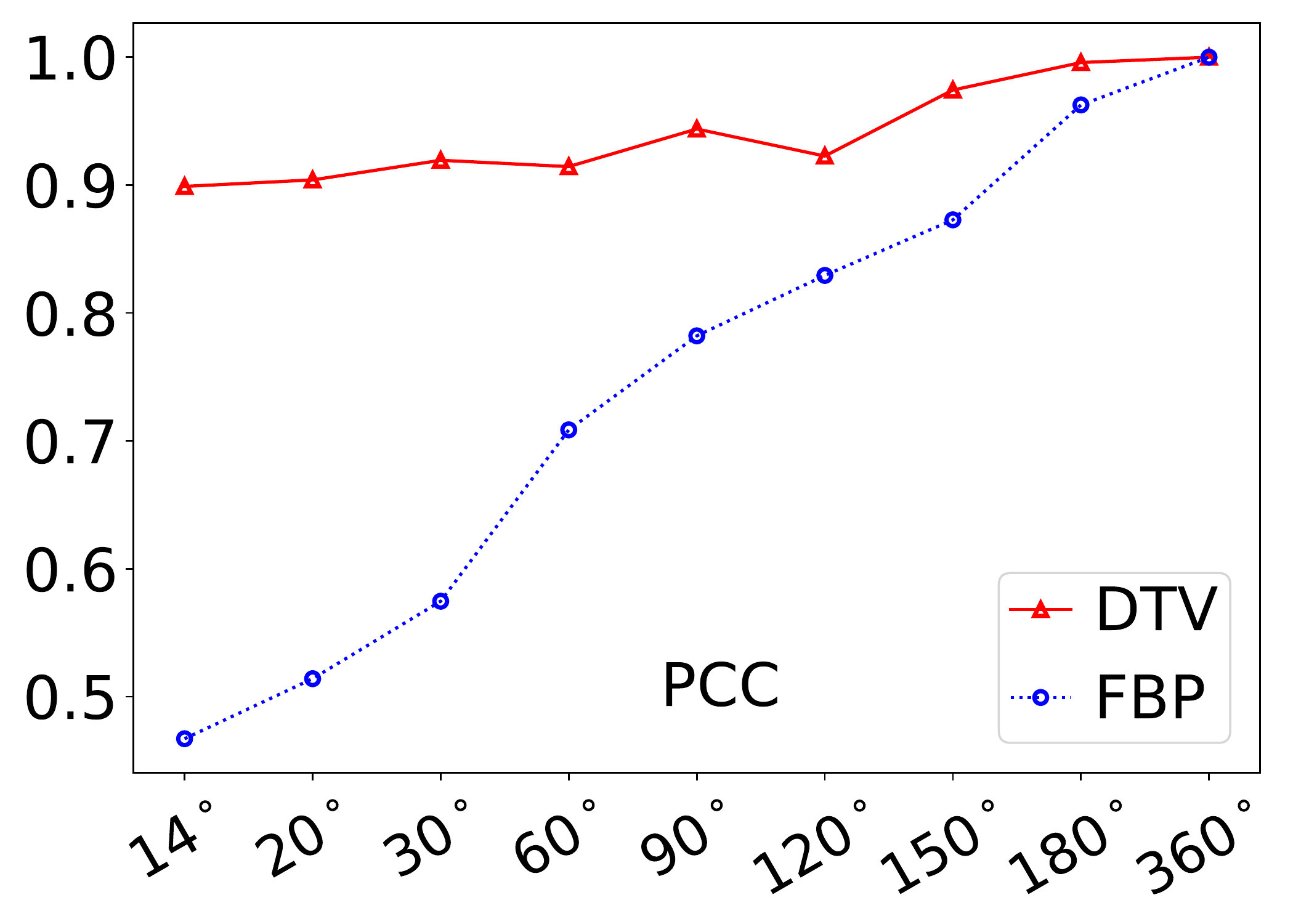}
		\includegraphics[width=0.24\textwidth]{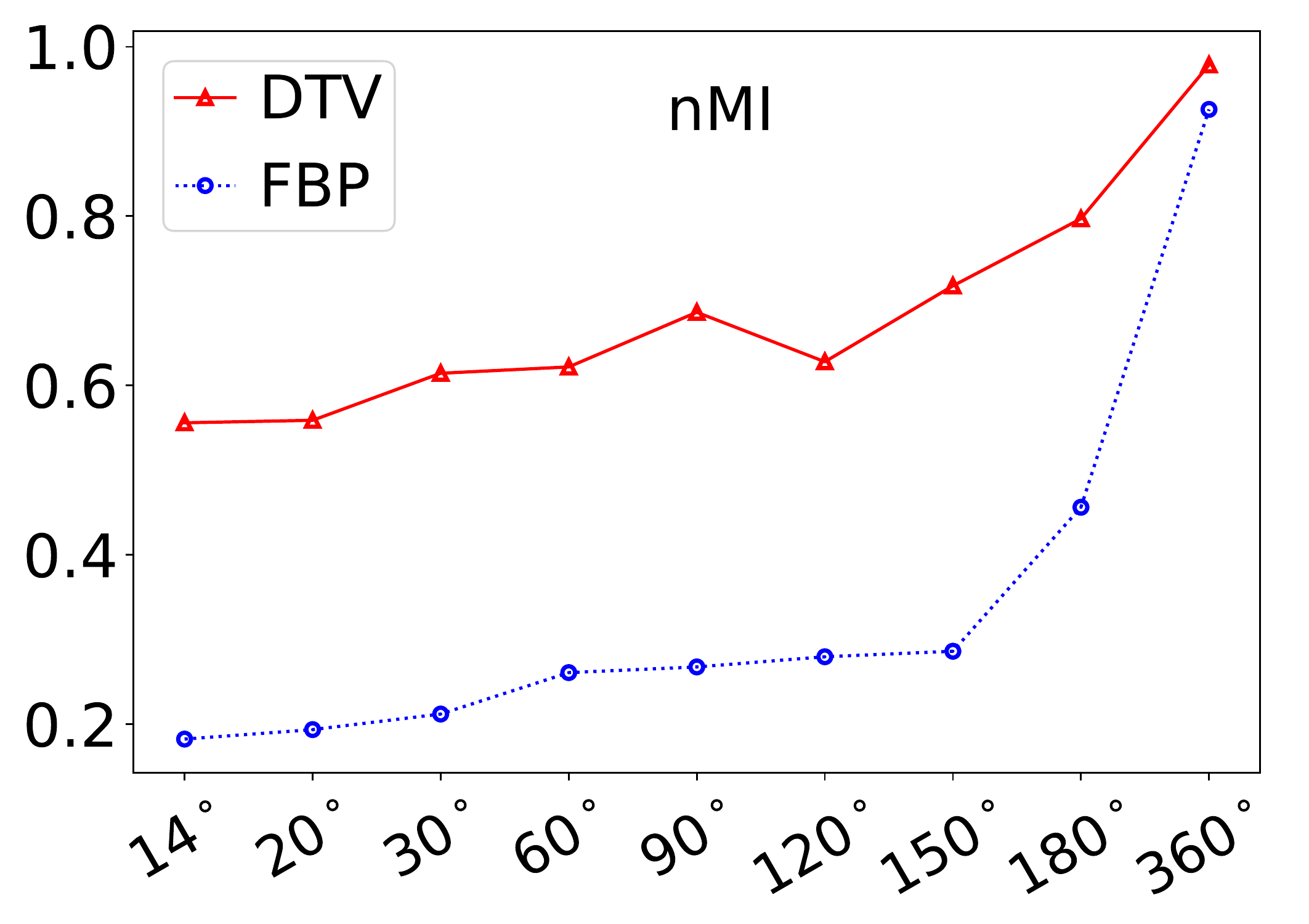}
	\caption{Metrics PCC and nMI, as functions of LAR $\alpha$, of monochromatic images of the suitcase at 40 keV obtained by use of the DTV (solid) and FBP (dotted) algorithms from noisy data.}
	\label{fig:suitcase-metrics-tech-noisy}
\end{figure}

Using the DTV-reference image (row 1, column 3) in Fig.~\ref{fig:suitcase-mono-60}, we also compute metrics PCC and nMI of the noisy DTV monochromatic images and show them in Fig.~\ref{fig:suitcase-metrics-tech-noisy} as functions of LAR $\alpha$. It can be observed that while the PCC and nMI drop understandably as $\alpha$ decreases, they remain generally above 0.9 and 0.5, respectively, suggesting that the DTV monochromatic images obtained with noisy LAR data correlate reasonably well with the DTV-reference image.  For providing a benchmark, we also obtain monochromatic images by use of the FBP algorithm for LARs $\alpha=14^\circ$, $20^\circ$, $30^\circ$, $60^\circ$, $90^\circ$, $120^\circ$, $150^\circ$, and $180^\circ$, but without showing them because the structures in the suitcase phantom are obscured by significant LAR artifacts in these FBP images, similar to those observed in the FBP image of $60^\circ$ shown in column 2 of Fig.~\ref{fig:suitcase-mono-60}.  Using the FBP-reference image (row 1, column 4) in Fig.~\ref{fig:suitcase-mono-60}, we compute metrics PCC and nMI of the FBP monochromatic images and plot them in Fig.~\ref{fig:suitcase-metrics-tech-noisy}. The results reveal that the FBP monochromatic images for $\alpha < 180^\circ$ correlate poorly with their reference image.

\paragraph{Estimation of atomic numbers}\label{sec:suitcase-noisy-atomic-numbers}

\begin{figure}[t!]
		\centering
		\includegraphics[width=0.24\textwidth]{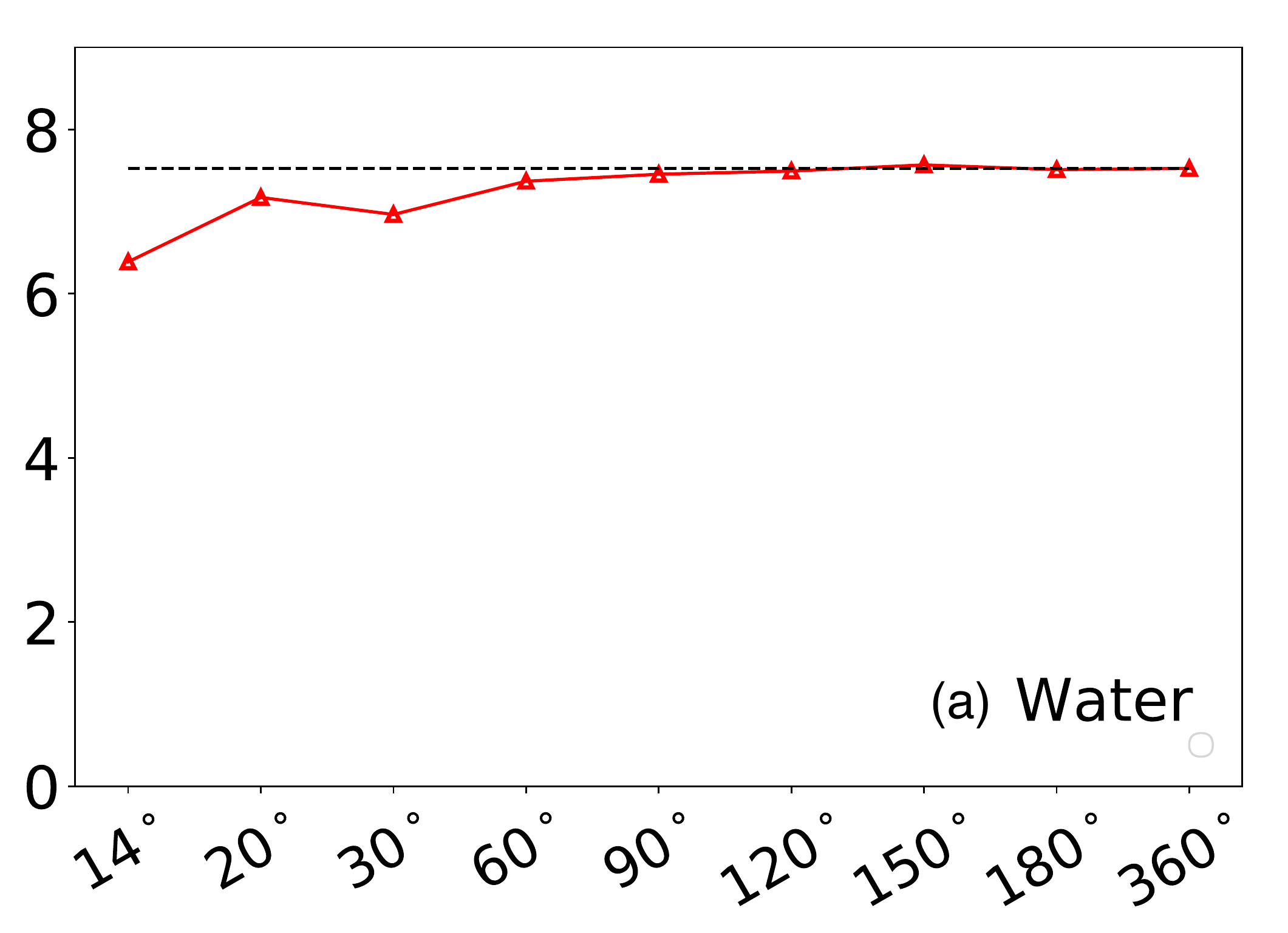}
		\includegraphics[width=0.24\textwidth]{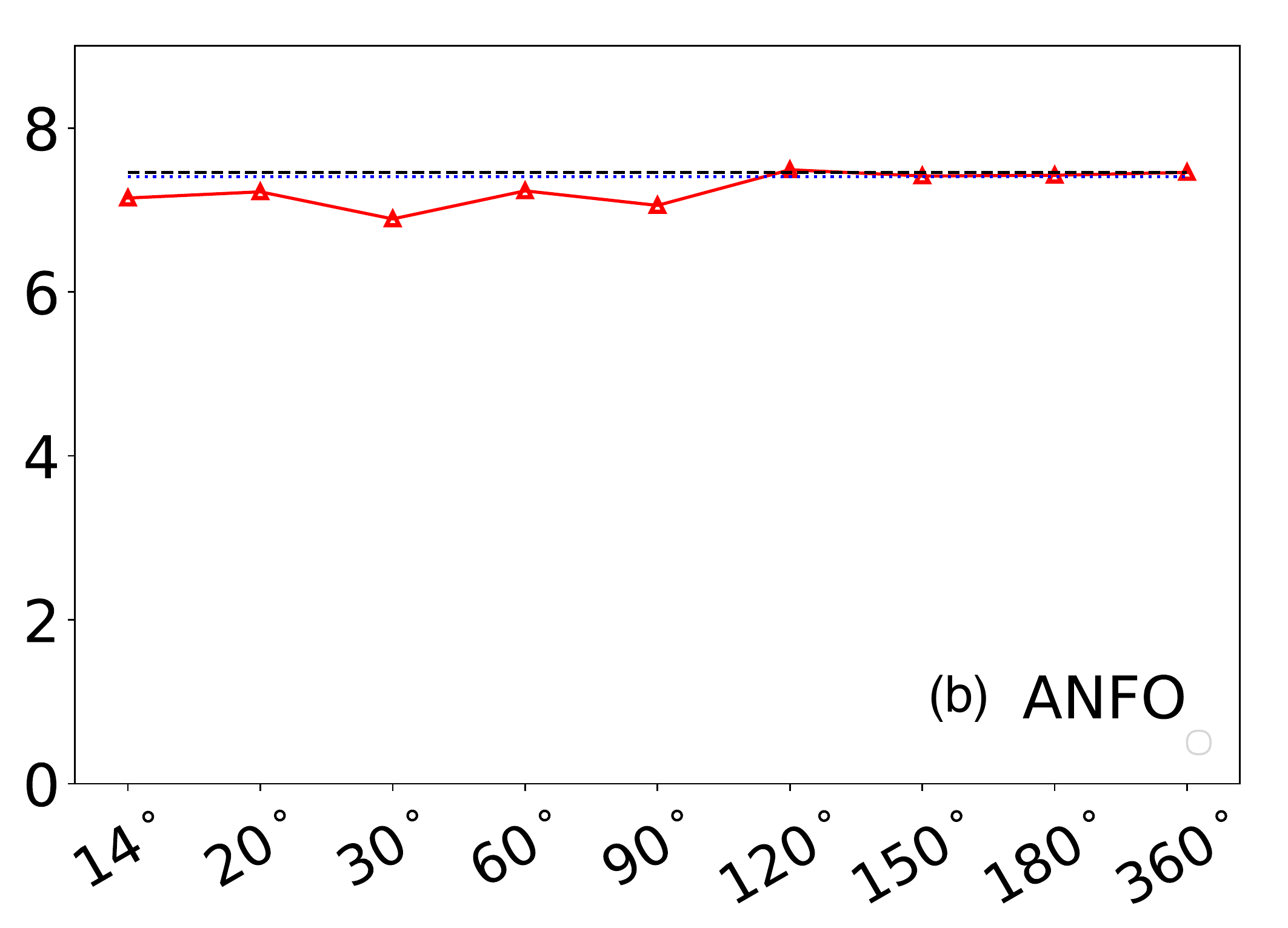}
		\includegraphics[width=0.24\textwidth]{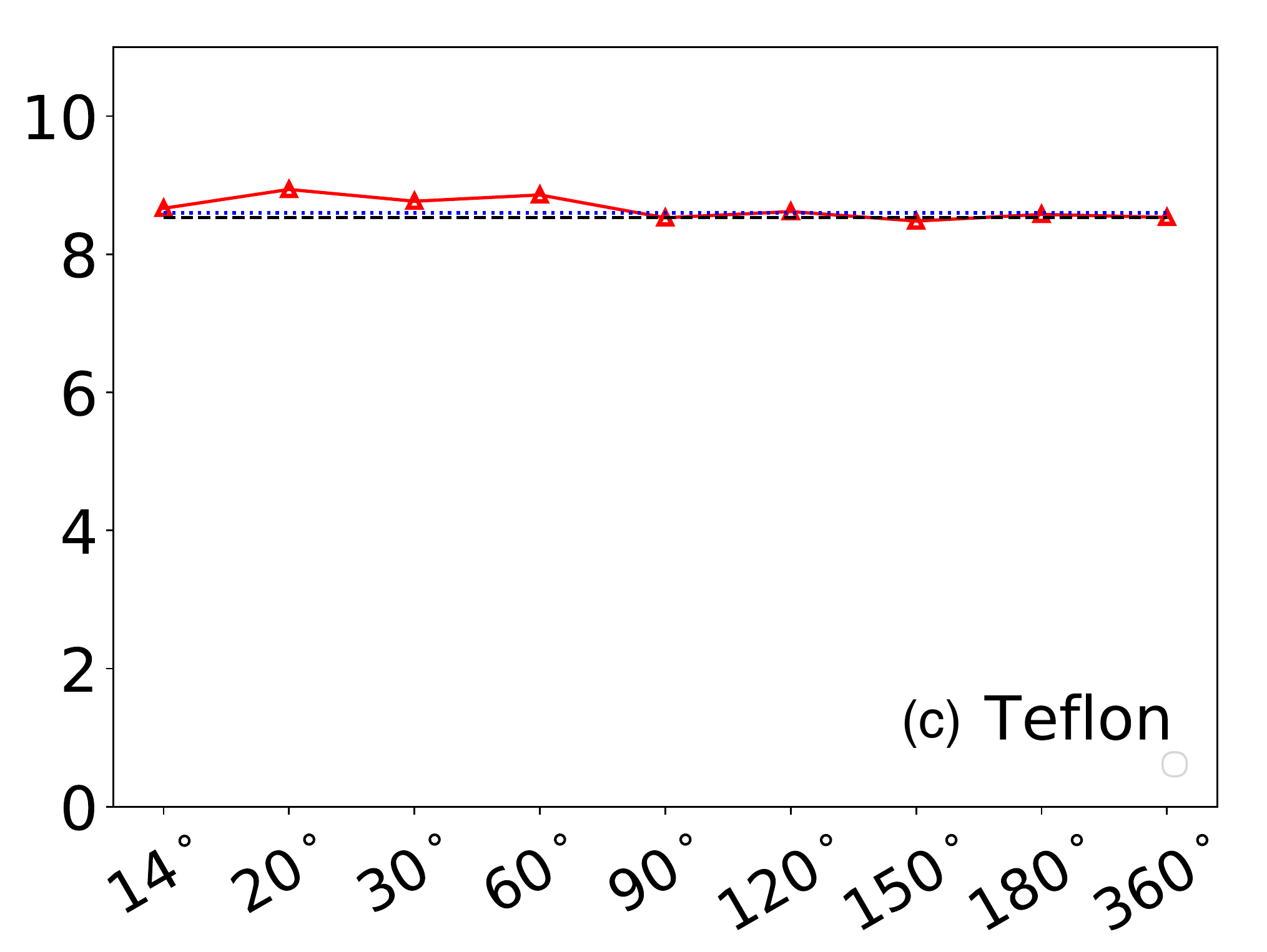}
		\includegraphics[width=0.24\textwidth]{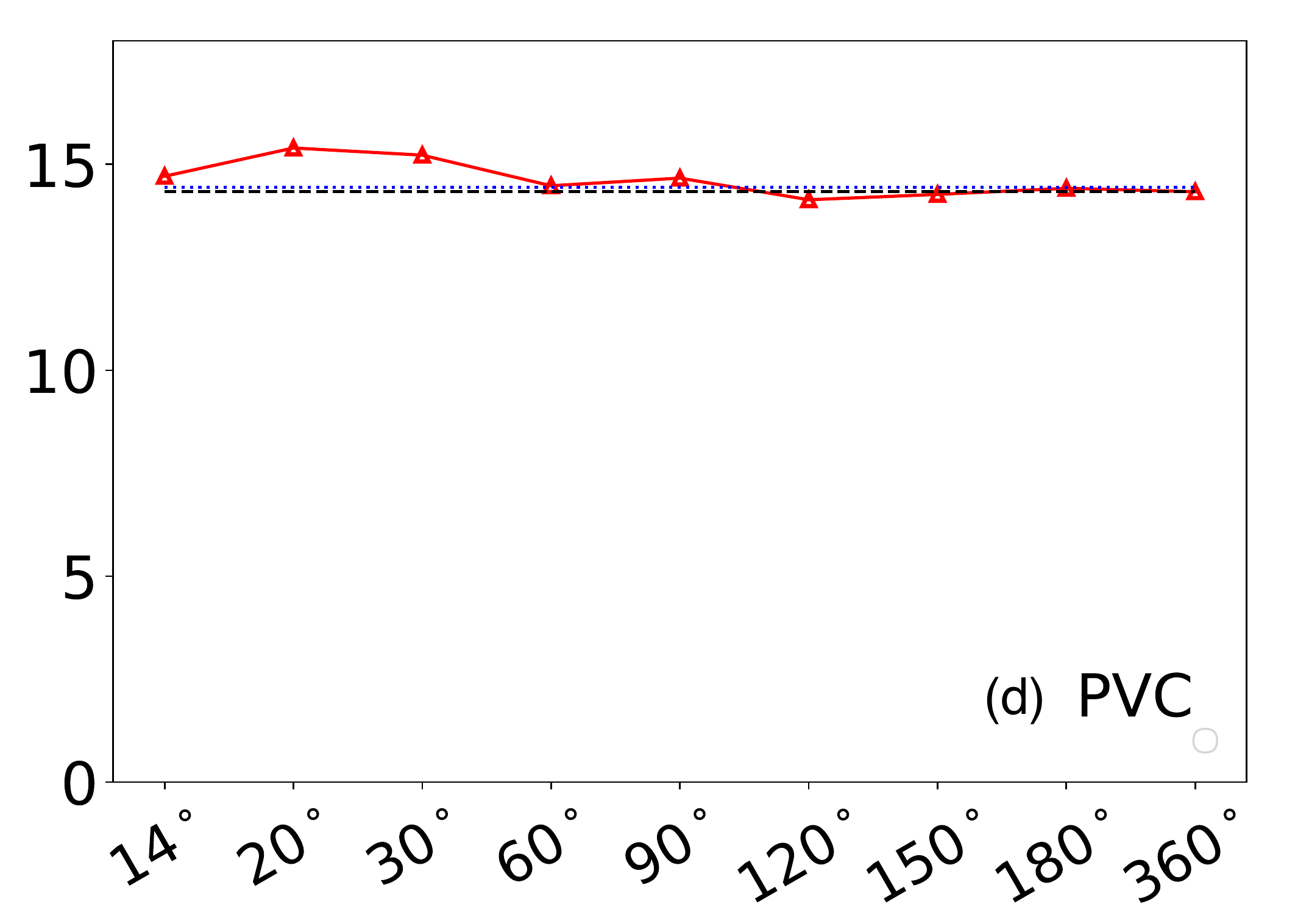}
	\caption{Atomic numbers (solid), as functions of LAR $\alpha$, for (a) water, (b) ANFO, (c) teflon, and (d) PVC in the suitcase phantom estimated from images reconstructed by use of the DTV algorithm from noisy data. The two horizontal lines, which are very close, indicate the atomic numbers estimated from the DTV-reference (dashed)  and FBP-reference (dotted) images.}
	\label{fig:suitcase-metrics-task-noisy}
\end{figure}

Using the interaction-based method described in \ref{app:decomp} and \ref{app:tasks}, we also compute atomic numbers for materials water, ANFO, teflon, and PVC, respectively, within ROIs 3-6 in the DTV images of the suitcase phantom, and plot them as functions of LAR $\alpha$ in Fig.~\ref{fig:suitcase-metrics-task-noisy}, along with the atomic numbers obtained from the DTV- and FBP-reference images. The results indicate that the atomic numbers obtained with the DTV algorithm for the LARs considered appear to agree well with those obtained from their reference images, only with slight deviations observed for LARs less than $90^\circ$. Due to the severe artifacts in images reconstructed by use of the FBP algorithm from LAR data, their basis images estimated can be negative, and Eq.~\eqref{eq:z-log} thus cannot be applied because it involves the computation of a logarithmic. Therefore, no atomic numbers can be estimated from images obtained by use of the FBP algorithm for a majority of the LARs considered in the work. 


%
\subsection{Image reconstruction from noiseless data of the  breast phantom}\label{sec:breast-noiseless}


We reconstruct images from noiseless low- and high-kVp data of the breast phantom generated over arcs of LARs $\alpha=14^\circ, 20^\circ, 30^\circ, 60^\circ, 90^\circ, 120^\circ$, $150^\circ$, and $180^\circ$ by use of the DTV algorithm, and then estimate basis images by using the material-based method with the DTV-reference image, as described in \ref{app:decomp}, from the images reconstructed. Subsequently, using the basis images estimated, we obtain monochromatic image at energy 34 keV for enhanced iodine contrast, and compute iodine concentrations within the selected ROIs as described in \ref{app:tasks}.

\begin{figure*}[t!]
		\centering
		\begin{tabular}{c c c c}
			DTV-$60^\circ$ & FBP-$60^\circ$  & DTV-$360^\circ$ & FBP-$360^\circ$ \\
			\includegraphics[width=0.22\textwidth,trim={10 0 10 0}, clip]{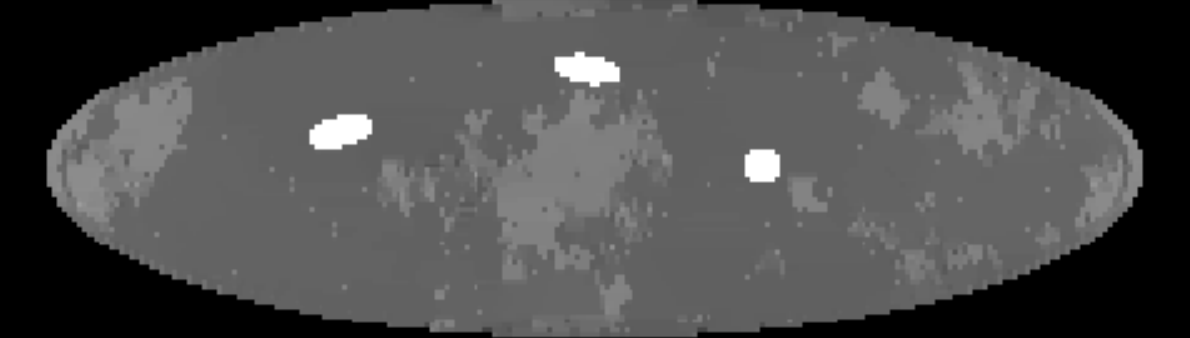}
			\hspace{-10pt} & 
			\includegraphics[width=0.22\textwidth,trim={10 0 10 0}, clip]{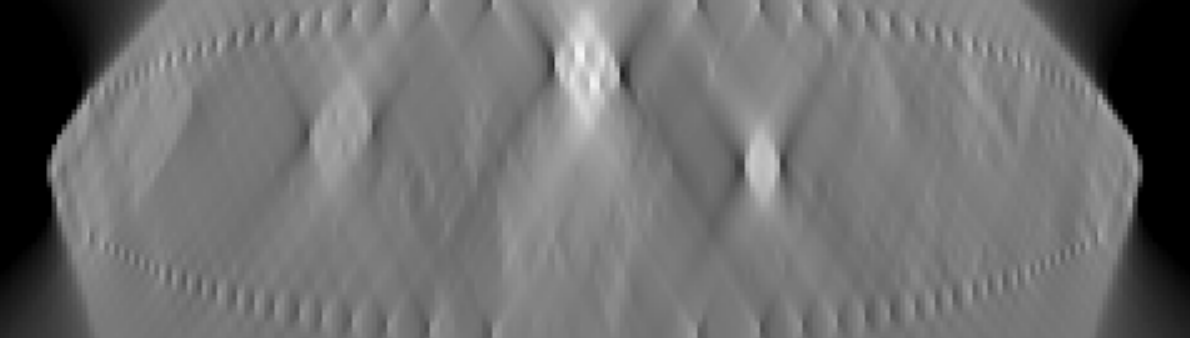}
			\hspace{-10pt} & 
			\includegraphics[width=0.22\textwidth,trim={10 0 10 0}, clip]{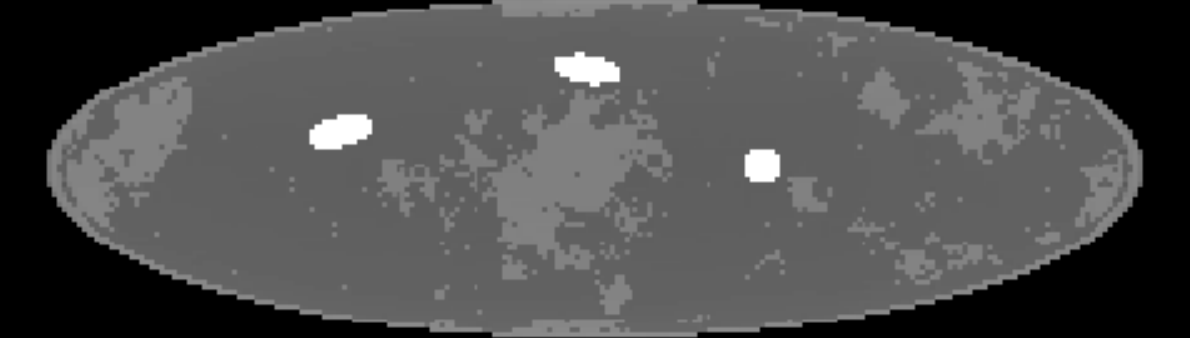}
			\hspace{-10pt} & 
			\includegraphics[width=0.22\textwidth,trim={10 0 10 0}, clip]{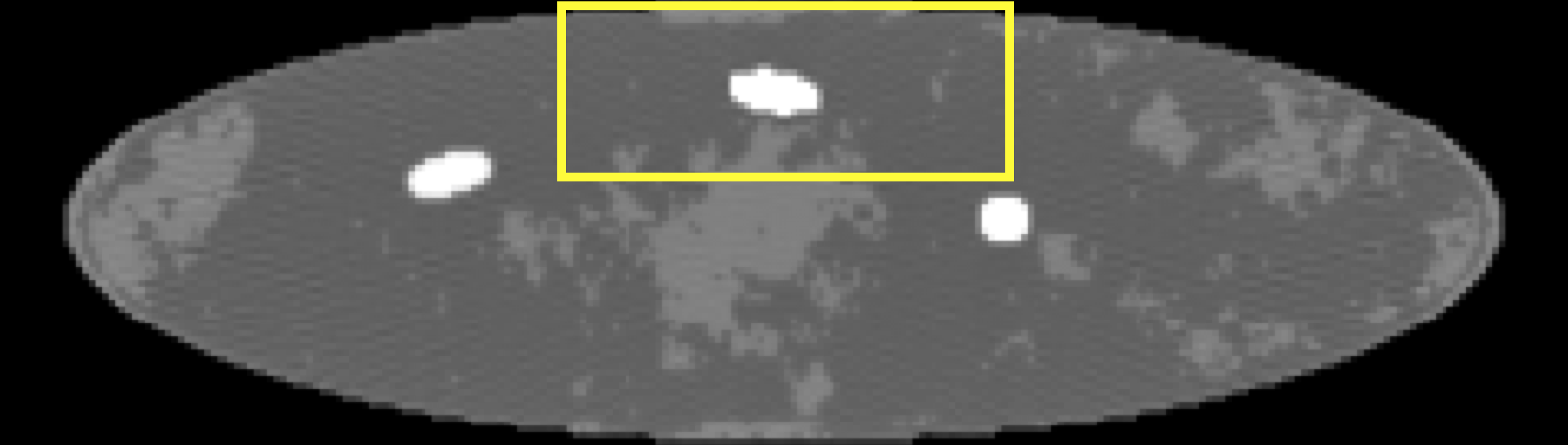}
			\\
			\includegraphics[width=0.22\textwidth,trim={110 60 100 0}, clip]{figures/breast2/Noiseless/XY_pdf/breast_XY_60D_34keV.pdf}
			\hspace{-10pt} & 
			\includegraphics[width=0.22\textwidth,trim={110 60 100 0}, clip]{figures/breast2/Noiseless/FBP_pdf/breast_FBP_60D_34keV.pdf}
			\hspace{-10pt} &
			\includegraphics[width=0.22\textwidth,trim={110 60 100 0}, clip]{figures/breast2/Noiseless/XY_pdf/breast_XY_360D_34keV.pdf}
			\hspace{-10pt} & 
			\includegraphics[width=0.22\textwidth,trim={110 60 100 0}, clip]{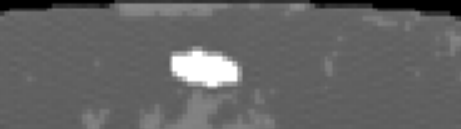} 
		\end{tabular}
	\caption{Monochromatic images at 34 keV (top row) and their respective zoomed-in ROI views (bottom row) within the breast phantom obtained by use of the DTV (column 1) and FBP (column 2) algorithms from noiseless data generated over an arc of LAR $60^\circ$, and the DTV-reference image (column 3) and FBP-reference image (column 4). The zoomed-in ROI is enclosed by the rectangular box depicted in the FBP-reference image. Display window: [0.2, 0.35] cm$^{-1}$.}
	\label{fig:breast-mono-60}
\end{figure*}

\begin{figure*}[t!]
		\centering
		\begin{tabular}{c c c c}
			DTV-$14^\circ$ & DTV-$20^\circ$ & DTV-$30^\circ$ & DTV-$60^\circ$ \\
			\includegraphics[width=0.22\textwidth,trim={10 0 10 0}, clip]{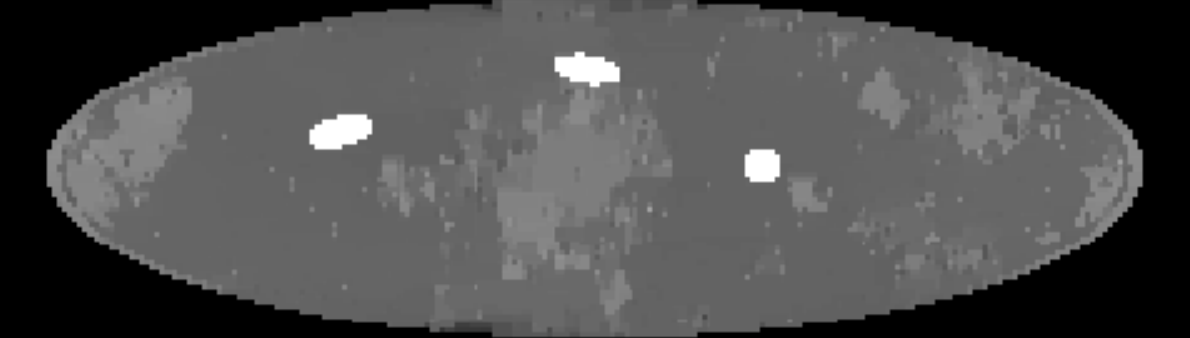}
			\hspace{-10pt} &
			\includegraphics[width=0.22\textwidth,trim={10 0 10 0}, clip]{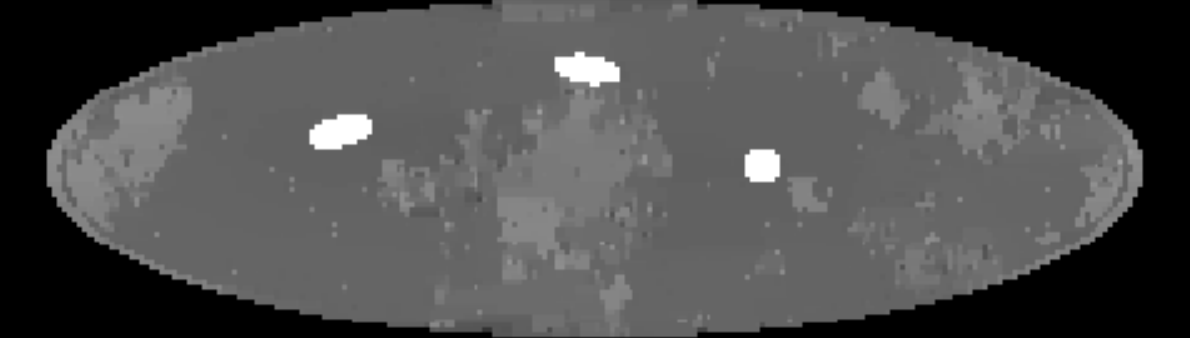}
			\hspace{-10pt} &
			\includegraphics[width=0.22\textwidth,trim={10 0 10 0}, clip]{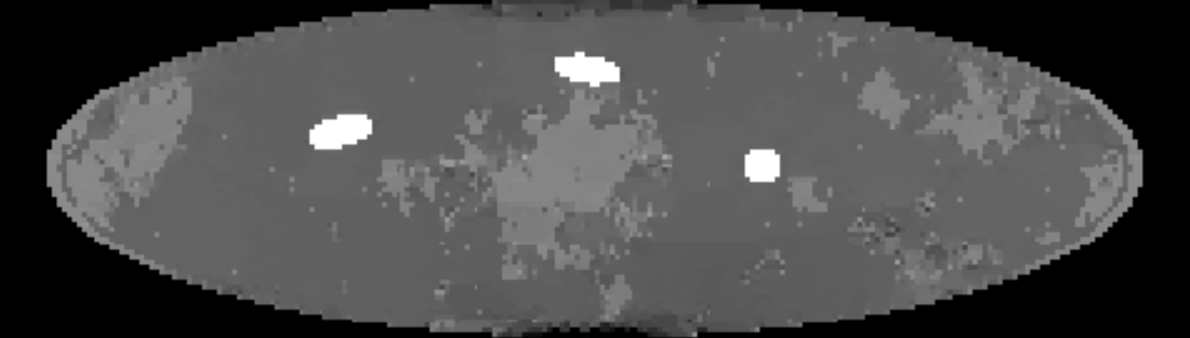}
			\hspace{-10pt} &
			\includegraphics[width=0.22\textwidth,trim={10 0 10 0}, clip]{figures/breast2/Noiseless/XY_pdf/breast_XY_60D_34keV.pdf}
			\\
			\includegraphics[width=0.22\textwidth,trim={110 60 100 0}, clip]{figures/breast2/Noiseless/XY_pdf/breast_XY_14D_34keV.pdf}
			\hspace{-10pt} &
			\includegraphics[width=0.22\textwidth,trim={110 60 100 0}, clip]{figures/breast2/Noiseless/XY_pdf/breast_XY_20D_34keV.pdf}
			\hspace{-10pt} &
			\includegraphics[width=0.22\textwidth,trim={110 60 100 0}, clip]{figures/breast2/Noiseless/XY_pdf/breast_XY_30D_34keV.pdf}
			\hspace{-10pt} &
			\includegraphics[width=0.22\textwidth,trim={110 60 100 0}, clip]{figures/breast2/Noiseless/XY_pdf/breast_XY_60D_34keV.pdf}
			\\
			DTV-$90^\circ$ & DTV-$120^\circ$ & DTV-$150^\circ$ & DTV-$360^\circ$ \\
			\includegraphics[width=0.22\textwidth,trim={10 0 10 0}, clip]{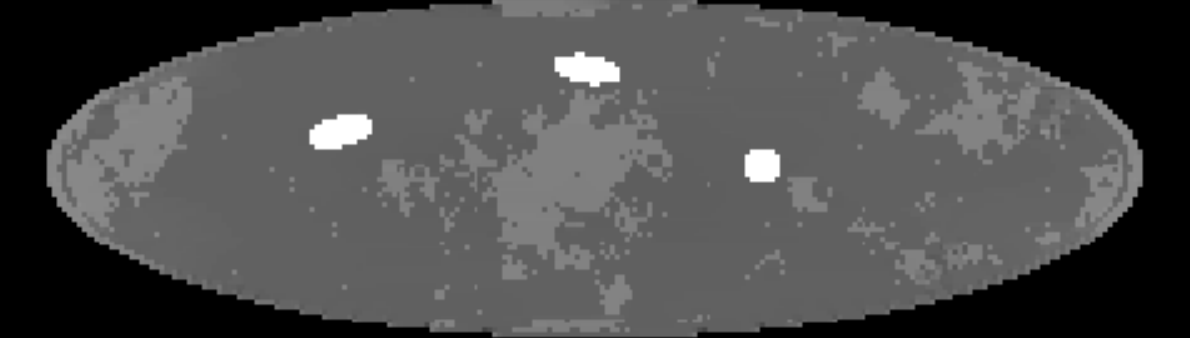}
			\hspace{-10pt} &
			\includegraphics[width=0.22\textwidth,trim={10 0 10 0}, clip]{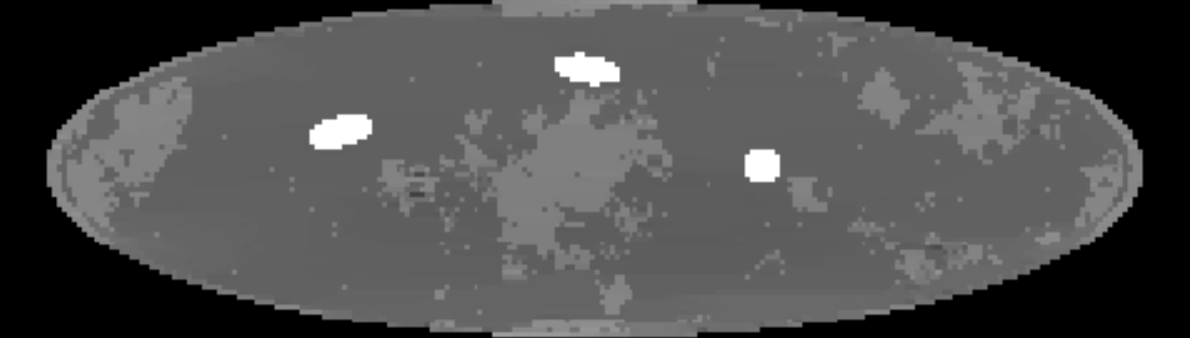}
			\hspace{-10pt} &
			\includegraphics[width=0.22\textwidth,trim={10 0 10 0}, clip]{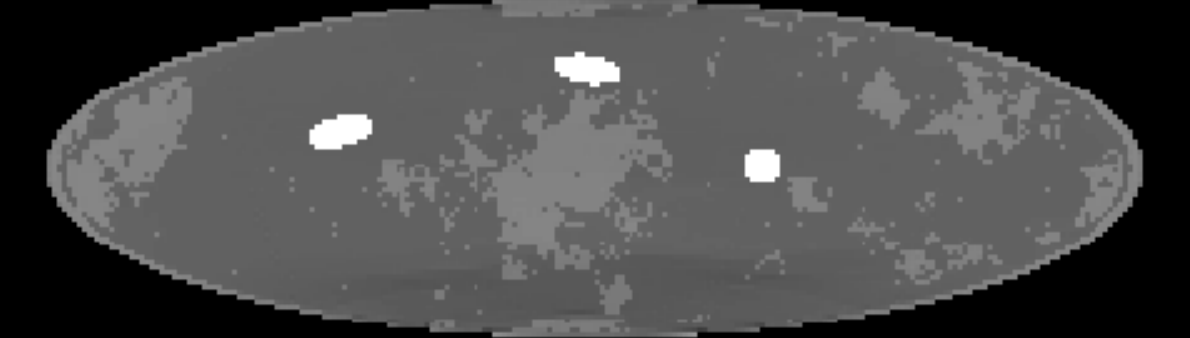}
			\hspace{-10pt} &
			\includegraphics[width=0.22\textwidth,trim={10 0 10 0}, clip]{figures/breast2/Noiseless/XY_pdf/breast_XY_360D_34keV.pdf}
			\\
			\includegraphics[width=0.22\textwidth,trim={110 60 100 0}, clip]{figures/breast2/Noiseless/XY_pdf/breast_XY_90D_34keV.pdf}
			\hspace{-10pt} &
			\includegraphics[width=0.22\textwidth,trim={110 60 100 0}, clip]{figures/breast2/Noiseless/XY_pdf/breast_XY_120D_34keV.pdf}
			\hspace{-10pt} &
			\includegraphics[width=0.22\textwidth,trim={110 60 100 0}, clip]{figures/breast2/Noiseless/XY_pdf/breast_XY_150D_34keV.pdf}
			\hspace{-10pt} &
			\includegraphics[width=0.22\textwidth,trim={110 60 100 0}, clip]{figures/breast2/Noiseless/XY_pdf/breast_XY_360D_34keV.pdf}
	\end{tabular}
	\caption{Monochromatic images (rows 1 and 3) of the breast phantom at 34 keV obtained from noiseless data generated over arcs of LARs $14^\circ$, $20^\circ$, $30^\circ$, $60^\circ$, $90^\circ$, $120^\circ$, $150^\circ$, and $360^\circ$ by use of the DTV algorithm, along with their respective  zoomed-in ROI views (rows 2 and 4). The zoomed-in ROI is enclosed by the rectangular box depicted in the FBP-reference image in Fig. \ref{fig:breast-mono-60}. Display window: [0.2, 0.35] cm$^{-1}$.}
	\label{fig:breast-mono-angles}
\end{figure*}

\paragraph{Visual inspection of monochromatic images}\label{sec:breast-noiseless-qual-vis}


We first show in Fig.~\ref{fig:breast-mono-60} monochromatic images and their respective zoomed-in ROI views at 34 keV obtained from noiseless data over $60^\circ$ by use of the DTV and FBP algorithms, along with the DTV- and FBP-reference images from the noiseless data over FAR of $360^\circ$. The zoomed-in ROI is enclosed by the rectangular box depicted in the FBP-reference image (row 1, column 4) in Fig.~\ref{fig:breast-mono-60}. It can be observed that the DTV image from $60^\circ$ data displays significantly reduced LAR artifacts, which are otherwise observed and severely obscuring structures in the FBP image from data over $60^\circ$. The DTV image is also visually comparable to the DTV-reference image and appears sharper than the FBP-reference image.

We display in Fig.~\ref{fig:breast-mono-angles} monochromatic images and their zoomed-in ROI views obtained by use of the DTV algorithm from noiseless data over, respectively, 7 arcs of LAR, i.e., $\alpha=14^\circ$, $20^\circ$, $30^\circ$, $60^\circ$, $90^\circ$, $120^\circ$, and $150^\circ$,  along with the DTV-reference image. It can be observed that while the DTV images of $\alpha \le 30^\circ$ contain some minor visible artifacts as a result of the compound  LAR and BH effects, the DTV images visually resemble the DTV-reference image.  

\paragraph{Quantitative analysis of monochromatic images}\label{sec:breast-noiseless-quan-vis}

\begin{figure}[t!]
		\centering
		\includegraphics[width=0.24\textwidth]{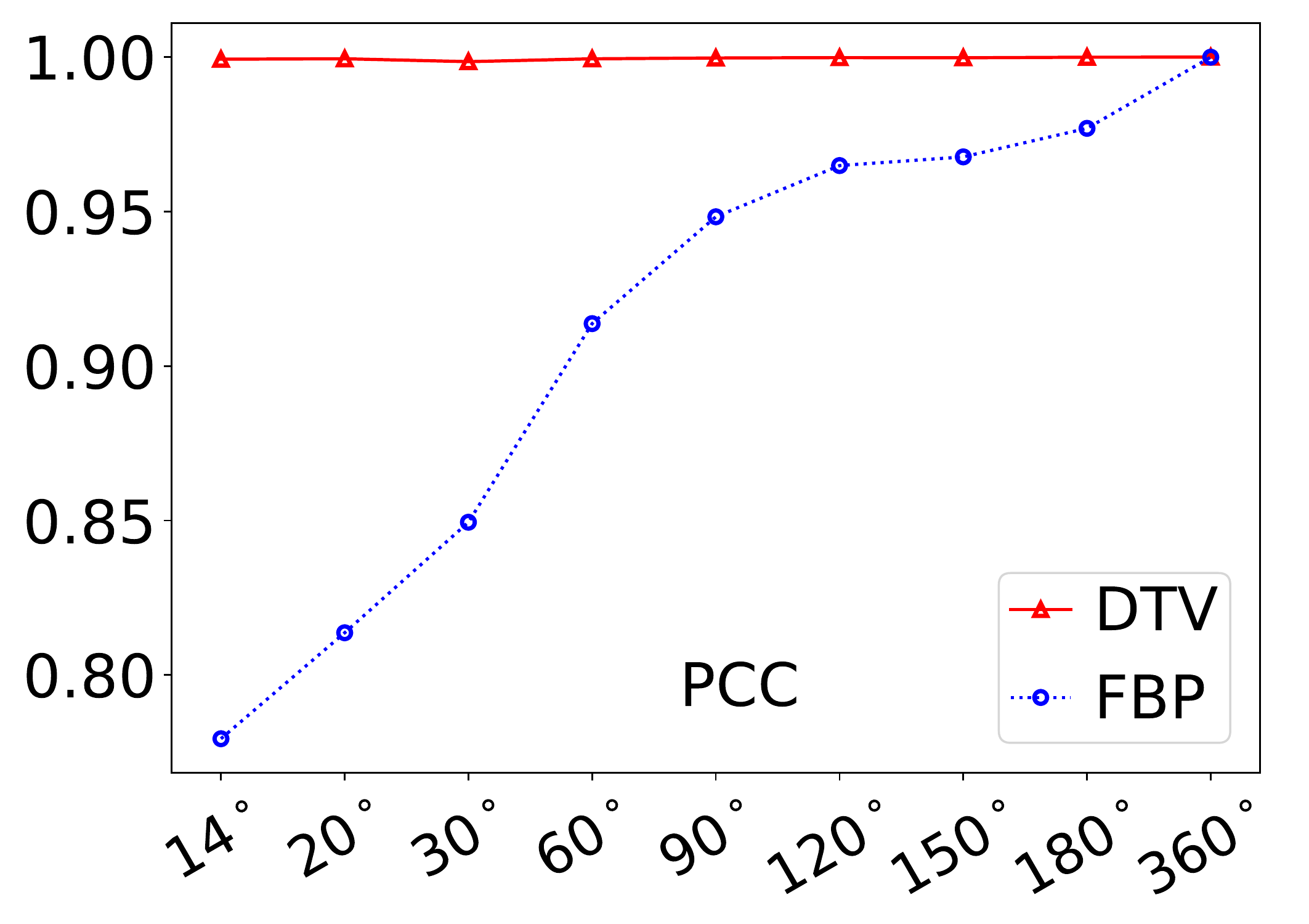}
		\includegraphics[width=0.24\textwidth]{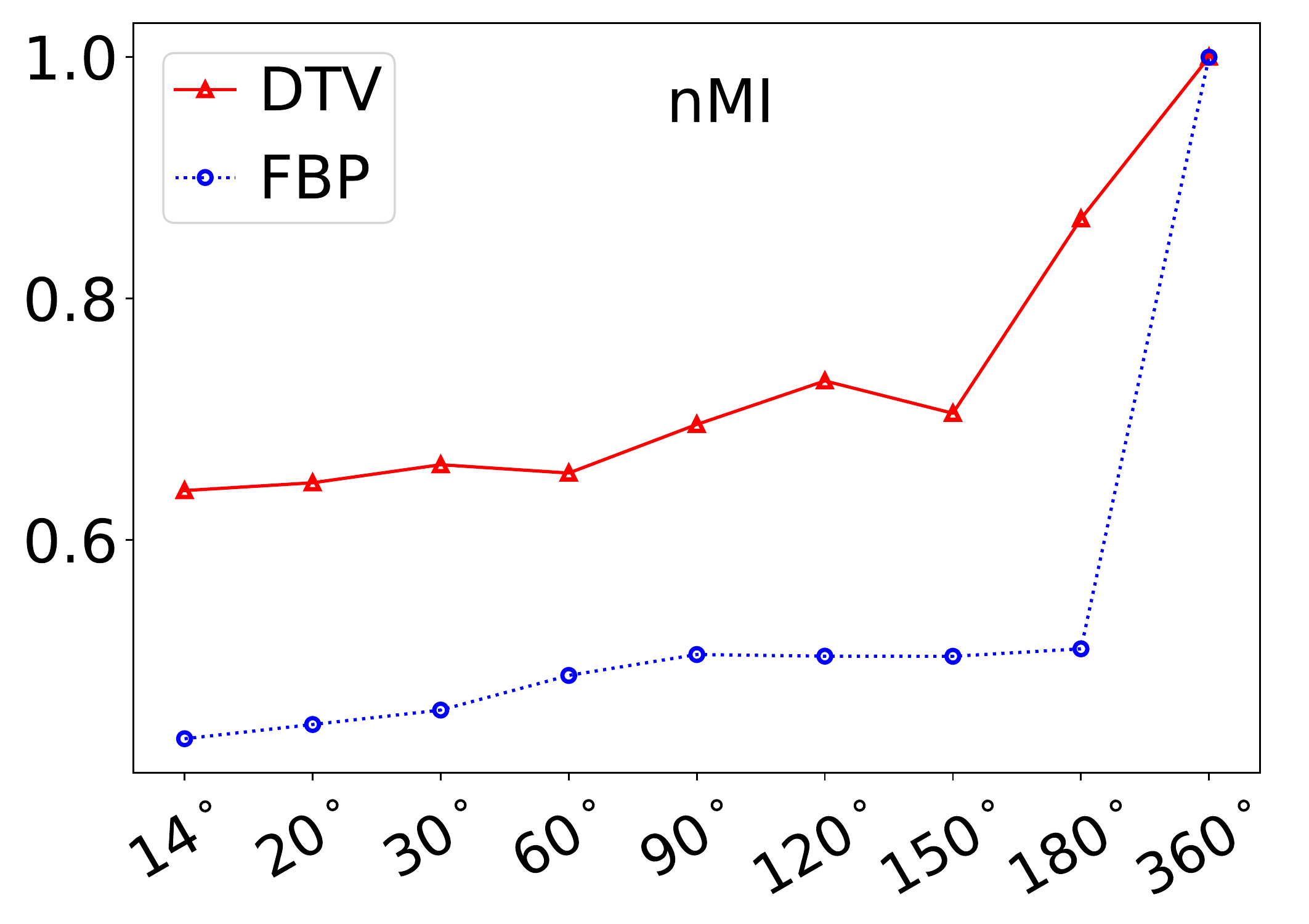}
	\caption{Metrics PCC and nMI, as functions of LAR $\alpha$, of monochromatic images at 34 keV obtained by use of the DTV (solid) and FBP (dotted) algorithms from noiseless data of the breast phantom.}
	\label{fig:breast-metrics-tech}
\end{figure}

Using the DTV-reference image (row 1, column 3) in Fig.~\ref{fig:breast-mono-60}, we compute metrics PCC and nMI of the DTV monochromatic images of the breast phantom, and display them in Fig.~\ref{fig:breast-metrics-tech} as functions of LAR $\alpha$. It can be observed that while the PCC and nMI drop understandably as $\alpha$ decreases, they remain generally close to 1 and above 0.6, respectively, suggesting that the DTV monochromatic images obtained from LAR data correlate well with the DTV-reference image.  For providing a benchmark, we also obtain monochromatic images by use of the FBP algorithm for $\alpha=14^\circ$, $20^\circ$, $30^\circ$, $60^\circ$, $90^\circ$, $120^\circ$, $150^\circ$, and $180^\circ$, but without showing them because the structures in the breast phantom are obscured by significant LAR artifacts in these FBP images, similar to those observed in the FBP image from data over $60^\circ$ shown in column 2 of Fig. \ref{fig:breast-mono-60}. Using the FBP-reference image (row 1, column 4) in Fig. \ref{fig:breast-mono-60}, we compute metrics PCC and nMI of the FBP monochromatic images and plot them in Fig.~\ref{fig:breast-metrics-tech}. The results reveal that the FBP monochromatic images for $\alpha < 180^\circ$ correlate poorly with their reference image.

\begin{figure}[t!]
		\centering
		\includegraphics[width=0.24\textwidth]{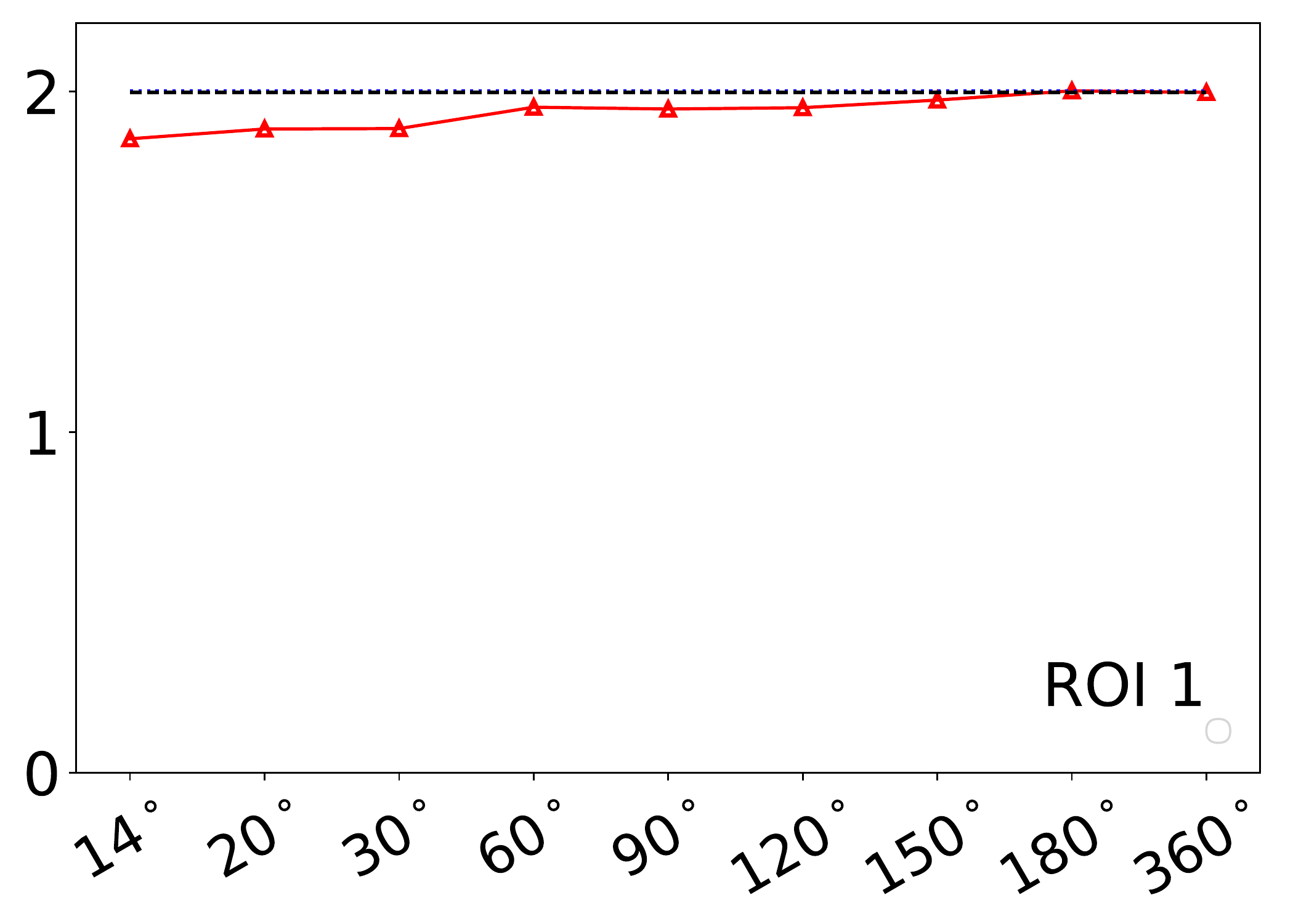}
		\includegraphics[width=0.24\textwidth]{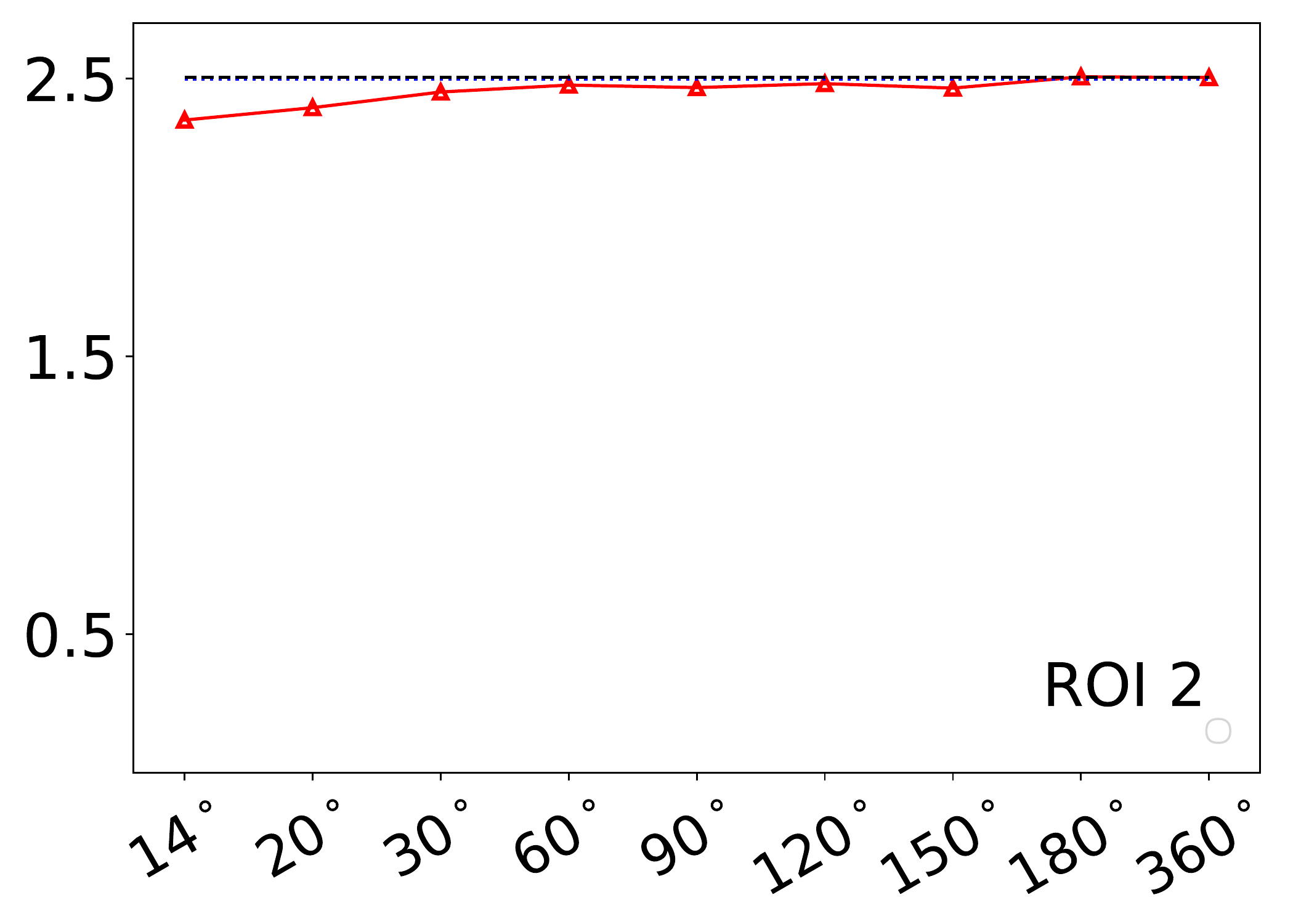}
		\includegraphics[width=0.24\textwidth]{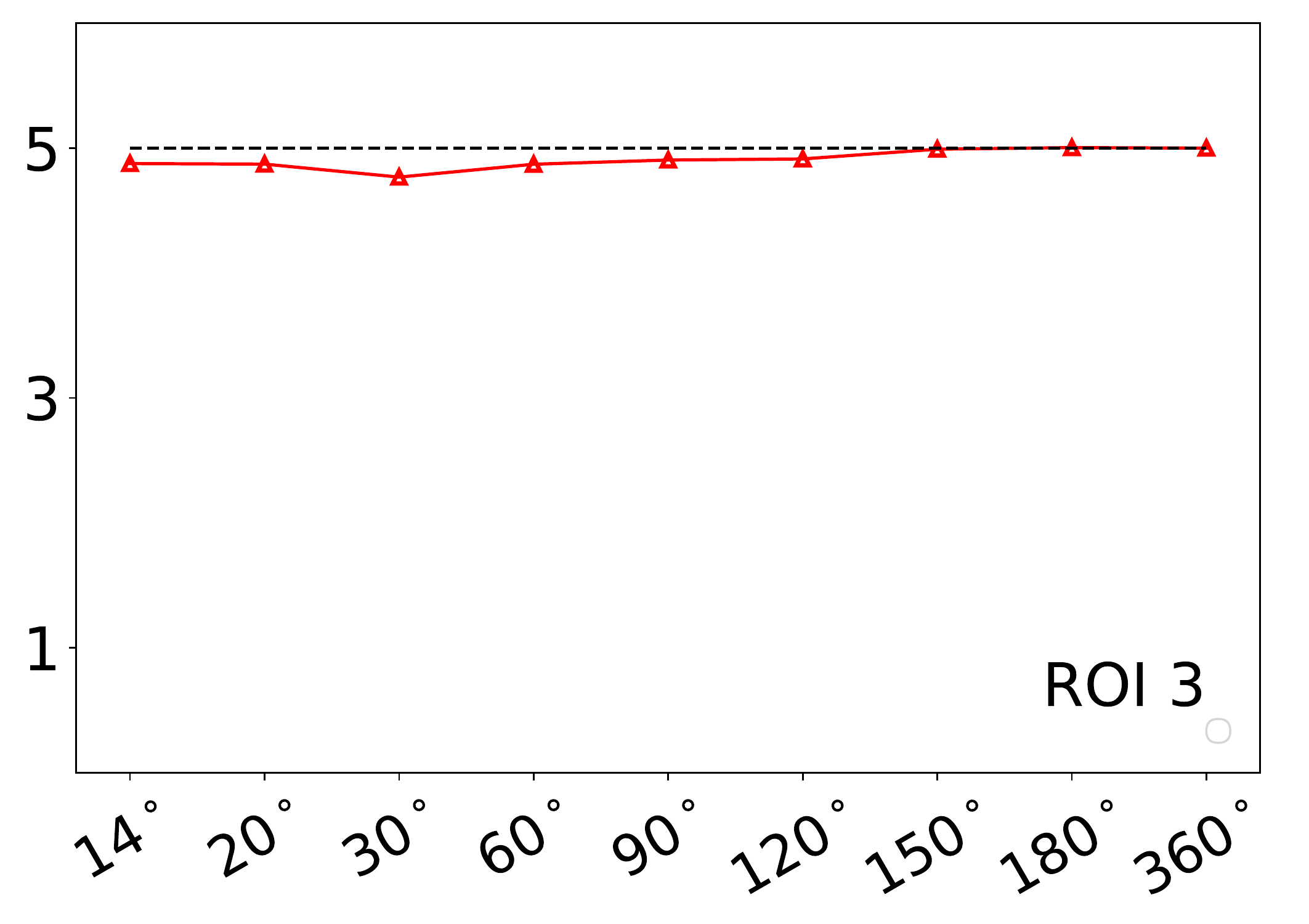}
	\caption{Estimated iodine concentrations (solid), as functions of LAR $\alpha$, for ROIs 1, 2, and 3 in the breast phantom from images reconstructed by use of the DTV algorithm from noiseless data. The two horizontal lines, which are very close, indicate the iodine concentrations estimated from the DTV-reference (dashed) and FBP-reference (dotted) images.}
	\label{fig:breast-metrics-task}
\end{figure}

\paragraph{Estimation of iodine concentrations}\label{sec:breast-noiseless-concentration}

Using the material-based method described in \ref{app:decomp} and \ref{app:tasks}, we estimate iodine concentrations within ROIs 1, 2, and 3 of the breast phantom, as shown in Fig.~\ref{fig:phan}b, using Eq.~\eqref{eq:beta-linear}. Specifically, using basis images estimated from the DTV images, along with constants $\gamma$ and $\tau$ in Eq.~\eqref{eq:beta-linear} fitted from the DTV-reference image, we obtain iodine concentrations for ROIs 1, 2, and 3 of the breast phantom and plot them as functions of LAR $\alpha$ in Fig.~\ref{fig:breast-metrics-task}, along with the iodine concentrations obtained from the DTV- and FBP-reference images. The results indicate that the iodine concentrations obtained with the DTV algorithm for the LARs considered appear to agree well with those obtained from their reference images, only with slight deviations observed for LARs less than $90^\circ$. Due to the severe LAR artifacts in images reconstructed by use of the FBP algorithm, the basis images estimated can be negative and cannot thus be interpreted meaningfully as iodine concentrations, which must physically be non-negative. Therefore, no iodine concentration is estimated from images reconstructed by use of the FBP algorithm for a majority of the LARs considered in the work.



\subsection{Image reconstruction from noisy data of the  breast phantom}\label{sec:breast-noisy}

We also repeat the study of Sec.~\ref{sec:breast-noiseless} except that noisy data are now used, which are obtained by addition of Poisson noise to the corresponding noiseless data, as described in Sec.~\ref{sec:methods-data}. 


\paragraph{Visual inspection of monochromatic images}\label{sec:breast-noisy-qual-vis}

\begin{figure*}[t!]
		\centering
		\begin{tabular}{c c c c}
			DTV-$14^\circ$ & DTV-$20^\circ$ & DTV-$30^\circ$ & DTV-$60^\circ$\\
			\includegraphics[width=0.22\textwidth,trim={10 0 10 0}, clip]{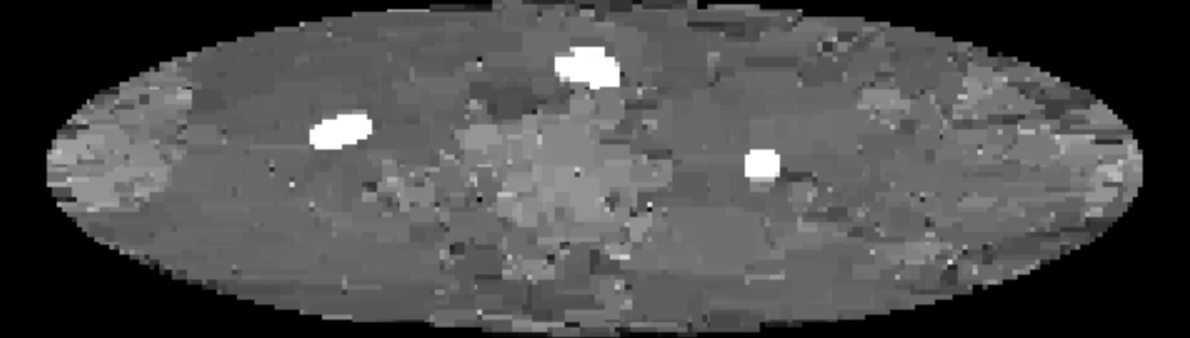}
			\hspace{-10pt} &
			\includegraphics[width=0.22\textwidth,trim={10 0 10 0}, clip]{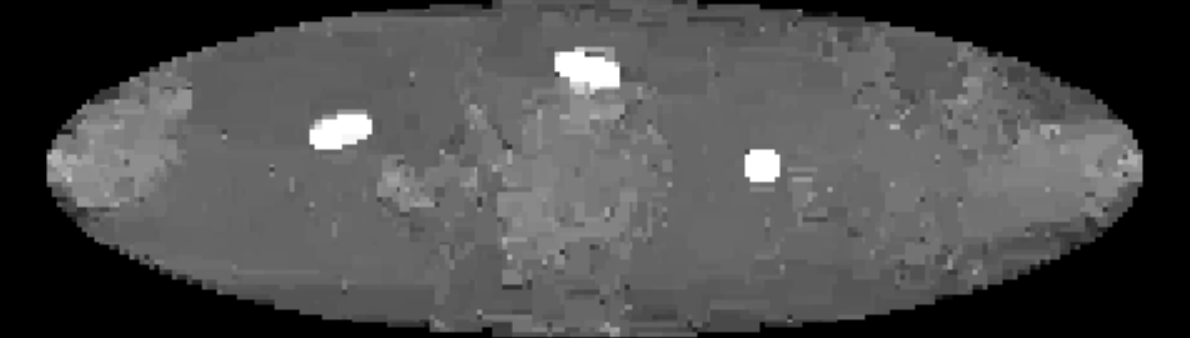}
			\hspace{-10pt} &
			\includegraphics[width=0.22\textwidth,trim={10 0 10 0}, clip]{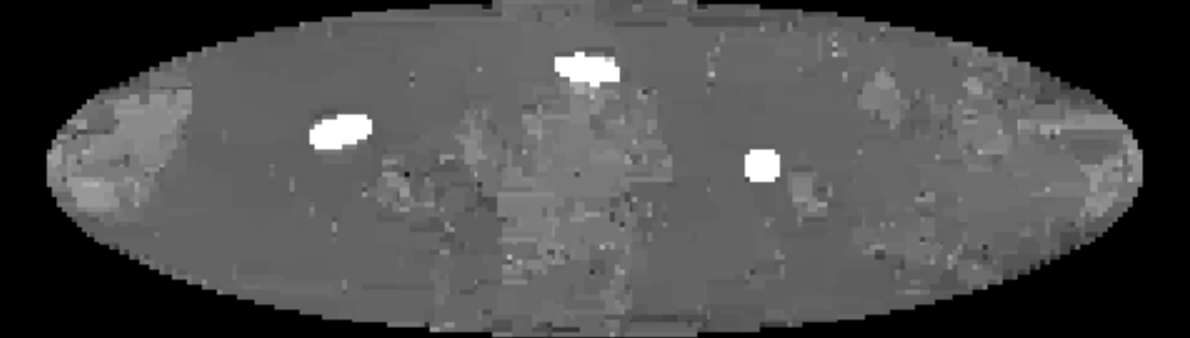}
			\hspace{-10pt} &
			\includegraphics[width=0.22\textwidth,trim={10 0 10 0}, clip]{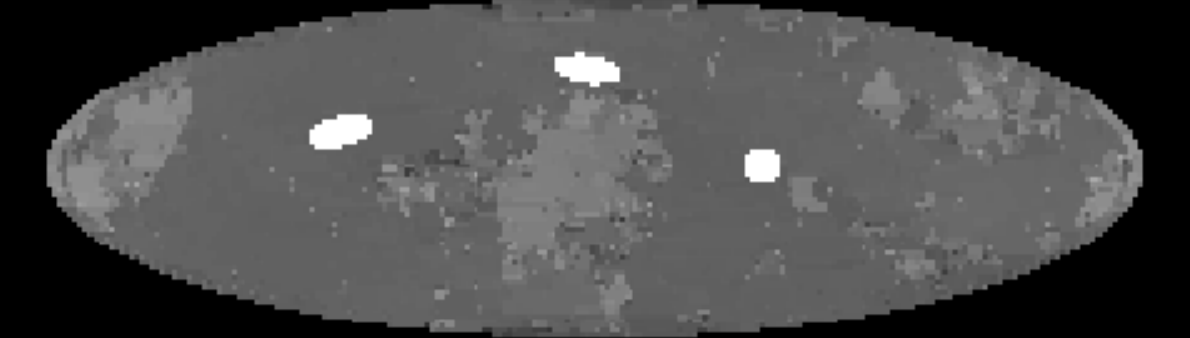}
			\\
			\includegraphics[width=0.22\textwidth,trim={110 60 100 0}, clip]{figures/breast2/Noise_1e7/XY_pdf/breast_XY_noise1e7_14D_34keV.pdf}
			\hspace{-10pt} &
			\includegraphics[width=0.22\textwidth,trim={110 60 100 0}, clip]{figures/breast2/Noise_1e7/XY_pdf/breast_XY_noise1e7_20D_34keV.pdf}
			\hspace{-10pt} &
			\includegraphics[width=0.22\textwidth,trim={110 60 100 0}, clip]{figures/breast2/Noise_1e7/XY_pdf/breast_XY_noise1e7_30D_34keV.pdf}
			\hspace{-10pt} &
			\includegraphics[width=0.22\textwidth,trim={110 60 100 0}, clip]{figures/breast2/Noise_1e7/XY_pdf/breast_XY_noise1e7_60D_34keV.pdf}
			\\
			DTV-$90^\circ$ & DTV-$120^\circ$ & DTV-$150^\circ$ & DTV-$360^\circ$ \\
			\includegraphics[width=0.22\textwidth,trim={10 0 10 0}, clip]{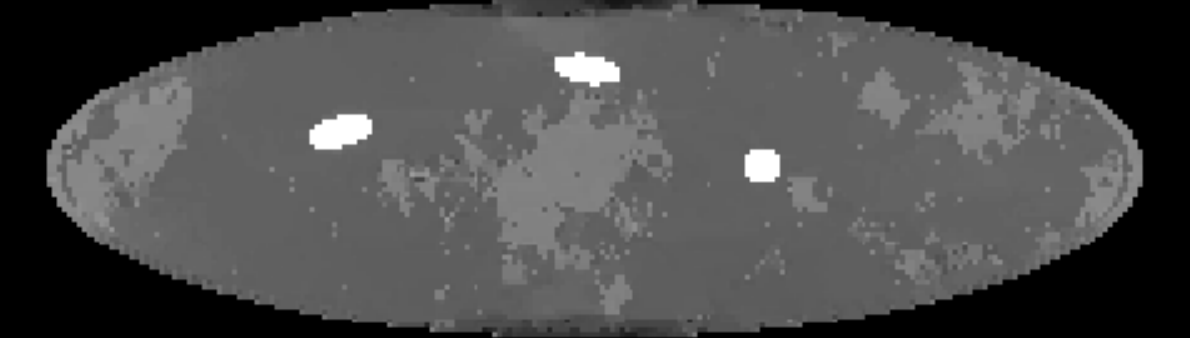}
			\hspace{-10pt} &
			\includegraphics[width=0.22\textwidth,trim={10 0 10 0}, clip]{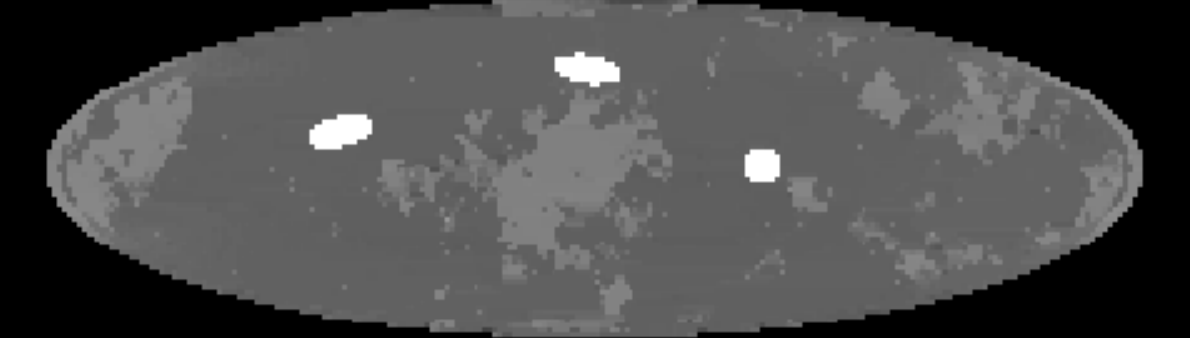}
			\hspace{-10pt} &
			\includegraphics[width=0.22\textwidth,trim={10 0 10 0}, clip]{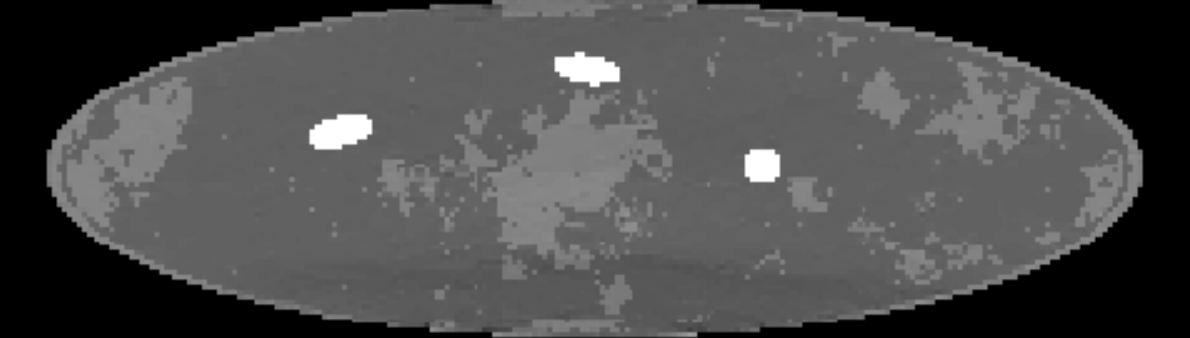}
			\hspace{-10pt} &
			\includegraphics[width=0.22\textwidth,trim={10 0 10 0}, clip]{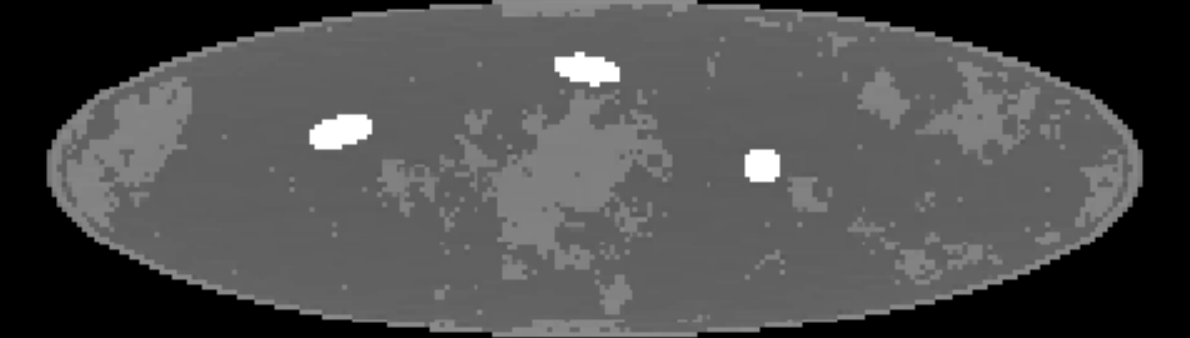}
			\\
			\includegraphics[width=0.22\textwidth,trim={110 60 100 0}, clip]{figures/breast2/Noise_1e7/XY_pdf/breast_XY_noise1e7_90D_34keV.pdf}
			\hspace{-10pt} &
			\includegraphics[width=0.22\textwidth,trim={110 60 100 0}, clip]{figures/breast2/Noise_1e7/XY_pdf/breast_XY_noise1e7_120D_34keV.pdf}
			\hspace{-10pt} &
			\includegraphics[width=0.22\textwidth,trim={110 60 100 0}, clip]{figures/breast2/Noise_1e7/XY_pdf/breast_XY_noise1e7_150D_34keV.pdf}
			\hspace{-10pt} &
			\includegraphics[width=0.22\textwidth,trim={110 60 100 0}, clip]{figures/breast2/Noise_1e7/XY_pdf/breast_XY_noise1e7_360D_34keV.pdf}
	\end{tabular}
	\caption{Monochromatic images (rows 1 and 3) of the breast phantom at 34 keV obtained from noisy data generated over arcs of LARs $14^\circ$, $20^\circ$, $30^\circ$, $60^\circ$, $90^\circ$, $120^\circ$, $150^\circ$, and $360^\circ$ by use of the DTV algorithm, along with their respective zoomed-in ROI views (rows 2 and 4). The zoomed-in ROI is enclosed by the rectangular box depicted in the FBP-reference image in Fig.~\ref{fig:breast-mono-60}. Display window: [0.2, 0.35] cm$^{-1}$.}
	\label{fig:breast-mono-angles-noisy}
\end{figure*}

We show in Fig.~\ref{fig:breast-mono-angles-noisy} monochromatic images and their zoomed-in ROI views obtained by use of the DTV algorithm from noisy data generated over, respectively, 7 arcs of LARs $\alpha=14^\circ$, $20^\circ$, $30^\circ$, $60^\circ$, $90^\circ$, $120^\circ$, and $150^\circ$, along with the FAR of $360^\circ$. 
It can be observed that the DTV images appear to visually resemble the DTV-reference image, while images from data over $\alpha \le 30^\circ$ contain some visible artifacts as a result of the compound effect of the LAR, BH, and noise.

\paragraph{Quantitative analysis of monochromatic images}\label{sec:breast-noisy-quan-vis}

\begin{figure}[t!]
		\centering
		\includegraphics[width=0.24\textwidth]{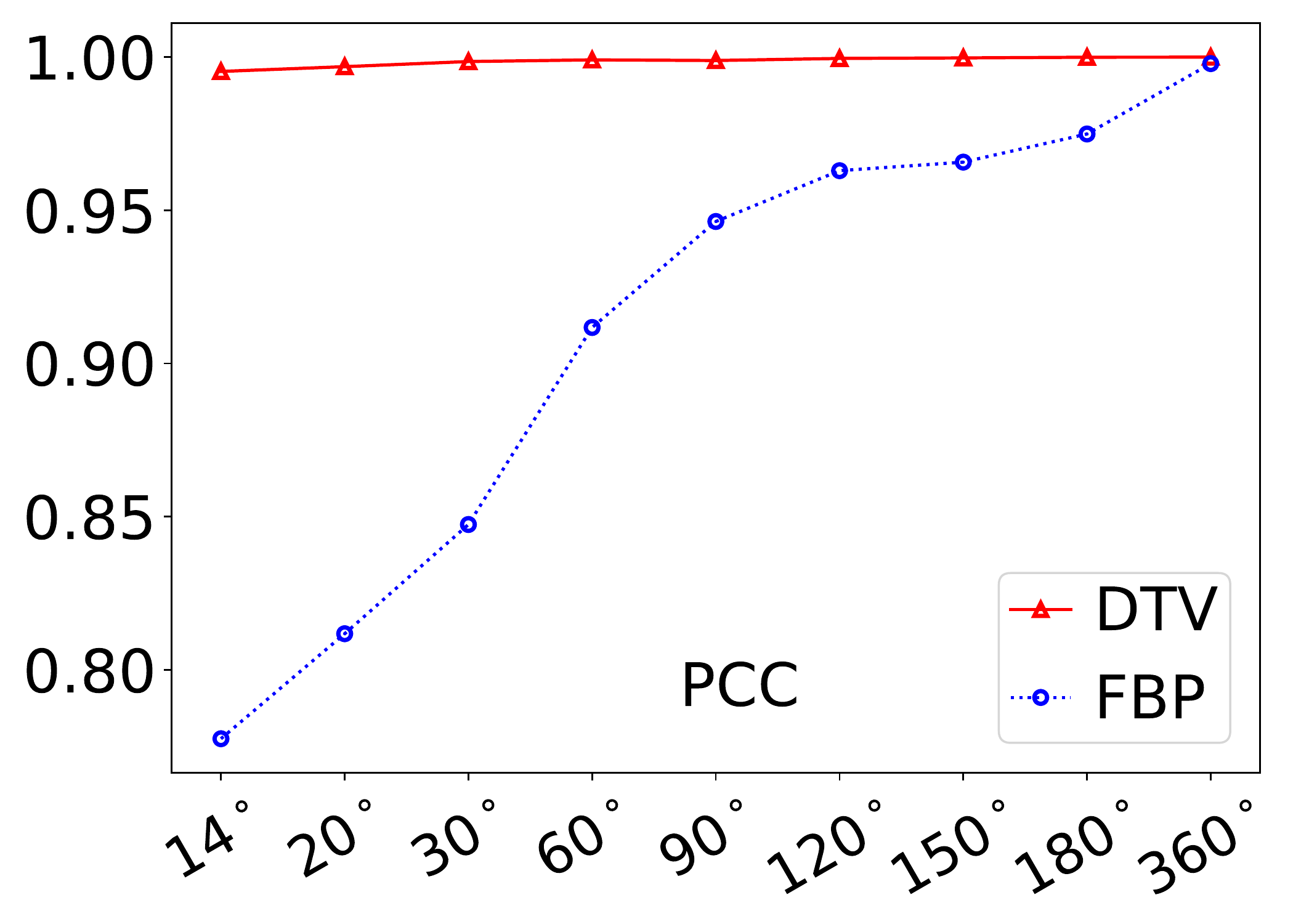}
    	\includegraphics[width=0.24\textwidth]{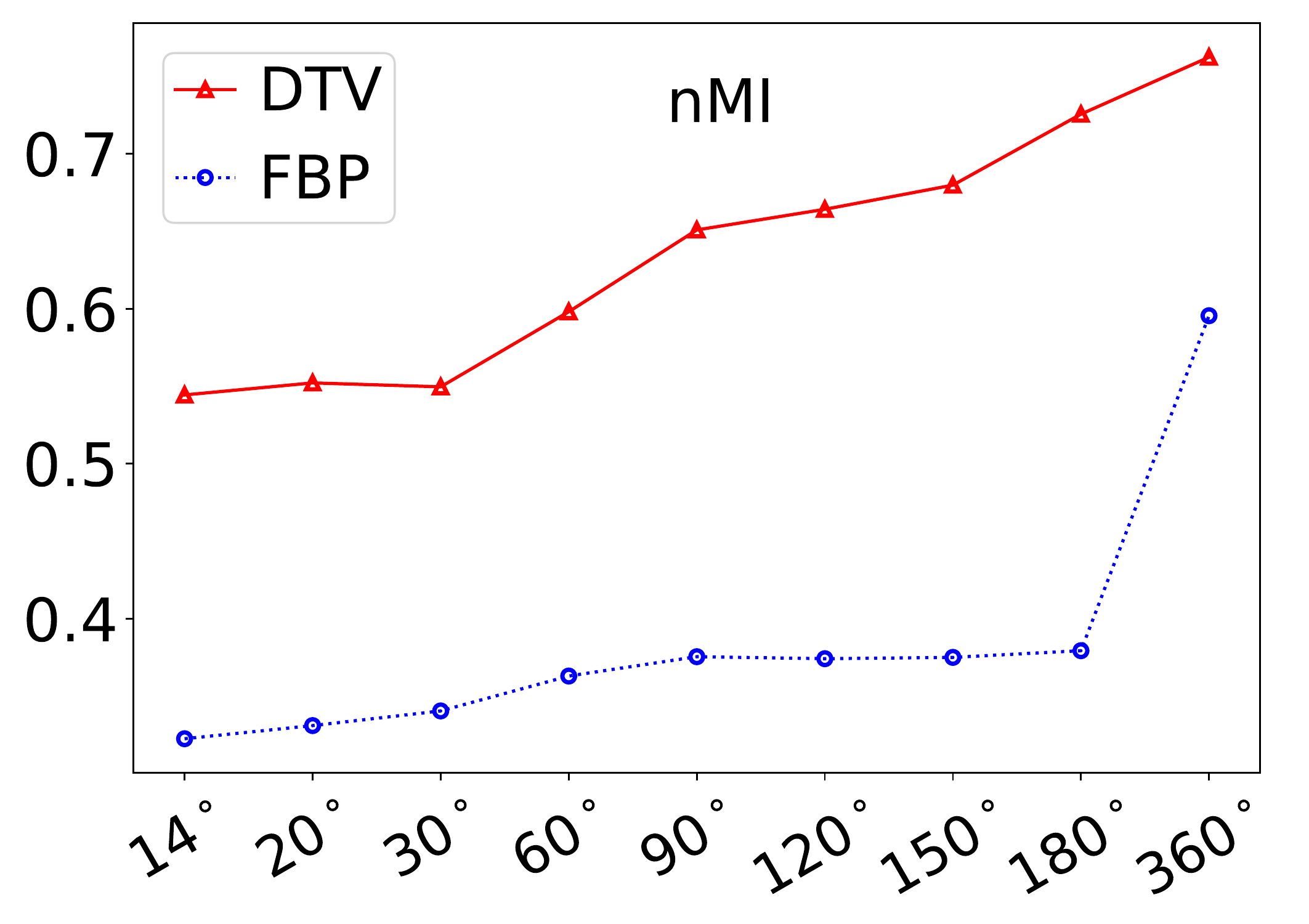}
	\caption{Metrics PCC and nMI, as functions of LAR $\alpha$, of monochromatic images at 34 keV obtained by use of the DTV (solid) and FBP (dotted) algorithms from noisy data of the breast phantom.}
	\label{fig:breast-metrics-tech-noisy}
\end{figure}

Using the DTV-reference image (row 1, column 3) in Fig.~\ref{fig:breast-mono-60}, we compute metrics PCC and nMI of the DTV monochromatic images obtained, and display them in Fig.~\ref{fig:breast-metrics-tech-noisy} as functions of LAR $\alpha$. It can be observed that while the PCC and nMI drop understandably as $\alpha$ decreases, they remain generally close to 1 and above 0.5, respectively, suggesting that the DTV monochromatic images from LAR data correlate reasonably well with the DTV-reference image.  For providing a benchmark, we also obtain monochromatic images by use of the FBP algorithm for $\alpha=14^\circ$, $20^\circ$, $30^\circ$, $60^\circ$, $90^\circ$, $120^\circ$,  $150^\circ$, and $180^\circ$, but without showing them because the structures in the breast phantom are obscured by significant LAR artifacts in these FBP images, similar to those observed in the FBP image of $60^\circ$ shown in column 2 of Fig. \ref{fig:breast-mono-60}.  Using the FBP-reference image (row 1, column 4) in Fig. \ref{fig:breast-mono-60}, we compute metrics PCC and nMI of the FBP monochromatic images and plot them in Fig.~\ref{fig:breast-metrics-tech-noisy}. The results reveal that the FBP monochromatic images for $\alpha < 180^\circ$ correlate poorly with their reference image.

\paragraph{Estimation of iodine concentrations}\label{sec:breast-noisy-concentration}

%

\begin{figure}[t!]
		\centering
		\includegraphics[width=0.24\textwidth]{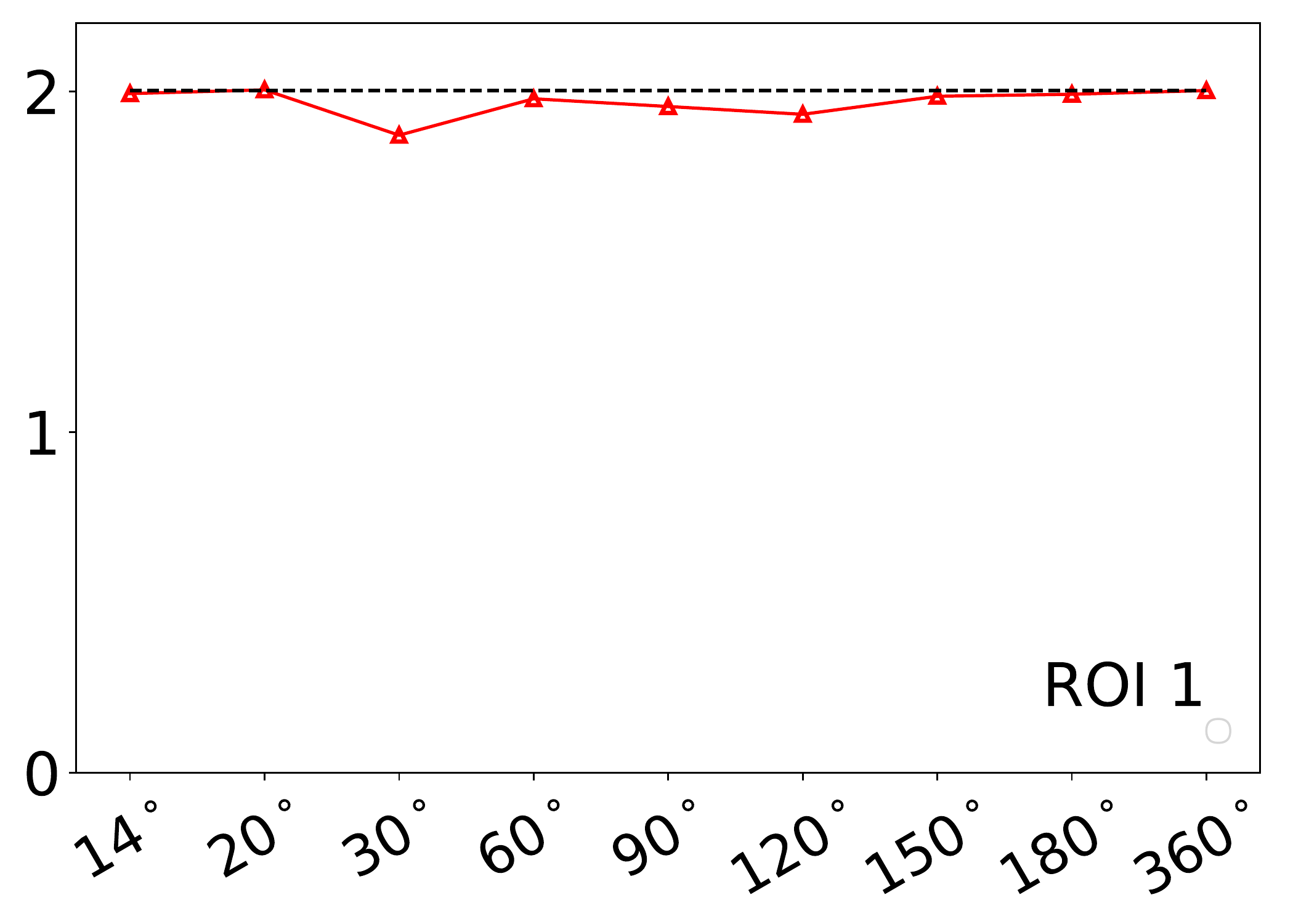}
		\includegraphics[width=0.24\textwidth]{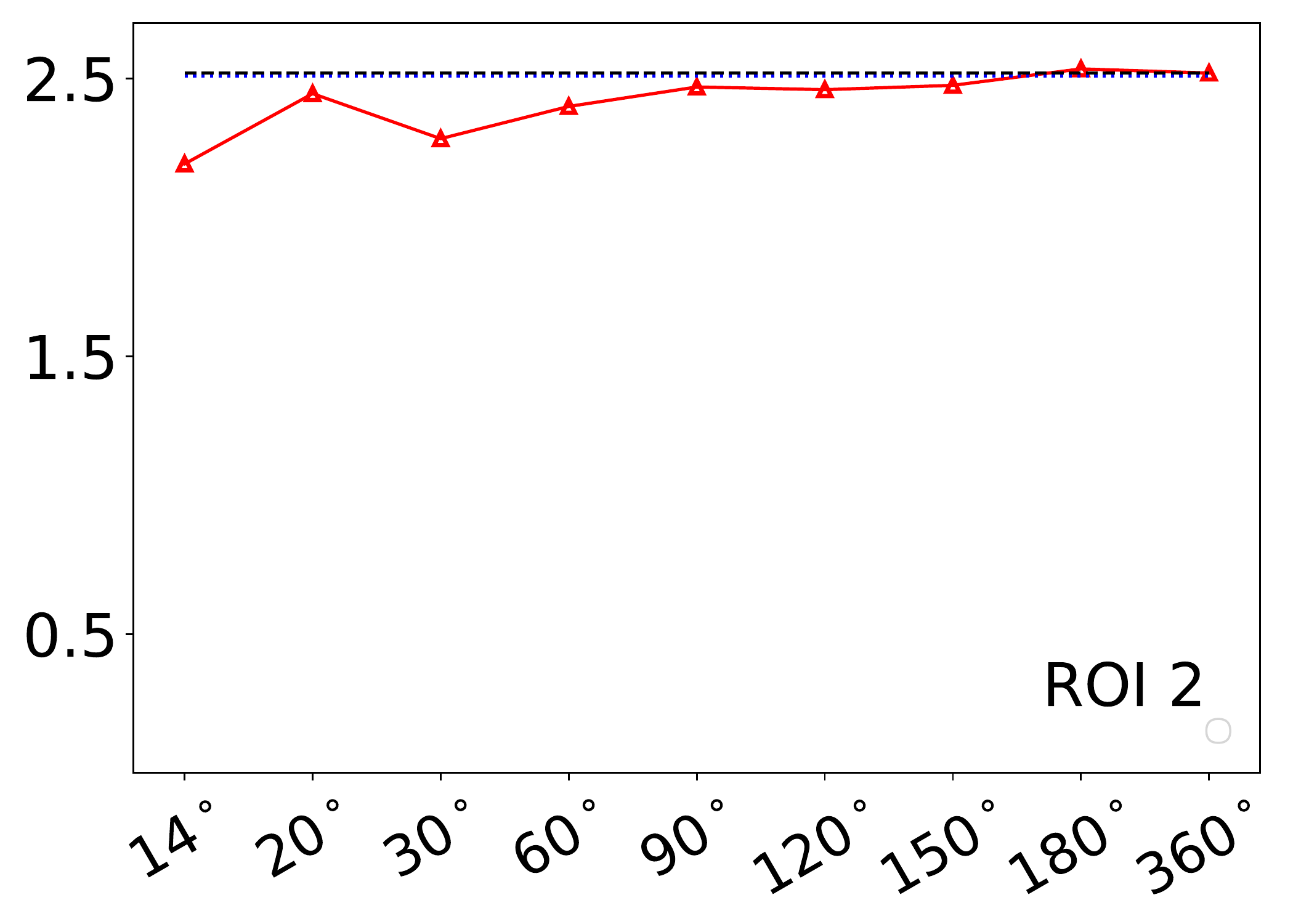}
		\includegraphics[width=0.24\textwidth]{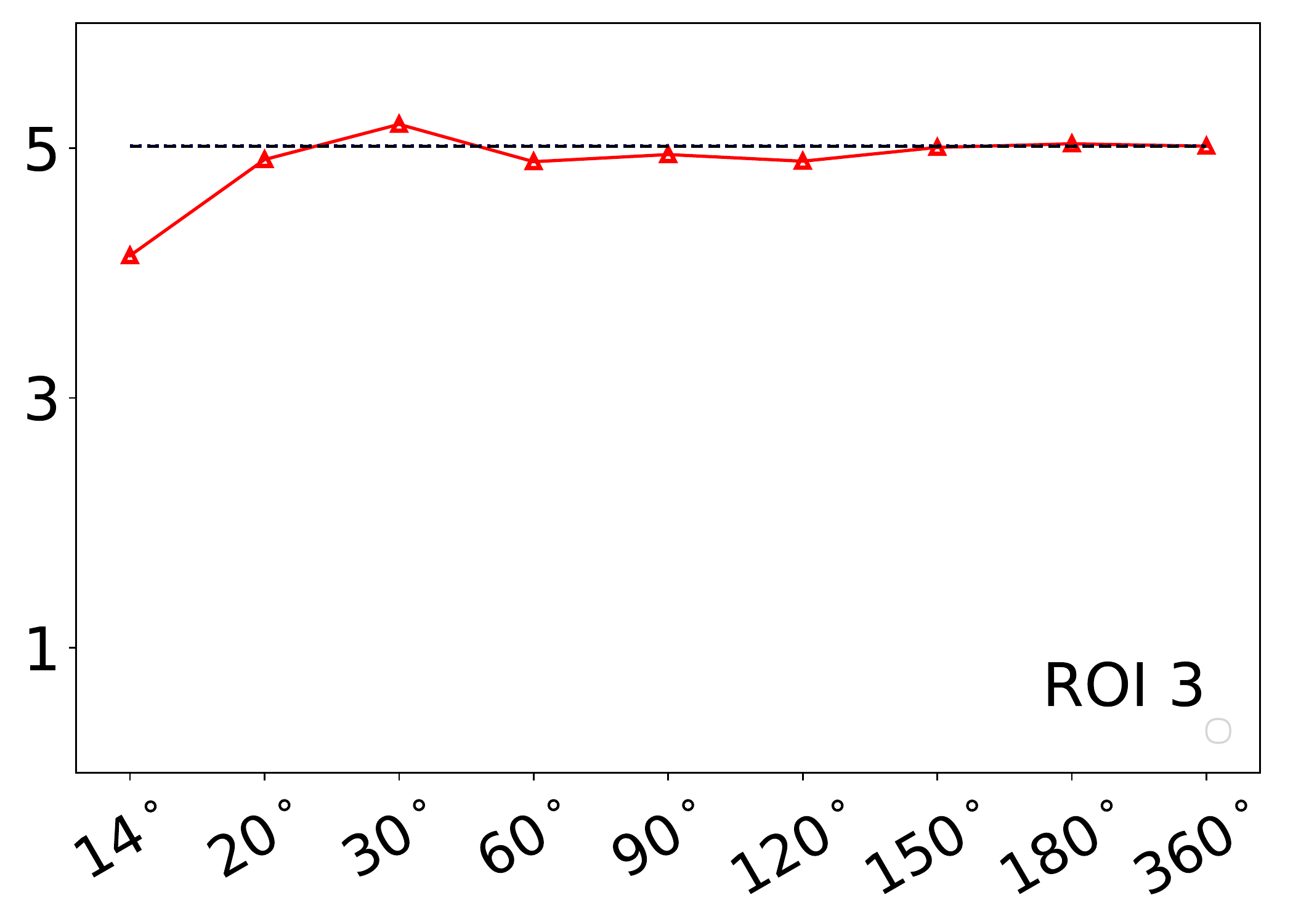}
	\caption{Estimated iodine concentrations (solid), as functions of LAR $\alpha$, for ROIs 1, 2, and 3 in the breast phantom from noisy images reconstructed by use of the DTV algorithm from noisy data. The two horizontal lines, which are very close, indicate the atomic numbers estimated from the DTV-reference (dashed) and FBP-reference (dotted) images.}
	\label{fig:breast-metrics-task-noisy}
\end{figure}

Using the material-based method described in \ref{app:decomp} and \ref{app:tasks}, we also estimate iodine concentrations within ROIs 1, 2, and 3 of the breast phantom using Eq.~\eqref{eq:beta-linear},
and plot them as functions of LAR $\alpha$ in Fig.~\ref{fig:breast-metrics-task-noisy}, along with the iodine concentrations obtained from the DTV- and FBP-reference images. The results suggest that the iodine concentrations obtained with the DTV algorithm appear to agree well with those obtained from their reference images, only with slight deviations observed for angular ranges less than $90^\circ$. Due to the severe LAR artifacts in images reconstructed by use of the FBP algorithm, the basis images estimated can be negative and cannot thus be interpreted meaningfully as iodine concentrations, which must physically be non-negative. Therefore, no iodine concentration is estimated from images reconstructed by use of the FBP algorithm for a majority of the LARs considered in the work.

\section{Discussions} \label{sec:discussion}

{\color{black}
In the work, we have tailored the DTV algorithm developed previously to investigating image reconstructions with minimized LAR artifacts from low- and high-kVp data in LAR DECT. In particular, the results of LAR DECT are compared against that of FAR DECT, which by definition is free of LAR artifacts.}
The reconstruction problem is formulated as a convex optimization problem involving separate DTV constraints along orthogonal directions designed for a given LAR configuration. The DTV algorithm was applied to solving the optimization problem for achieving image reconstruction from low- and high-kVp data generated over arcs of LARs, ranging from $14^\circ$ to $180^\circ$. From the images reconstructed, we estimate basis images and then obtain monochromatic images at energies of interest.   
Monochromatic images obtained have been visually inspected and quantitatively analyzed through comparison with their reference images in FAR DECT, revealing that monochromatic images obtained with the DTV algorithm are with substantially reduced artifacts that are observed often in monochromatic images obtained with existing algorithms in DECT. Additionally, using the basis images estimated, we have computed atomic numbers and iodine concentrations, and again compared them with their respective references obtained in FAR DECT. The study results reveal that the DTV algorithm can yield physical quantities such as atomic number and iodine concentration comparable to that estimated in FAR DECT. 
{\color{black}
For comparison, we have applied also a standard isotropic total variation (ITV) algorithm~\cite{zhang2021dtv} to reconstructing images from low- and high-kVp LAR data and show them in Fig.~\ref{fig:itv}. It can be observed that the ITV images contain conspicuous LAR artifacts, which are, however, largely reduced in the corresponding DTV images in Figs.~\ref{fig:suitcase-mono-angles} and~\ref{fig:breast-mono-angles}.
\begin{figure*}[t!]
		\centering
		\begin{tabular}{c c c c}
			ITV-$14^\circ$ & ITV-$20^\circ$ & ITV-$30^\circ$ & ITV-$60^\circ$
			\\
			\includegraphics[width=0.22\textwidth]{figures/ITV_bar/ITV_14deg_mono_150.pdf}
			\hspace{-10pt} &
			\includegraphics[width=0.22\textwidth]{figures/ITV_bar/ITV_20deg_mono_150.pdf}
			\hspace{-10pt} &
			\includegraphics[width=0.22\textwidth]{figures/ITV_bar/ITV_30deg_mono_150.pdf}
			\hspace{-10pt} &
			\includegraphics[width=0.22\textwidth]{figures/ITV_bar/ITV_60deg_mono_200.pdf}
			\\
			\includegraphics[width=0.22\textwidth,trim={10 0 10 0}, clip]{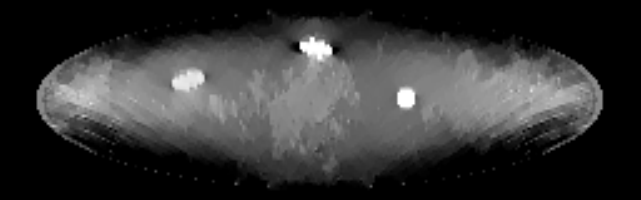}
			\hspace{-10pt} &
			\includegraphics[width=0.22\textwidth,trim={10 0 10 0}, clip]{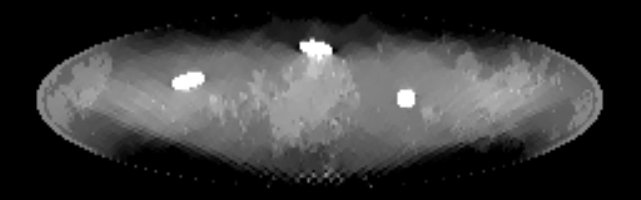}
			\hspace{-10pt} &
			\includegraphics[width=0.22\textwidth,trim={10 0 10 0}, clip]{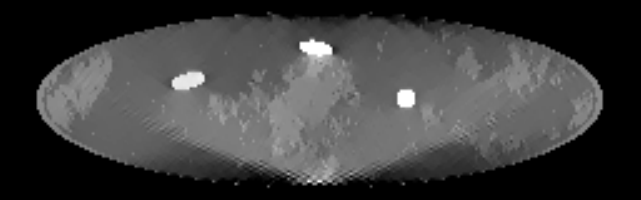}
			\hspace{-10pt} &
			\includegraphics[width=0.22\textwidth,trim={10 0 10 0}, clip]{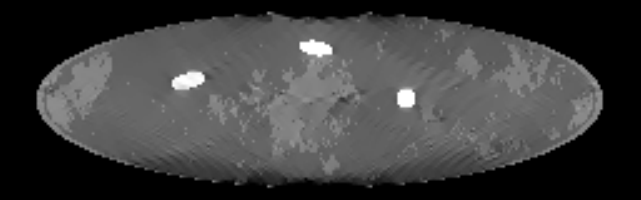}
			\end{tabular}
	\caption{\color{black}Monochromatic images of the suitcase phantom at 40 keV (row 1) and of the breast phantom at 34 keV (row 2) obtained from noiseless data generated over arcs of LARs $14^\circ$ (columns 1), $20^\circ$ (columns 2), $30^\circ$ (columns 3), and $60^\circ$ (columns 4) by use of the  ITV algorithm. Display windows: [0.1, 0.65] cm$^{-1}$ and [0.2, 0.35] cm$^{-1}$ for the suitcase and breast phantoms, respectively.}
	\label{fig:itv}
\end{figure*}
}

The scanning configuration considered in the work includes completely overlapping arcs of LARs for the low- and high-kVp scans. However, our proposed approach  can readily be applied to DECT scanning configuration that consists of two partially- or non-overlapping arcs of LARs of either same or different spans, because the DTV algorithm allows images to be reconstructed separately for low- and high-kVp data collected over LARs. We are investigating currently image reconstruction from low- and high-kVp data collected over two arcs of LARs that are not completely overlapping with each other~\cite{chen2021-90}. 

While the optimization problem and DTV algorithm are designed in the work for two-dimensional (2D) DECT with a fan-beam scanning configuration, it is conceptually and mathematically straightforward to extend them to three-dimensional (3D) DECT with a cone-beam scanning configuration. The key to the extension is to design DTV constraints along orthogonal axes specific to a given 3D scanning geometry. We are investigating currently the design of a DTV algorithm for 3D image reconstruction from data collected over a circular segment in DBT with a cone-beam projection geometry.

Similar to many of the existing algorithms for image reconstruction in DECT,  the DTV and FBP algorithms are based upon a linear-data model, i.e., the DXT that does not model the non-linear BH effect.  As such, BH artifacts may be observed also in monochromatic images obtained with the DTV and FBP algorithms from FAR data. The BH effect may also result in estimation errors in physical quantities such as atomic number and iodine concentration. 
{\color{black} The work is not intended to correct for the BH artifacts; instead, it focuses on investigating the LAR effect on monochromatic images and physical quantities estimated relative to those obtained in FAR DECT without explicit BH-artifact correction.} 
One can develop algorithms by basing upon the non-linear data model (see Eq.~\eqref{eq:nonlinear-model}) to correct for the BH artifacts and to improve the accuracy of physical quantity estimation~\cite{zou_analysis_2008,barber_algorithm_2016, chen2021non}. For DECT with completely overlapping arcs of LARs for collecting low- and high-kVp data, we are investigating currently the application of a data-domain method for BH-effect correction~\cite{zou_analysis_2008} to low- and high-kVp LAR data. From the corrected basis projections, the DTV algorithm can then be tailored to reconstruct basis images and monochromatic images with BH artifacts corrected. Furthermore, one may develop a one-step algorithm~\cite{chen2021non} with DTV constraints basing upon the non-linear data model in Eq.~\eqref{eq:nonlinear-model} for reconstructing basis images without the BH artifacts directly from low- and high-kVp data in LAR DECT, leading to monochromatic images and physical quantities free from BH artifacts.

{\color{black}
In the work, we have focused on investigating image reconstructions and physical-quantity estimation in LAR DECT with computer-simulated data generated from phantoms of practical relevance. However, it would be necessary and important to evaluate the approach to image reconstruction in LAR DECT by use of real data collected in research or practical applications. Knowledge acquired in the work can be exploited to design and conduct extensive studies on image reconstructions from real LAR data collected in research and clinical DECT.
}

\section{Conclusion} \label{sec:conclusion}
In this work, using the DTV algorithm developed previously for conventional LAR CT, we have investigated image reconstruction from low- and high-kVp data in LAR DECT.
Results of our studies reveal that monochromatic images obtained from data collected over arcs of LARs as low as $60^\circ$ appear visually comparable to their corresponding reference images obtained in FAR DECT and that the accuracy of atomic numbers and iodine concentrations estimated from data of LARs across $14^\circ$ to $180^\circ$ is quantitatively comparable to that obtained in FAR DECT. The results acquired in the work may engender insights into the design of DECT with LAR scanning configurations of application significance.

\section*{Acknowledgment}

This work was supported in part by NIH R01 Grant Nos. EB026282 and EB023968, and the Grayson-Jockey Club Research. 
The contents of this article are solely the responsibility of the authors and do not necessarily represent the official views of the National Institutes of Health.

\appendix

\section{Parameter selection for \lowercase{$t^s_x$} and \lowercase{$t^s_y$}} \label{app:para-selection}

Like any reconstruction algorithms, the DTV algorithm investigated in the work for LAR DECT involves constraint parameters, such as $t^s_x$ and $t^s_y$, that can impact the reconstruction. Without loss of generality, we consider an example of DTV reconstruction from low- and high-kVp data of the breast phantom collected over an arc of $60^\circ$, and illustrate how $t^s_x$ and $t^s_y$ are selected by basing upon visual inspection of monochromatic images obtained for artifacts reduction. 

\begin{figure*}[t!]
	\centering
	\begin{tabular}{c c c c}
			\includegraphics[angle=180,width=0.22\textwidth]{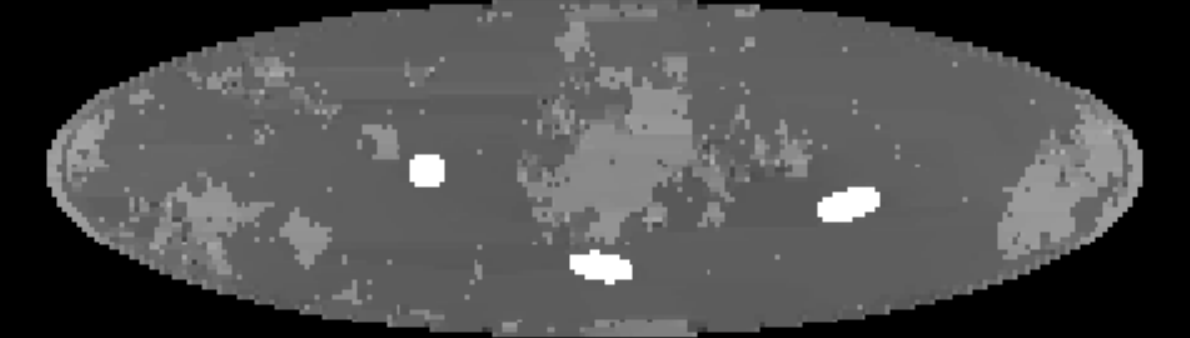}
			\hspace{-10pt} & 
			\includegraphics[angle=180,width=0.22\textwidth]{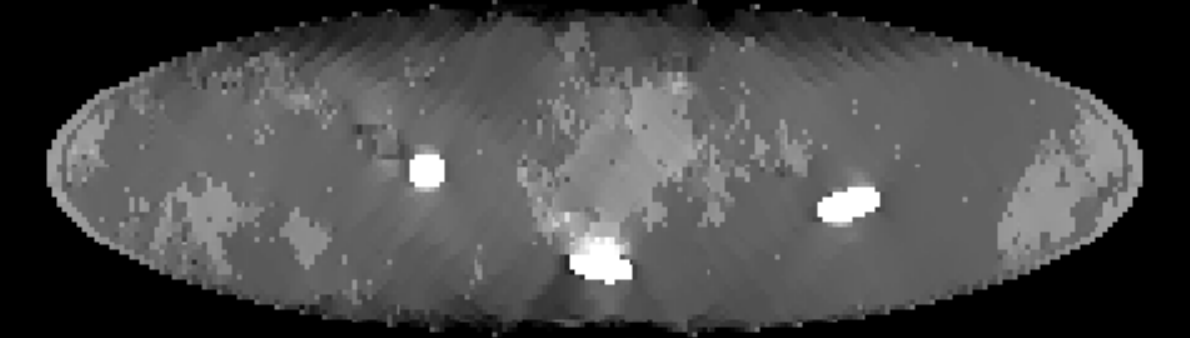}
			\hspace{-10pt} & 
			\includegraphics[angle=180,width=0.22\textwidth]{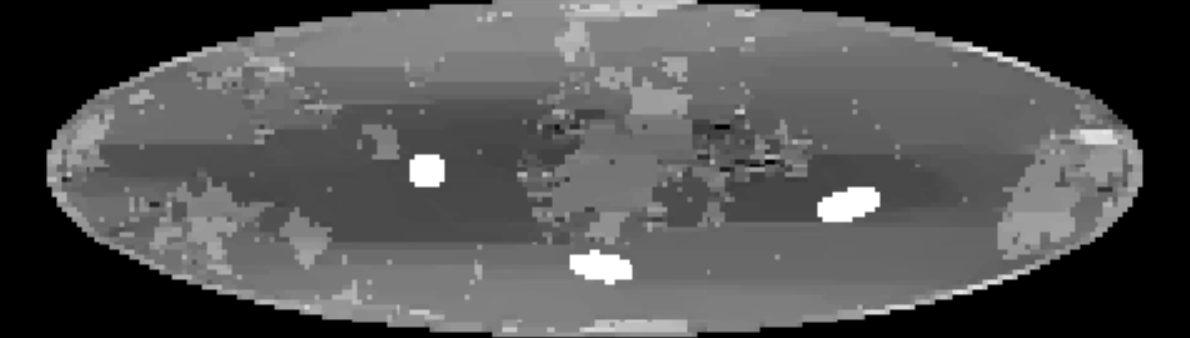}
			\hspace{-10pt} &
			\includegraphics[angle=180,width=0.22\textwidth]{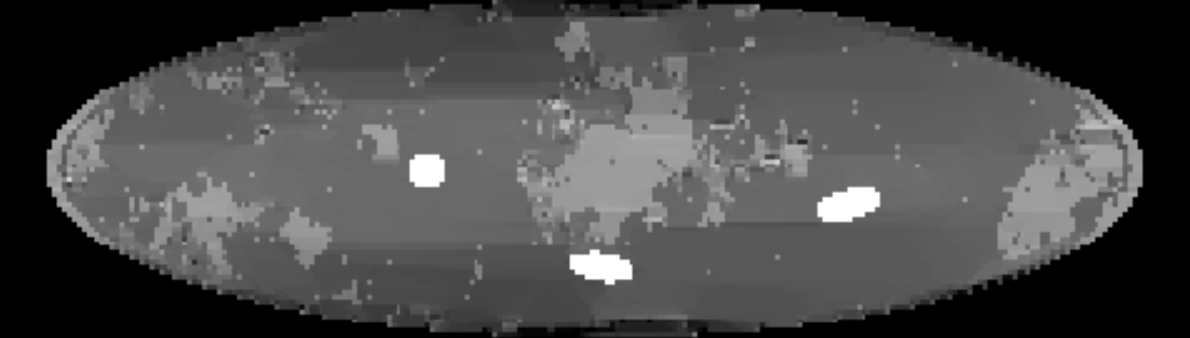} \vspace{0pt}
			\\ 
			\includegraphics[angle=180,width=0.22\textwidth,trim={120 0 110 60}, clip]{figures/para-select/breast_60D_mono_4_4.pdf}
			\hspace{-10pt} & 
			\includegraphics[angle=180,width=0.22\textwidth,trim={120 0 110 60}, clip]{figures/para-select/breast_60D_mono_0_0.pdf}
			\hspace{-10pt} & 
			\includegraphics[angle=180,width=0.22\textwidth,trim={120 0 110 60}, clip]{figures/para-select/breast_60D_mono_2_5.pdf}
			\hspace{-10pt} &
			\includegraphics[angle=180,width=0.22\textwidth,trim={120 0 110 60}, clip]{figures/para-select/breast_60D_mono_3_2.pdf}
			\\
			(a) & (b) & (c) & (d)	
	\end{tabular}
	\caption{Monochromatic images at 34 keV of the breast phantom obtained by use of the DTV algorithm from $60^\circ$ data with $(t_x^L, t^L_y, t^H_x, t^H_y)=(123.5, 224.0, 80.8, 156.0)$ (a), $(169.0, 224.0, 102.0, 156.0)$ (b), $(130.0, 224.0, 80.8, 165.8)$ (c), and $(130.0, 238.0, 85.0, 175.5)$ (d), respectively. Display window: [0.2, 0.35] cm$^{-1}$.}
	\label{fig:para-selection}
\end{figure*}

From data of the breast phantom over $60^\circ$, we perform image reconstructions for multiple values of $t^s_x$ and $t^s_y$  selected  and show in Fig.~\ref{fig:para-selection} monochromatic images at 34 keV obtained with four different sets of selected values of $t^s_x$ and $t^s_y$ .
It can be observed that the DTV-constraint parameters can impact image reconstruction. 
After sweeping through a range of the DTV parameters, we select empirically $t^s_x$ and $t^s_y$ that yield an image of minimized artifacts, as shown in column 1 of Fig.~\ref{fig:para-selection}, for image reconstruction in the case. The parameters for the reconstructions with the DTV algorithm are selected consistently in this way throughout this work.

\section{Basis and monochromatic images} \label{app:decomp}

Letting $\mathbf{f}_{m_L}$ and $\mathbf{f}_{m_H}$ denote monochromatic images at two different energy levels $m_L$ and $m_H$, we obtain from Eq.~\eqref{eq:decomp-mono}
\begin{eqnarray} \label{eq:decomp-image}
	\begin{pmatrix}
		{f}_{m_L i} \\
		{f}_{m_H i} \\
	\end{pmatrix}
	=
	\begin{pmatrix}
		{\mu}_{m_L 0} & {\mu}_{m_L 1} \\
		{\mu}_{m_H 0} & {\mu}_{m_H 1}
	\end{pmatrix}
	\begin{pmatrix}
		{b}_{0i}\\
		{b}_{1i}\\
	\end{pmatrix},
\end{eqnarray}
where ${f}_{m_L i}$ and ${f}_{m_H i}$ denote values at pixel $i$ of $\mathbf{f}_{m_L}$ and $\mathbf{f}_{m_H}$, and ${b}_{ki}$ the value at pixel $i$ of basis image $k$. 
In the work, images $\mathbf{f}^L$ and $\mathbf{f}^H$ are reconstructed directly from low- and high-kVp spectral data by use of either the FBP or DTV algorithm, and basis images can be estimated approximately from $\mathbf{f}^L$ and $\mathbf{f}^H$ (or from their corrected versions~\cite{maass_image-based_2009}.) 
Using $f^L_i$ and $f^H_i$ at pixel $i$ of $\mathbf{f}^L$ and $\mathbf{f}^H$ to {\it approximate} monochromatic images ${f}_{m_L i}$ and ${f}_{m_H i}$ in Eq.~\eqref{eq:decomp-image}, we obtain estimates of basis images as 
\begin{eqnarray} \label{eq:decomp-basis}
	\begin{pmatrix}
		{b}_{0i}\\
		{b}_{1i}\\
	\end{pmatrix}
	\approx
	\begin{pmatrix}
		{\mu}_{m_L 0} & {\mu}_{m_L 1} \\
		{\mu}_{m_H 0} & {\mu}_{m_H 1}
	\end{pmatrix}^{-1}
	\begin{pmatrix}
		{f}^L_{i} \\
		{f}^H_{i}
	\end{pmatrix}. 
\end{eqnarray}
As a result, if ${\mu}_{m_L k}$ and ${\mu}_{m_H k}$ ($k=0$ and 1) can be estimated, they can be used in Eq.~\eqref{eq:decomp-basis} for obtaining basis images $\mathbf{b}_k$ and subsequently monochromatic image $\mathbf{f}_m$ using Eq.~\eqref{eq:decomp-mono}.  
We consider in the work the interaction- and material-based methods, as described below, for computing $\mu_{m_L k}$ and ${\mu_{m_H k}}$ in estimation tasks of atomic number and iodine concentration, respectively.

In the interaction-based method, the photoelectric effect and Compton scattering are the basis functions of choice, which cannot be mutually exclusively present within a ROI in the image. As a result, we estimate $\mu_{m_L k}$ and $\mu_{m_H k}$ by estimating the energy levels $m_L$ and $m_H$ of a common material within pre-determined calibration ROIs from knowledge of $\mathbf{f}^L$ and $\mathbf{f}^H$. For a calibration ROI containing a common material, e.g., water, the mean values are first obtained by averaging $f_i^L$ or $f_i^H$ over pixels within the ROI. Using the mean values, we then determine effective energy levels $m_L$ and $m_H$ that yield the closest values of linear attenuation coefficients for the material within the ROI by using the NIST table~\cite{hubbell_tables_2004}. Plugging $\mu_{m_L k}$ and $\mu_{m_H k}$ estimated into Eq.~\eqref{eq:decomp-basis}, we obtain basis images within the ROI. Moreover, while $\mu_{m_L k}$ and $\mu_{m_H k}$ are estimated only from knowledge of $\mathbf{f}^L$ and $\mathbf{f}^H$ within the calibration ROI, they are used immediately to compute the basis images at pixels that are not within the calibration ROI. Combination of the basis images estimated within and outside the calibration ROI yields the basis images for the entire image array.
Finally, using the basis images estimated and $\mu_{mk}$ at energy level $m$, calculated using $1/E^3$ (where $E$ is the energy corresponding to energy bin $m$) and the Klein-Nishina formula~\cite{attix2008introduction}, we then obtain a monochromatic image for all of the pixels in the image array by using Eq.~\eqref{eq:decomp-mono}.

In the material-based method, two basis images $\mathbf{b}_0$ and $\mathbf{b}_1$ represent two different materials, which are breast tissue and iodine in our study with the breast phantom. Two calibration ROIs 0 and 1, corresponding to these two basis materials, are first selected in the breast phantom. For calibration ROI 0 in the breast phantom containing breast tissue, the pixel values of $\mathbf{b}_0$ and $\mathbf{b}_1$ are 1 and 0, respectively. Conversely, for calibration ROI 1 in the breast phantom containing iodine, the pixel values of $\mathbf{b}_0$ and $\mathbf{b}_1$ are thus 0 and 1, respectively. Therefore, for calibration ROIs 0 and 1, individual estimates of $\mu_{m_L k}$ and $\mu_{m_H k}$ for $k=0$ and $1$ are obtained simply as the corresponding pixel values of $\mathbf{f}^L$ and $\mathbf{f}^H$ within calibration ROIs 0 and 1. We then compute the final estimates by averaging the individual estimates over pixels within the calibration ROIs, e.g., $\mu_{m_L k} = \sum_{i \in U_k} f^L_i / I_k$, where $U_k$, with size $I_k$, denotes the set of those pixels within ROI $k$ ( $k=0$ or $1$). Plugging the final estimates of $\mu_{m_L k}$ and $\mu_{m_H k}$ into Eq. \eqref{eq:decomp-basis}, we obtain estimates of basis images from knowledge of $\mathbf{f}^L$ and $\mathbf{f}^H$ for pixels outside the ROIs. Combination of the basis images estimated within and outside the calibration ROIs yields the basis images for the entire image array. Finally, using the basis images estimated and $\mu_{mk}$ at energy level $m$, looked up from the NIST table, we then obtain a monochromatic image for all of the pixels in the image array by using Eq.~\eqref{eq:decomp-mono}.

\section{Estimation of physical quantities} \label{app:tasks}

{\it Estimation of atomic number:}
For the energy range of our interest, monochromatic image $\mathbf{f}_m$ in Eq.~\eqref{eq:decomp-mono} can be interpreted as a linear combination of two interaction-based basis images, i.e., the photoelectric effect (PE) ($k=0$) and Compton scattering (KN) ($k=1$)~\cite{alvarez_energy-selective_1976,attix2008introduction}.
As a result, from the two basis images, we can estimate atomic number $\hat{z}_i$ at pixel $i$ as~\cite{alvarez_energy-selective_1976, ying2006dual}
\begin{eqnarray} \label{eq:z-log}
	\ln \hat{z}_i = \ln c + n \ln \left(\frac{b_{0i}}{b_{1i}}\right),
\end{eqnarray}
where $b_{0i}$ and $b_{1i}$ are values at pixel $i$ for basis images $\mathbf{b}_0$ and $\mathbf{b}_1$ corresponding to the PE and KN interaction types, respectively, which are estimated from knowledge of $\mathbf{f}^L$ and $\mathbf{f}^H$ by use of the interaction-based method described in \ref{app:decomp}, and constants $c$ and $n$ are related to the physical modeling of the cross sections for PE and KN~\cite{alvarez_energy-selective_1976,attix2008introduction}.
As Eq.~\eqref{eq:z-log} depicts an affine relationship between the atomic number and the ratio of the basis image values in the log-log domain, 
constants $c$ and $n$ can be estimated using a linear regression in the log-log domain and calibration materials with known atomic numbers, such as single-element materials, in the imaged subject. For each material, a single atomic number is computed as the mean value of estimated atomic numbers within the ROI.


{\it Estimation of iodine concentration:}
In the study of the breast phantom, we use the material-based method described in \ref{app:decomp} to estimate 
$\mathbf{b}_0$ and $\mathbf{b}_1$ for breast tissue and iodine contrast agent from knowledge of $\mathbf{f}^L$ and $\mathbf{f}^H$. An affine relationship is assumed between basis image $\mathbf{b}_1$ of the iodine contrast agent and the estimated iodine concentration $\hat{\beta}_i$ at pixel $i$ as
\begin{eqnarray} \label{eq:beta-linear}
	\hat{\beta}_i = \gamma b_{1i} + \tau,
\end{eqnarray}
where $b_{1i}$ is the image value at pixel $i$ of the iodine basis, and $\gamma$ and $\tau$ are the linear coefficients,
which can be estimated using a linear regression in a calibration scan with known iodine concentrations.
For each iodine contrast agent ROI, a single concentration level is computed as the mean value of estimated iodine concentrations in the ROI.

\section*{References}

\bibliographystyle{IEEEtran}
\bibliography{journals-abrv,limAngDECT}

\end{document}